\documentclass[]{scrartcl}

\usepackage{amsmath,amssymb,graphicx,xcolor}

\usepackage{dutchcal}
\usepackage{dsfont}
\usepackage{float}
\usepackage{hyperref}
\usepackage{graphicx}
\usepackage[numbers,sort&compress,square]{natbib}
\usepackage{authblk}

\newcommand \hmu {\hat{\mu}}
\newcommand \imag {\mathrm{i}}
\newcommand \pbp {\langle\bar{\psi}\psi\rangle}

\begin{document}

\title{Overview of the QCD phase diagram - Recent progress from the lattice}

\author[]{Jana N. Guenther\thanks{jana.gunther@univ-amu.fr}}
\affil[]{Aix Marseille Univ., Université de Toulon, CNRS, CPT, Marseille, France}

\date{12/17/2020}

\maketitle

\begin{abstract}
 In recent years there has been much progress on the investigation of the QCD phase diagram with lattice QCD simulations. In this review I focus on the developments in the last two years. Especially the addition of external influences or new parameter ranges yield an increasing number of interesting results. I discuss the progress for small, finite densities from both extrapolation based methods (Taylor expansion and analytic continuation for imaginary chemical potential) and Complex Langevin simulations, for heavy quark bound states (quarkonium), the dependence on the quark masses (Columbia plot) and the influence of a magnetic field. Many of these conditions are relevant for the understanding of both the QCD transition in the early universe and heavy ion collision experiments which are conducted for example at the LHC and RHIC.
\end{abstract}

\section{Introduction\label{sec:Introduction}}
The strong interaction between quarks and gluons in the standard model is described by Quantum Chromo Dynamics (QCD).
The investiagation of the phase diagram of QCD has been an active subject for many years. A special focus of this research is on the transition between hadrons at low temperature and the quark gluon plasma at high temperature. The nature of the transition as an analytic crossover (Ref.~\cite{Aoki:2006we,Aoki:2006br,Aoki:2009sc,Borsanyi:2010bp,Bhattacharya:2014ara,Bazavov:2011nk}) has been known since 2006 (Ref.~\cite{Aoki:2006we,Aoki:2006br}). However, different external influences may change the nature of this transition. One famous example is the addition of a chemical potential. The crossover has been established for vanishing baryon chemical potential only, which means for a setting with the same number of quarks and anti-quarks. This setting is a good approximation for the QCD transition in the early universe. However, the experimental investigations of the QCD transition rely on heavy ion collision experiments which produce the quark gluon plasma with a finite density. In addition, the collided matter goes through several stages and is not permanently in an equilibrium. Since, at the moment, full QCD cannot be solved under that conditions one has to take other paths. Certain parameter ranges can be described efficiently by effective theories. For other situations we depend on phenomenological models, which inturn require input information that can be gained from lattice simulations.

The progress on the phase diagram form lattice QCD corresponds well with, and is in part triggered by, an increasing number of available experimental results. At the moment most erxperimental results for heavy ion collision experiments are generated either by the Large Hadron Collider (LHC) at Cern in Genf, Switzerland or the Relativistic Heavy Ion Collider (RHIC) at the Brookhaven national laboratory in New York, USA. While the results form the LHC are at low densities, RHIC accesses information for lager chemical potential. However, there are several upcoming facilities that will allow inside to even higher densities. Some of them are for example the Nuclotron-based Ion Collider fAcility (NICA) at the Joint Institute for Nuclear Research (JINR) in Dubna, Russia, the Compressed Baryonic Matter Experiment (CBM) at the Facility for Antiproton and Ion Research in Europe (FAIR) at Darmstadt, Germany and the J-PARC heavy ion project (J-PARC-HI) at the Japan Proton Accelerator Research Complex (J-PARC) in Tokai, Japan.

After this introduction, I will briefly describe the stages of a heavy ion collision experiment in section~\ref{sec:HIC}. Linking lattice QCD to  heavy ion collision experiments requires results at low finite density which I discuss in section~\ref{sec:lowDensity}. The main challenge in this parameter range is the infamous sign problem which is discussed in section~\ref{sec:signProblem}. For continuum extrapolated, physical results, one therefore has to rely on extrapolations (section~\ref{sec:anaCont}) form vanishing (Taylor method) or imaginary chemical potential. One method that is getting relatively close to direct lattice simulations at finite density are Complex Langevin simulations, which are therefore reviewed in section~\ref{sec:CL}. Other methods that are not discussed in this review are for example reweighting techniques \cite{Barbour:1997ej,Fodor:2001au,Fodor:2001pe,Csikor:2002ic}, density of state methods \cite{Fodor:2007vv,Alexandru:2014hga}, using the canonical ensemble \cite{Alexandru:2005ix,Kratochvila:2005mk,Ejiri:2008xt}, formulations with dual variables \cite{Gattringer:2014nxa} or Lefschetz thimbles \cite{Scorzato:2015qts, Alexandru:2015xva}.

Another way to gain information on finite density QCD is discussed in section~\ref{sec:ELT}, the lattice simulations of effective field theories. Here, I focus, with section~\ref{sec:Quarkonium} on the results for heavy quark bound states the so called heavy quarkonium.

Despite the importance at finite density, also for zero density there are interesting influences to consider. One of them is the dependence of the transition type on the quark masses. It is often summarized in the Columbia plot, which is discussed in section~\ref{sec:Columbia}. The most interesting areas of the Columbia plot are the upper right (section~\ref{sec:upperCorner}) and lower left corner (section~\ref{sec:lowerCorner}). Due to the low quark masses, and the resulting expensive computations, different tactics are employed in the study of the lower left corner of the Columbia plot. In this review I discuss the use of imaginary chemical potential in section~\ref{sec:immuColumbia} and the variation of the number of quark flavors in section~\ref{sec:Nf}. Due to the lack of continuum extrapolated results in this parameter range, I provide an overview over the numbers obtained from different lattice actions and lattice spacings in section~\ref{sec:overview}.

There have been many results for yet another region. The addition of a magnetic field to finite temperature QCD does not suffer from a sign problem and is also relevant for the understanding of heavy ion collision experiments. The progress recently made on our understanding of the phenomena triggered by a magnetic field is discussed in section~\ref{sec:magneticFields}.

Finally, as common practice, this review closes with a conclusion. Here I hope to convince you that finite temperature lattice QCD and especially the investigation of the QCD phase diagram is a fascinating, thriving topic, of which I can only cover parts in this review.

\section{Heavy ion collisions\label{sec:HIC}}
\begin{figure}
\begin{center}
 \includegraphics[width=.9\textwidth]{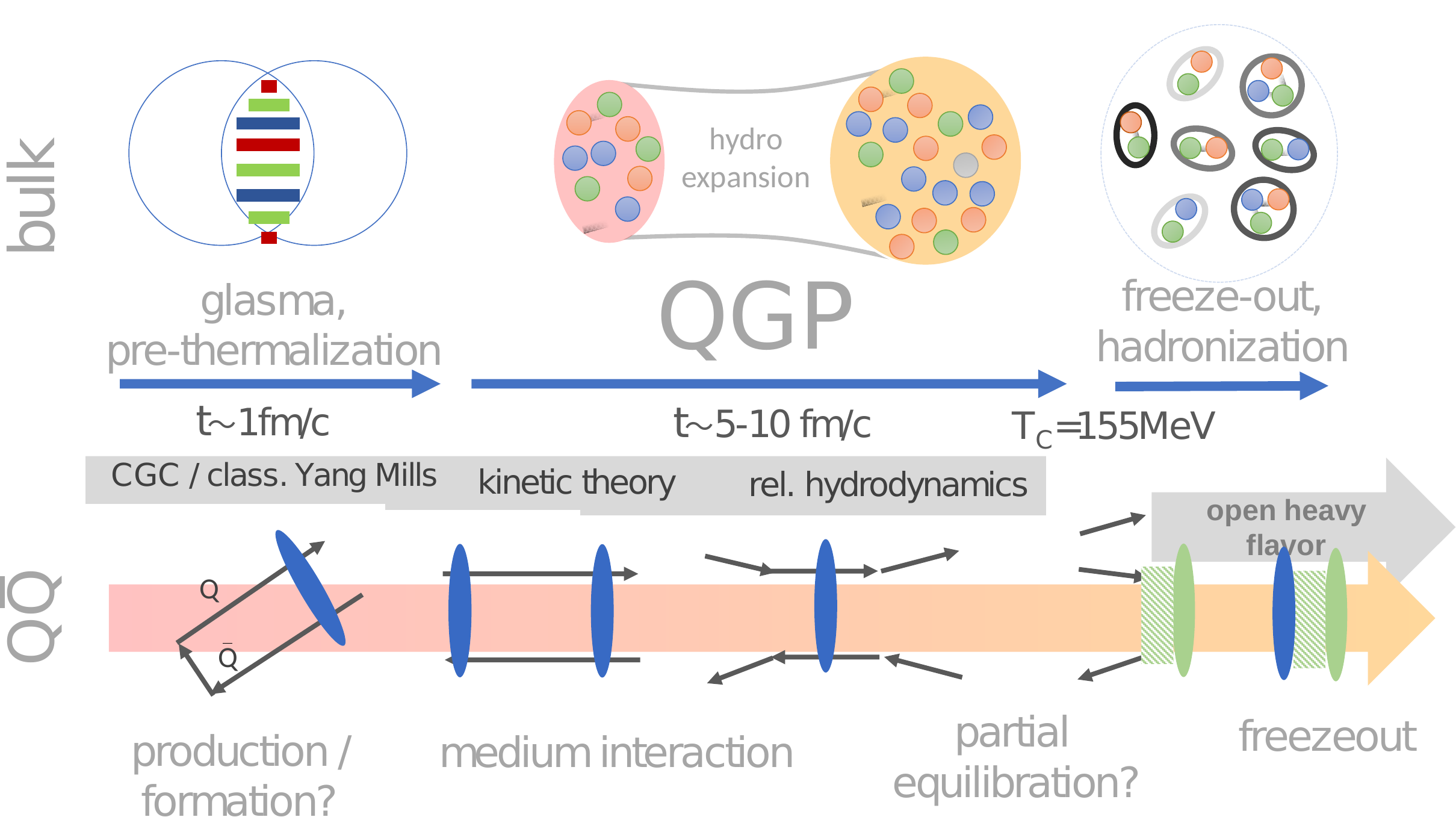}
\end{center}
\caption{(Ref.~\cite{Rothkopf:2019ipj}) Overview of of the different stages of a heavy ion collision. In grey the different effective theories are shown which are used to describe the different stages (see section \ref{sec:Quarkonium}).  \label{fig:HIC}}
\end{figure}

As the experimental realization of QCD thermodynamics is strongly linked with heavy ion collision experiments, I will take a small detour to explain the different stages occurring during such a collision which are depicted in figure~\ref{fig:HIC}. In a heavy ion collision various nuclei can be used. Prominent examples are gold and lead which have been used at RHIC and LHC. Two beams of nuclei are accelerated to relativisic velocities. 

When the two beams collide, they form a non thermaliesed state with strongly interacting fields, which is called a glasma and commonly treated in the color glass condensate framework (for details on this framework see Ref.~\cite{Gelis:2010nm,Gelis:2012ri}). Here the glasma is seen in the infinite momentum framework where there are partons of the nuclei, valence quarks and pairs of sea quarks. A main challenge  is the description of different momentum scales. Due to the un-thermaliesed nature of this state, it is not accessible for lattice QCD simulations.

In the next stage, the further fragmentation of the partons into quarks and gluons leads to the quark gluon plasma. This is a strongly interacting state of deconfined quarks and gluons. While first it was expected to be similar to an electromagnetic plasma, by now it is well established that is more similar to a strongly interacting fluid. Therefore, it is often described by relativistic hydrodynamics. 

The quark gluon plasma expands and cools down at the same time. When it reaches temperatures close to the QCD transition temperature, the deconfined quarks and gluons have to recombine to colour-neutral hadrons. At this time, the chemical aboundance of the hadrons is determined. The corresponding temperature is called the chemical freeze-out temperature. However, the hadrons can still exchange energy and momentum until the point where the kinetic freeze-out takes place. 

An important question is, whether at the stage of the Quark-Gluon-plasma and the following transition to hadrons, the states thermalize. Only if this is the case, results from lattice QCD, which simulates thermal equilibrium, are directly applicable.

\clearpage
\section{Low Finite density\label{sec:lowDensity}}

\begin{figure}
 \centering
 \includegraphics[width=0.7\textwidth]{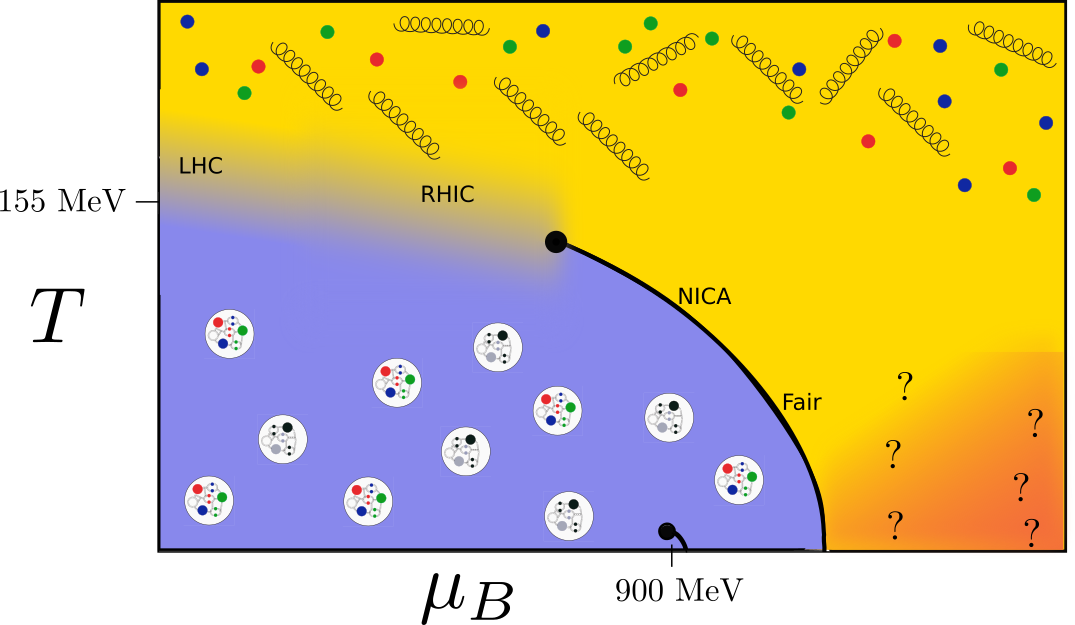}
 \caption{A schematic view on the $T$-$\mu_B$-plane of the QCD phase diagram. \label{fig:phaseDiagram}}
\end{figure}

One representation of the QCD phase diagram is in the $T$-$\mu_B$-plane (figure \ref{fig:phaseDiagram}). The $\mu_B = 0$ axis is well investigated by lattice QCD. For low chemical potential and temperature, there is a hadronic phase of colour neutral bound states. At high temperatures the effective degrees of freedom are the quarks and gluons. This phase is called the quark gluon plasma. These two phases are separated by a crossover at low and zero chemical potential. The transition is expected to change into a first order transition for higher $\mu_B$ with a critical second order point in between. The position of the critical point is under active investigation both by heavy ion collision experiments and by theoretic calculations. 
For very large chemical potentials futher fascinating phenomena, like a colour super conducting phase \cite{Rischke:2003mt}, are expected.

To analyse the quark gluon plasma that is created in heavy ion collision experiments at the LHC or RHIC, a theoretical understanding of the quark gluon plasma in QCD is needed. In the region of the deconfinement transition, QCD can not be studied with perturbative methods. To get non-perturbative results with a controlled error, one has to turn to lattice QCD. At the moment, direct simulations that are continuum extrapolated and at physical quark masses are restricted to vanishing or imaginary chemical potential due to the infamous sign problem. On the other hand, the collisions especially at RHIC and upcoming heavy ion collision facilities like Fair and NICA take place away from the axis of zero $\mu_B$. Therefore, information in that region are needed. 

\subsection{The sign problem \label{sec:signProblem}}
For simulation without chemical potential  there are on average the same number of particles and antiparticles. Therefore, the expectation value of the  overall baryon number density $\langle n_B \rangle$ is zero. To describe a system with finite baryon density, we need to introduce a finite quark chemical potential $\mu_q$ to the Lagrangian. In the continuum this is relatively simple to achieve by adding a term of the form $\overline{\psi} \gamma_4 \psi$. However, on the lattice, it is not that simple. Adding a similar term to the Dirac operator leads to a divergent energy density in the continuum limit, which is clearly unphysical. Instead one follows the idea of Hasenfratz and Karsch in \cite{Hasenfratz:1983ba} where the chemical potential is understood as the temporal component of  a  vector field.  The temporal hopping term has then the form
\begin{eqnarray}
 &-&\frac{1}{2a} \sum_{n\in \Lambda} \left( e^{\frac{\mu}{T}} (\eta_4)_{\alpha\beta} U_{\hat 4}(n)_{ab}\delta_{n+\hat 4,m} \right. \nonumber\\
 &+& \left.  e^{-\frac{\mu}{T}} (\eta_4)_{\alpha\beta} U_{\hat 4}^\dagger(n-\hat4)_{ab}\delta_{n-\hat 4,m}\right).
\end{eqnarray}
It recovers the original action for $\mu = 0$ and with $\frac{\mu}{T} = a \mu N_t$  it reproduces the   correct density term at linear order in $a\mu$. However, this term breaks the $\gamma_5$-hermiticity of the Dirac operator
\begin{equation}
 \gamma_5 D \gamma_5 = D^\dagger
\end{equation}
and leads  to a complex fermion determinant for  real $\mu$.
The fermion determinant enters in the Boltzmann weight factor which has to be positive to allow a Monte Carlo simulation. 
There have been several ideas on how to obtain results at real finite chemical potential like reweighting techniques \cite{Barbour:1997ej,Fodor:2001au,Fodor:2001pe,Csikor:2002ic}, Taylor expansion \cite{Allton:2002zi,Allton:2005gk,Gavai:2008zr,Basak:2009uv,Kaczmarek:2011zz}, density of state methods \cite{Fodor:2007vv,Alexandru:2014hga}, using the canonical ensemble \cite{Alexandru:2005ix,Kratochvila:2005mk,Ejiri:2008xt}, formulations with dual variables \cite{Gattringer:2014nxa}, Lefschetz thimbles \cite{Scorzato:2015qts, Alexandru:2015xva} or complex Langevin \cite{Seiler:2012wz,Sexty:2013ica}.

\subsection{Extrapolation to finite $\mu$ \label{sec:anaCont}}
Instead of conducting direct simulations, for small densities it is possible to obtain results with physical quark masses by extrapolation. Since at $\mu_B=0$ the transition is a crossover (Ref.~\cite{Aoki:2006we,Aoki:2006br,Aoki:2009sc,Borsanyi:2010bp,Bhattacharya:2014ara,Bazavov:2011nk}), certain observables can be parameterized by an analytic function in the vicinity of zero as well. This fact can be exploited, by using results at zero (Ref.~\cite{Allton:2002zi,Allton:2005gk,Gavai:2008zr,Basak:2009uv,Borsanyi:2011sw,Borsanyi:2012cr,Bellwied:2015lba,Ding:2015fca,Bazavov:2017dus,Bazavov:2018mes,Bazavov:2020bjn,Bonati:2018nut,Kaczmarek:2011zz,Endrodi:2011gv}) or imaginary chemical potential (Ref.~\cite{deForcrand:2002hgr,DElia:2002tig,DElia:2009pdy,Cea:2014xva,Bonati:2014kpa,Cea:2015cya,Bonati:2015bha,Bellwied:2015rza,DElia:2016jqh,Gunther:2016vcp,Alba:2017mqu,Vovchenko:2017xad,Bonati:2018nut,Borsanyi:2018grb,Borsanyi:2020fev}), to find a function that can be extrapolated to real, positive chemical potential. If one aims at gaining information from zero chemical potential, one uses the higher order derivatives of the partition function to calculate the Taylor coefficients of an expansion that can be continued to finite $\mu_B$. This method is therefore often called Taylor method.

\begin{figure}
 \centering
 \includegraphics[width=0.7\textwidth]{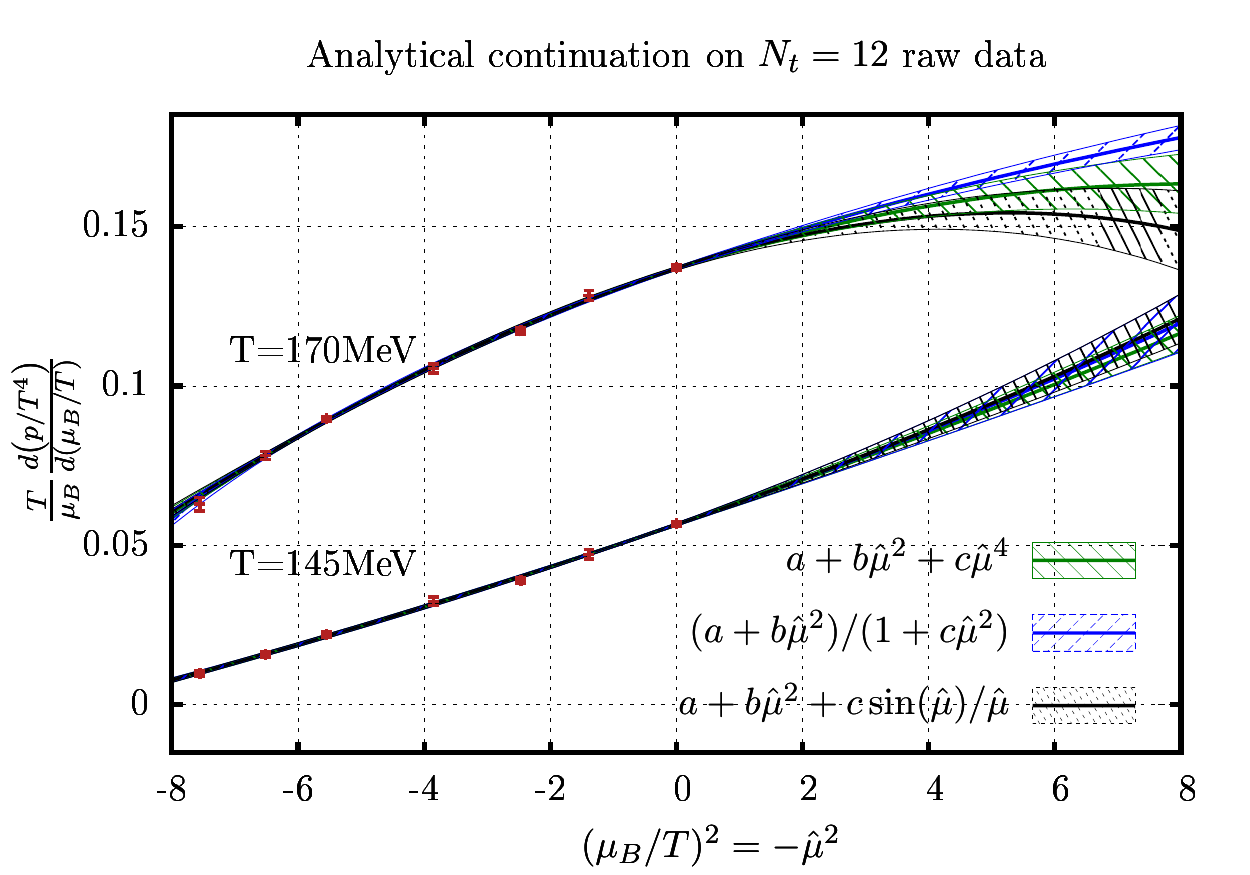}
 \caption{(Ref.~\cite{Pasztor:2016iqd}) Illustration of the analytic continuation from imaginary chemical potential. Data points generated with purley imagniary $\mu_B$ can be fitted as a function in $\mu_B^2$ and then extrapolated from $\mu_B^2\leq0$ to $\mu_B>0$.\label{fig:imuCartoon}}
\end{figure}
The other possibility to determine a function that can be continued to finite chemical potential is the use of simulations at purely imaginary chemical potential, where there is no sign problem. As illustrated in figure~\ref{fig:imuCartoon}, the data points can be described by a function of $\mu_B^2$. For imaginary chemical potential $\mu_B^2<0$ and this is the area where simulation results are available. The resulting data points can then be fitted with a function which can be extrapolated to $\mu_B>0$ and therefore to real, positive chemical potential. Since not much knowledge of the concrete shape of the describing function is available, the choice of the fit function influences the result at real chemical potential. Figure~\ref{fig:imuCartoon} shows three different fit functions, namely
\begin{eqnarray}
 f_1(\mu,T) &=&a+b\left( \frac{\mu}{T} \right)^2+c\left( \frac{\mu}{T} \right)^4,\\
 f_2(\mu,T) &=& \frac{a+b\left( \frac{\mu}{T} \right)^2}{1+c\left( \frac{\mu}{T} \right)^2},\\
 f_3(\mu,T) &=& a+b\left( \frac{\mu}{T} \right)^2+c \frac{\sin\left(  \frac{\mu}{T} \right)}{\frac{\mu}{T}},
 \end{eqnarray}
which each having three fit parameter $a$, $b$, $c$. All three functions describes the available data well. However their continuation to $\mu_B>0$ varies, especially for a temperature of $T=170$~MeV. This variation has, therefore, to be taken into account as a systematic error.

When utilising simulations at imaginary chemical potential, one exploits the fact that the $\gamma_5$-hermicity of the Dirac operator with a chemical potential reads
\begin{equation}
 \gamma_5 D(\mu) \gamma_5 = D^\dagger(-\mu).
\end{equation}
This means that in the action, the term adding the chemical potential $e^{-\frac{\mu}{T}} = f$ is replaced by $(e^{\frac{\mu}{T}})^\ast = \frac{1}{f^\ast}$ when switching to purely imaginary $\mu$. For the determinant this yields
\begin{equation}
 \det\left( D(f) \right) = \det\left( D\left( \frac{1}{f^\ast} \right) \right),
\end{equation}
meaning that the determinant is only real if 
\begin{equation}
 f =  \frac{1}{f^\ast}.
\end{equation}
For real $f$ this is only fulfilled if $f = e^{-\frac{\mu}{T}} = 1$, and, therefore, $\mu = 0$. However, if $\mu = \imag \mu^I$ is chosen to be purely imaginary, this yields
\begin{equation}
 f = e^{-\imag \frac{\mu^I}{T}} = \left( e^{\imag \frac{\mu^I}{T}} \right)^\ast = \frac{1}{f^\ast}.
\end{equation}
Thus the determinant is real for a purely imaginary chemical potential.
When using simulations at imaginary chemical potential, less derivatives are needed than for the Taylor method. Instead, one uses different fit functions to describe a data set for several imaginary chemical potentials. This method is sometimes called analytical continuation, but, since also the Taylor method relies on an analytical continuation, I will refer to it as imaginary chemical potential method. The different fit functions can lead to different results which leads to the requirement of a careful systematic analysis. The behaiviour of one type of fit functions, the Padé approximation, is discussed in Ref.~\cite{Pasztor:2020dur}.

\paragraph{The transition temperature:}
\begin{figure}
 \centering
 \includegraphics[width=0.7\textwidth]{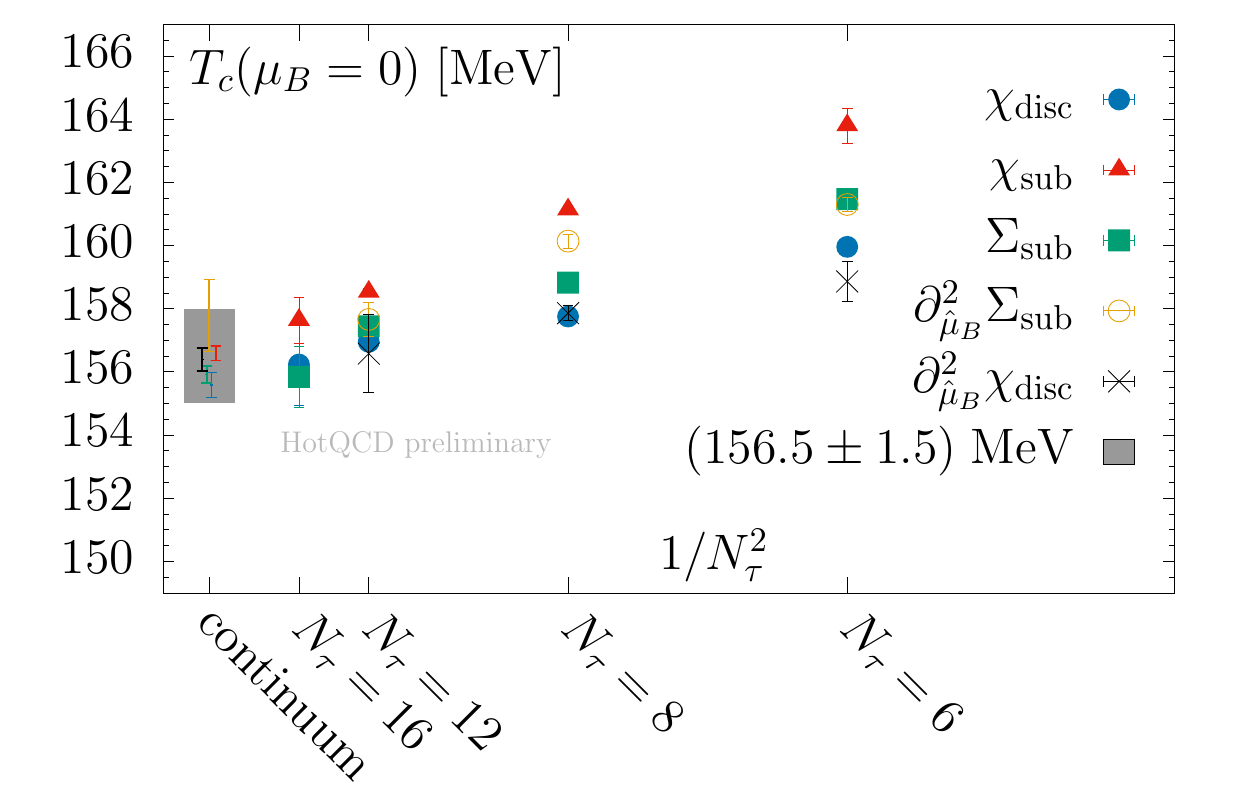}
 \caption{(Ref.~\cite{Steinbrecher:2018phh}) The continuum extrapolation of five different definitions of the transition temperature. Towards the continuum limit the different definitions converge, yielding a combined result of $T_c=(156\pm1.5)$~MeV.\label{fig:Tc0HotQCD}}
\end{figure}

A common observable to extrapolate is the transition temperature. Because of the crossover nature of the transition, its definition is ambiguous. Typical definitions are, for example, the peak of the chiral susceptibility as a function of temperature or the inflection point of the chiral condensate. Often different definitions yield consistent results within the available precision. However for an analytic transition this is not guaranteed in contrast to the situation for phase transition. 

A high precision determination of the transition temperature at $\mu_B=0$ from five different observables was done in Ref.~\cite{Steinbrecher:2018phh}. As can be seen in figure~\ref{fig:Tc0HotQCD}, all five definitions have the same continuum limit within the available precision. This yields a combined value of $T_c=(156\pm1.5)$~MeV.

\begin{figure}
 \centering
 \includegraphics[width=0.7\textwidth]{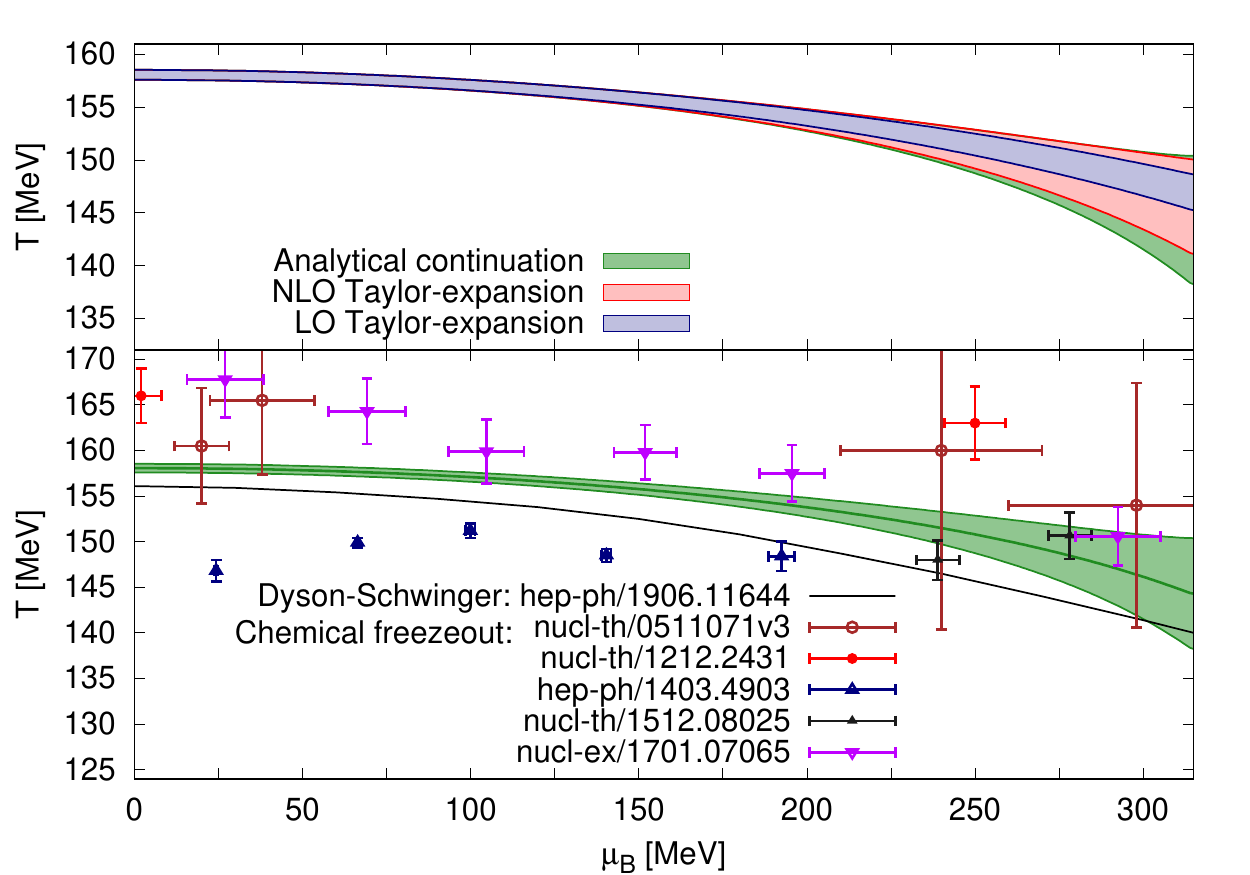}
 \caption{(Ref.~\cite{Borsanyi:2020fev}) Top: A comparison between the extrapolated transition temperature obtained from analytical continuation from imaginary chemical potential with different functions, compared to the result from extrapolating with the leading order (LO) order next to leading order (NLO) Taylor coefficients. Bottom: A comparison of the extrapolated transition temperature to recent results from calculations with Dyson-Schwinger-Equations \cite{Isserstedt:2019pgx} and the freeze-out temperature from heavy ion collision experiments  \cite{Andronic:2005yp,Becattini:2012xb,Alba:2014eba,Vovchenko:2015idt,Adamczyk:2017iwn}.   \label{fig:tcline}}
\end{figure}

A new definition as the peak of the chiral susceptibility as a function of the chiral condensate (instead of the more common definition as function of the temperature) was introduced this year in Ref.~\cite{Borsanyi:2020fev}. It allows for a precise extraction of the transition temperature and, therefore, also for an improvement in the extrapolation. The results obtained from simulations at imaginary chemical potential are shown in figure~\ref{fig:tcline}.  They are continuum extrapolated from the three lattices with sizes $40^3\times10$, $48^3\times12$ and $64^3\times16$. The extrapolation was done with two different functions,
\begin{equation}
 T_c=1 + \hmu_B^2 \left( a+ \frac{d}{N_t^2}\right) + \hmu_B^4 \left( b+ \frac{e}{N_t^2}\right) + \hmu_B^6 \left( c+ \frac{f}{N_t^2}\right)
\end{equation}
and
\begin{equation}
 T_c=\frac{1}{1 + \hmu_B^2 \left( a+ \frac{d}{N_t^2}\right) + \hmu_B^4 \left( b+ \frac{e}{N_t^2}\right) + \hmu_B^6 \left( c+ \frac{f}{N_t^2}\right)}.
\end{equation}

The transition temperature from imaginary chemical potential (\emph{Analytical continuation} in the top of figure~\ref{fig:tcline}) is compared to the Taylor expansion at leading and next to leading order, as well as to results from truncated Dyson-Schwinger equations (Ref.~\cite{Isserstedt:2019pgx}) and various determination of the chemical freeze-out temperature in heavy ion collision experiments~(Ref.~\cite{Andronic:2005yp,Becattini:2012xb,Alba:2014eba,Vovchenko:2015idt,Adamczyk:2017iwn}).

\begin{figure}
 \centering
 \includegraphics[width = 0.7\textwidth]{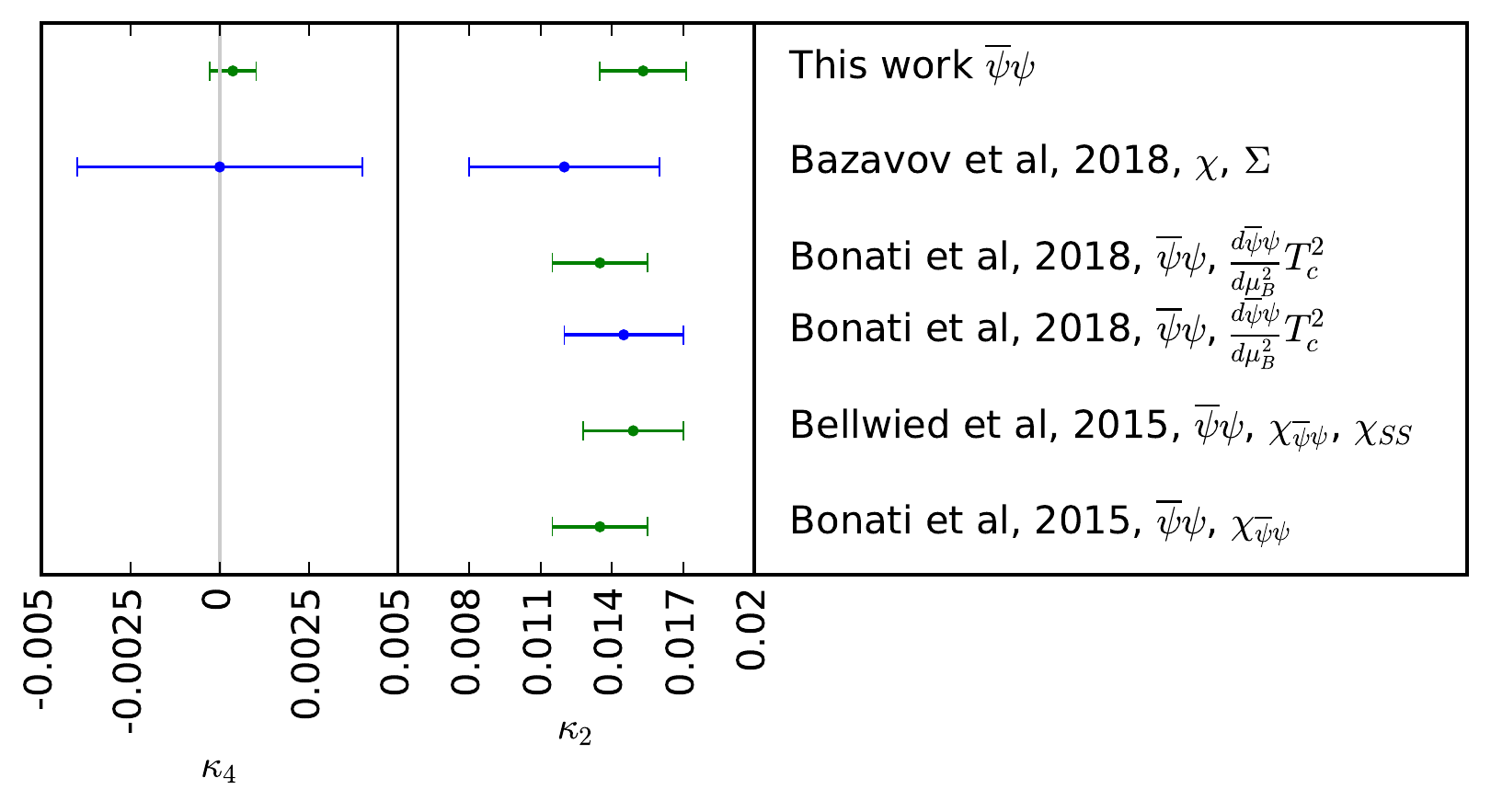}
 \caption{(Ref.~\cite{Borsanyi:2020fev}) A comparison of recent results for $\kappa_2$ and $\kappa_4$ as defined in equation~(\ref{eq:kappa}). Ref.~\cite{Bazavov:2018mes} and the lower point of Ref.~\cite{Bonati:2018nut} use the Taylor method and are depicted in blue. The green points, which are the result from Ref.~\cite{Borsanyi:2020fev}, the upper point of Ref.~\cite{Bonati:2018nut}, Ref.~\cite{Bellwied:2015rza} and Ref.~\cite{Bonati:2015bha} used lattice data at imaginary chemical potential. \label{fig:kappaLambda}}
\end{figure}

The transition temperature at finite chemical potential $T_c(\mu_B)$ normalized by the transition temperature at zero chemical potential $T_c(0)$ is often parameterized by:
\begin{equation}
 \frac{T_c(\mu_B)}{T_c(0)} = 1 - \kappa_2 \left(\frac{\mu_B}{T_c} \right)^2  - \kappa_4 \left(\frac{\mu_B}{T_c} \right)^4   + \mathcal O(\mu_B^6).\label{eq:kappa}
\end{equation}
While there is a long history of the determination of $\kappa_2$ \cite{deForcrand:2002hgr, Kaczmarek:2011zz, Endrodi:2011gv, Bonati:2015bha, Bellwied:2015rza, Bonati:2018nut}, results for $\kappa_4$ \cite{Bazavov:2018mes, Borsanyi:2020fev} only recently became available, both from the Taylor and imaginary potential method. It emerges that $\kappa_4$ is significantly smaller than $\kappa_2$. Within its error it is still compatible with zero. A comparison of the recent determinations of both $\kappa_2$ and $\kappa_4$ is shown in figure~\ref{fig:kappaLambda}. 

Instead of an expansion in the baryon chemical potential, an expansion in isospin, strangeness or charge chemical potential is also an option.  They were studied in detail in Ref.~\cite{Bazavov:2018mes}. There are two very similar possibilities to chose $\mu_S$ and $\mu_Q$ for the expansion in $\mu_B$ that are commonly used. The first is the choice of  a purely baryon chemical potential ($\mu_S = \mu_Q = 0$). This conditions have been used in Ref.~\cite{Bonati:2015bha}. The second possible choice is the set up at the strangeness neutral point. In this case $\mu_S$ and $\mu_Q$ are chosen in a way that $\langle n_S \rangle = 0$ and $0.4\langle n_B \rangle=\langle n_Q \rangle$ to match the conditions in heavy ion collision experiments. This was done in Ref.~\cite{Bellwied:2015rza,Bonati:2018nut}. In all cases the two values agree within the error. 

\begin{figure}
 \begin{center}
 \includegraphics[width = 0.7\textwidth]{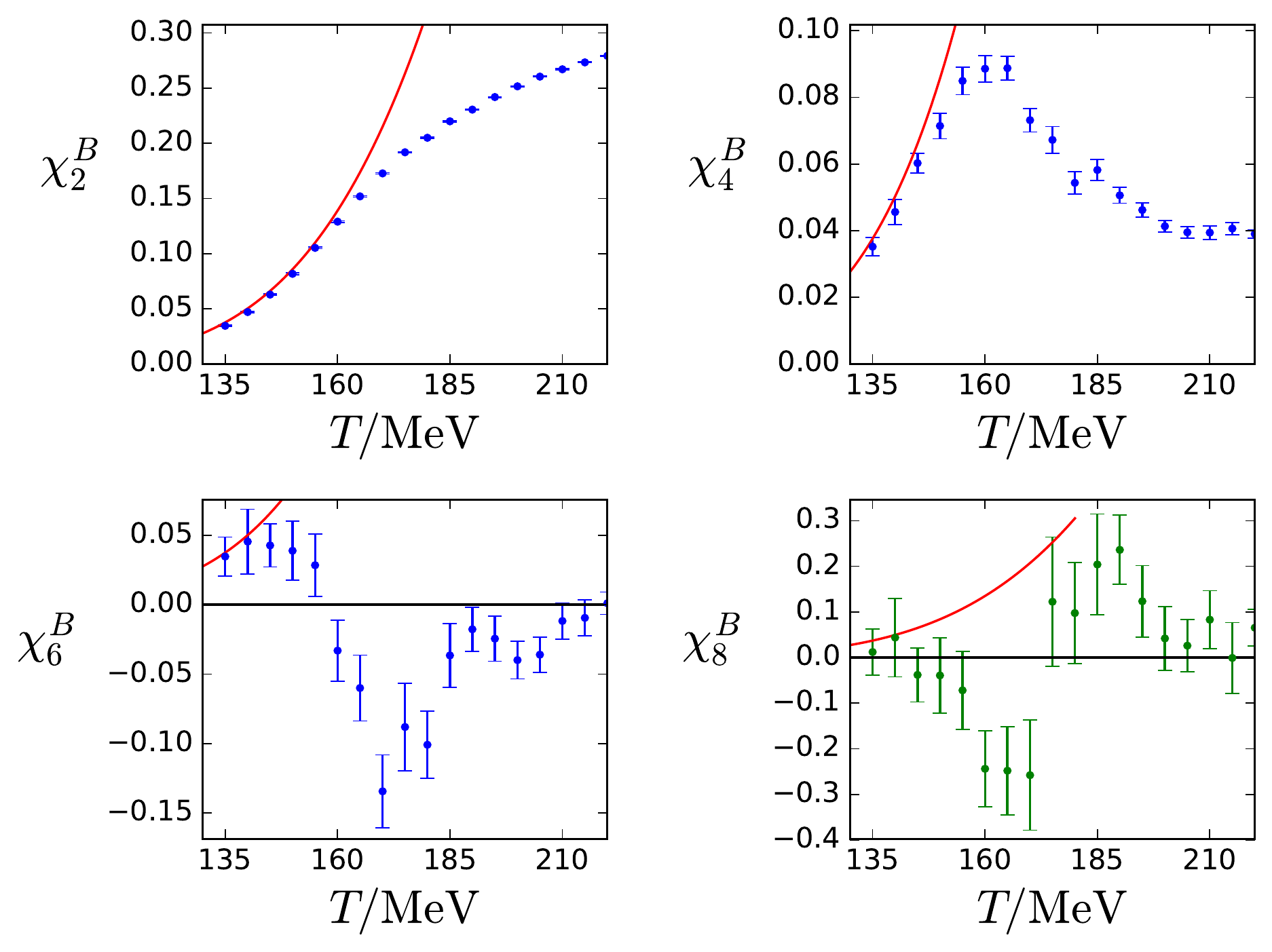}
 \end{center}
 \caption{(Ref.~\cite{Borsanyi:2018grb}) Results for $\chi^{B}_{2}$, $\chi^{B}_{4}$, $\chi^{B}_{6}$ and an estimate for
$\chi^{B}_{8}$ on a $N_t = 12$ lattice as functions of the temperature, obtained from the
single-temperature analysis (see text). We plot $\chi^{B}_{8}$ in green to point out that its
determination is guided by a prior, which is linked to $\chi^{B}_{4}$. The red curve in each panel corresponds to the results from the Hadron Resonance Gas model. 
\label{fig:fluctuations}} 
\end{figure}

\paragraph{Fluctuations:}
Higher order fluctuations of conserved char\-ges are calculated as the partial derivatives of the pressure (or the QCD partition function) with respect to the chemical potentials. Here,
\begin{equation}
     \chi^{B,Q,S}_{i,j,k}= \frac{\partial^{i+j+k} (p/T^4)}{
(\partial \hat\mu_B)^i
(\partial \hat\mu_Q)^j
(\partial \hat\mu_S)^k
}
\end{equation}
with $\hat \mu = \frac{\mu}{T} $. These fluctuations are of great interest in the search for the critical endpoint, both for heavy ion experiments and theoretical calculations (Ref.~\cite{Hatta:2003wn,Stephanov:2008qz,Friman:2011pf}). They are proportional to powers of the correlation length and, therefore, expected to diverge in the vicinity of a critical end point (Ref.~\cite{Gavai:2008zr,Halasz:1998qr, Stephanov:1999zu,Cheng:2007jq}). Up to the 8th order fluctuations  have been calculated on finite lattices in Ref.~\cite{Borsanyi:2011sw,DElia:2016jqh,Bazavov:2017tot,Borsanyi:2018grb}. The baryon fluctuations up to 8th order from Ref.~\cite{Borsanyi:2018grb} are shown in figure~\ref{fig:fluctuations}.

In principal, fluctuations can also be measured both on the lattice and experiments. A comparison between
theoretical and experimental results can be used to extract the chemical freeze-out
temperature $T_f$ and the corresponding chemical potential $\mu_{Bf}$ as functions of the
collision energy (Ref.~\cite{Karsch:2012wm,Bazavov:2012vg,Borsanyi:2013hza,Borsanyi:2014ewa,Ratti:2018ksb}). To compare to experiments, one uses ratios of fluctuations to cancel out the explicit volume dependence. These can be matched to the ratios of cumulants of particle number distributions in experiments. However, not all baryons can be measured in experiments. Therefore, the cumulants of the proton number distribution are considered a usefull  proxy to compare to the baryon number cumulants calculated on the lattice. Some combinations are:
\begin{align}
R_{12}^B(T,\mu_B) &= \frac{M_B}{\sigma^2_B} = \frac{\chi_1^B(T,\mu_B)}{\chi_2^B(T,\mu_B)} \\
 R_{31}^B(T,\mu_B) &= \frac{S_B \sigma_B^3}{M_B} = \frac{\chi_3^B(T,\mu_B)}{\chi_1^B(T,\mu_B)} \\
R_{42}^B(T,\mu_B) &= \kappa_B \sigma_B^2 = \frac{\chi_4^B(T,\mu_B)}{\chi_2^B(T,\mu_B)}\\
R_{51}^B(T,\mu_B) &= \frac{S^H_B\sigma_B^5}{M_B} 
	=\frac{\chi_5^B(T,\mu_B)}{\chi_1^B(T,\mu_B)}\\
R_{62}^B(T,\mu_B) &= \kappa^H_B \sigma_B^4 
=\frac{\chi_6^B(T,\mu_B)}{\chi_2^B(T,\mu_B)},
\end{align}
where the mean $M_B$, the variance $\sigma_B^2$, the skewness $S_B$, the kurtosis $\kappa_B$, the hyper-skewness $S^H_B$ and the hyper-kurtosis $\kappa^H_B$ are the 1st to 6th order moments of the baryon number distribution. 

\begin{figure}
 \centering
 \includegraphics[width=0.7\textwidth]{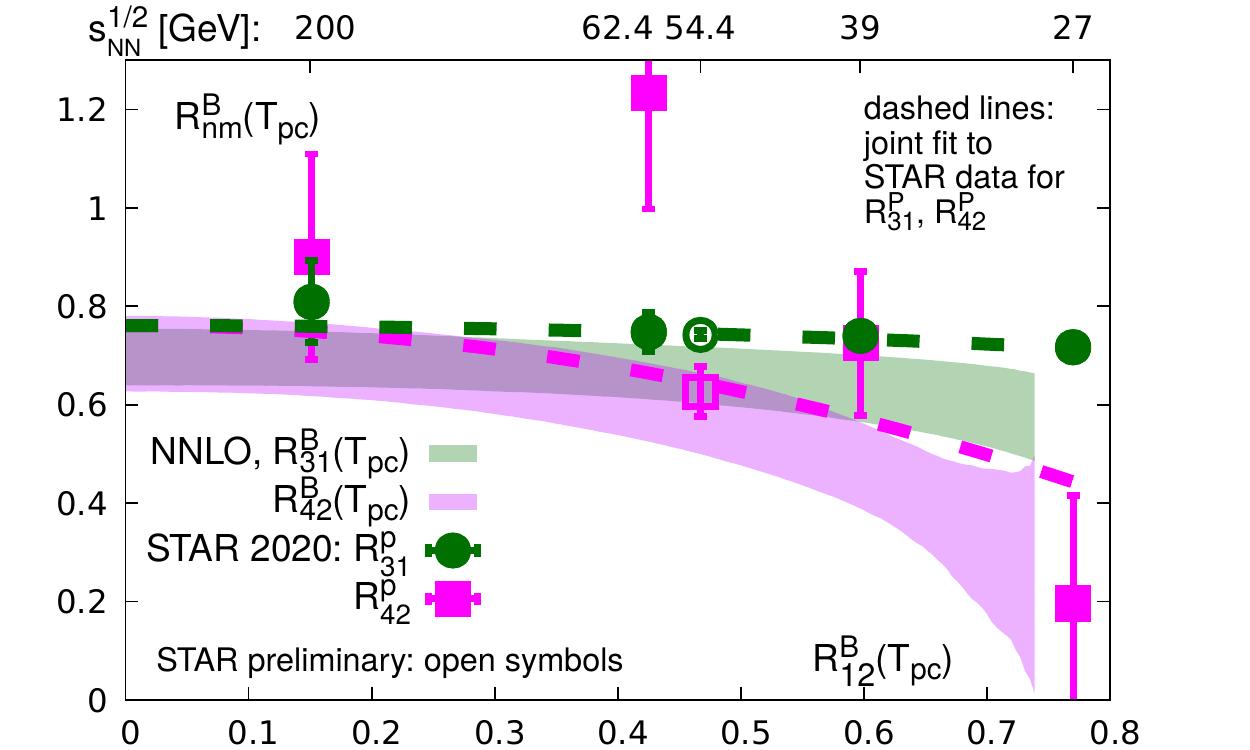}\\
 \vspace{12pt}
 \includegraphics[width=0.7\textwidth]{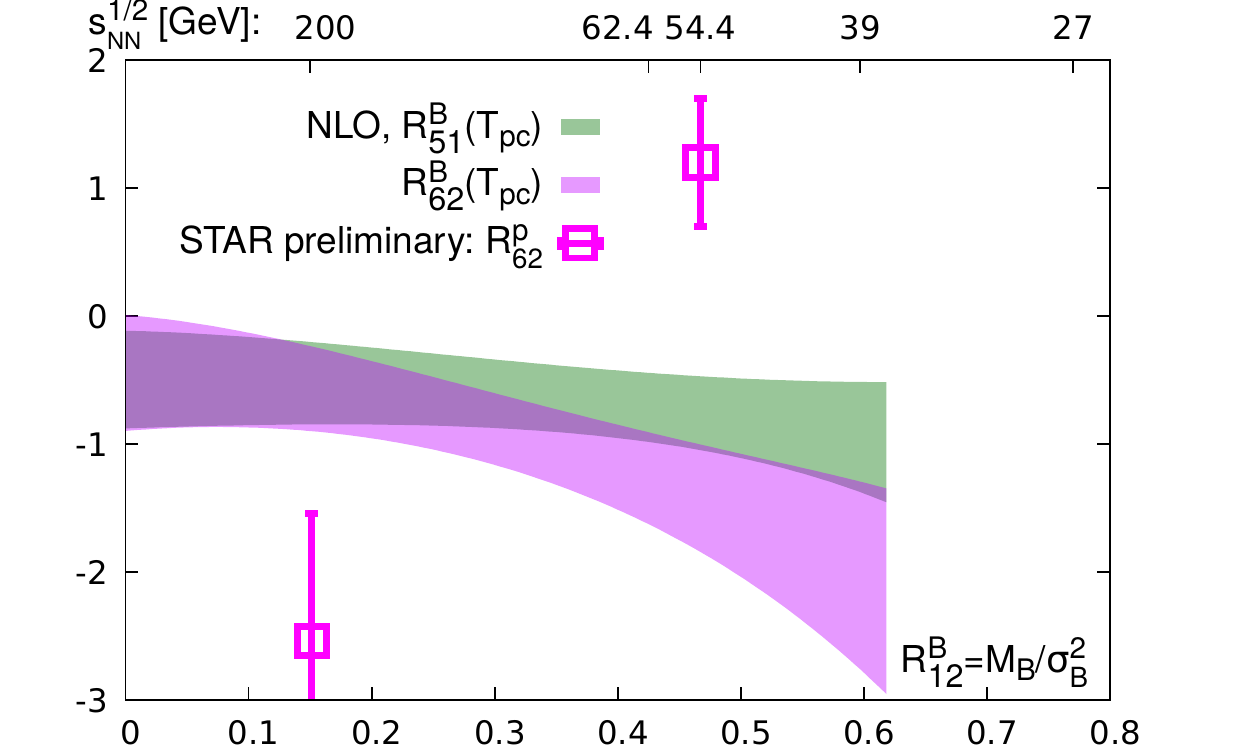}
 \caption{(Ref.~\cite{Bazavov:2020bjn}) Top: The ratios $R_{31}^B(T,\mu_B)= S_B \sigma_B^3/M_B$ and $R_{42}^B(T,\mu_B)\equiv \kappa_B\sigma_B^2$ as a function of $R_{12}^B(T,\mu_B) = M_B/\sigma_B^2$ evaluated along the transition line in comparison to the data from the STAR collaboration (Ref.~\cite{Adam:2020unf,Nonaka:2020crv}). The lattice calculation is an continuum estimate from $N_t=8$ and $N_t=12$ lattices. Bottom: The ratios $R_{51}^B(T,\mu_B)$  and $R_{62}^B(T,\mu_B)$ as a function of
$R_{12}^B(T,\mu_B)$ evaluated along the pseudo-critical line in comparison to the data from the STAR collaboration (Ref.~\cite{Nonaka:2020crv}). The lattice determination was done on an $N_t=8$ lattice. \label{fig:HotQCD}}
\end{figure}

New results on continuum estimates for $R_{12}^B$, $R_{31}^B(T,\mu_B)$ and $R_{42}^B(T,\mu_B)$ from lattice sizes $32^3\times 8$ and $48^3\times12$ have been recently presented in Ref.~\cite{Bazavov:2020bjn}. For the comparison with experimental data along the transition line, $R_{31}^B(T,\mu_B)$ and $R_{42}^B(T,\mu_B)$ are shown (see top of figure~\ref{fig:HotQCD}) as a function of $R_{12}^B$. Therefore, it is not necessary to determine the temperature and chemical potential in a heavy ion collision separately. This allows for a comparison with data from the STAR collaboration (Ref.~\cite{Adam:2020unf,Nonaka:2020crv}). The experimental datapoints fit the lattice results well.

First calculations of the 5th and 6th order cumulants from a $32^3\times 8$ are presented in the same way at the bottom of Fig.~\ref{fig:HotQCD}. Here only preliminary experimental data (Ref.~\cite{Nonaka:2020crv}) is available for a comparison with $R_{62}^B(T,\mu_B)$. While errors are still large, both on the lattice and the experimental side, there seems to be a tension between both data sets. Changing the lattice results such that they can describe the two available experimental points would require unexpectedly large higher order corrections.

\begin{figure}
 \centering
 \includegraphics[width=0.7\textwidth]{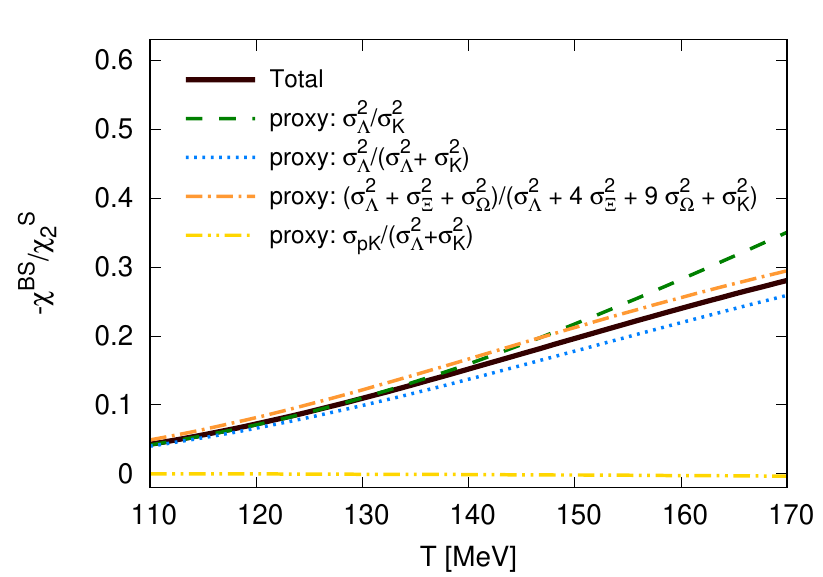}\\
 \caption{(Ref.~\cite{Bellwied:2019pxh}) Comparison between different possible combinations of particle number cumulants that could be measured experimentally (proxies) and the total of $-\frac{\chi^{BS}_{11}}{\chi^S_2}$ that could be determined in lattice QCD calculations. Both calculated in the Hadron Rasonance Gas model. \label{fig:PaoloPaper1}}
\end{figure}

\begin{figure}
 \centering
 \includegraphics[width=0.7\textwidth]{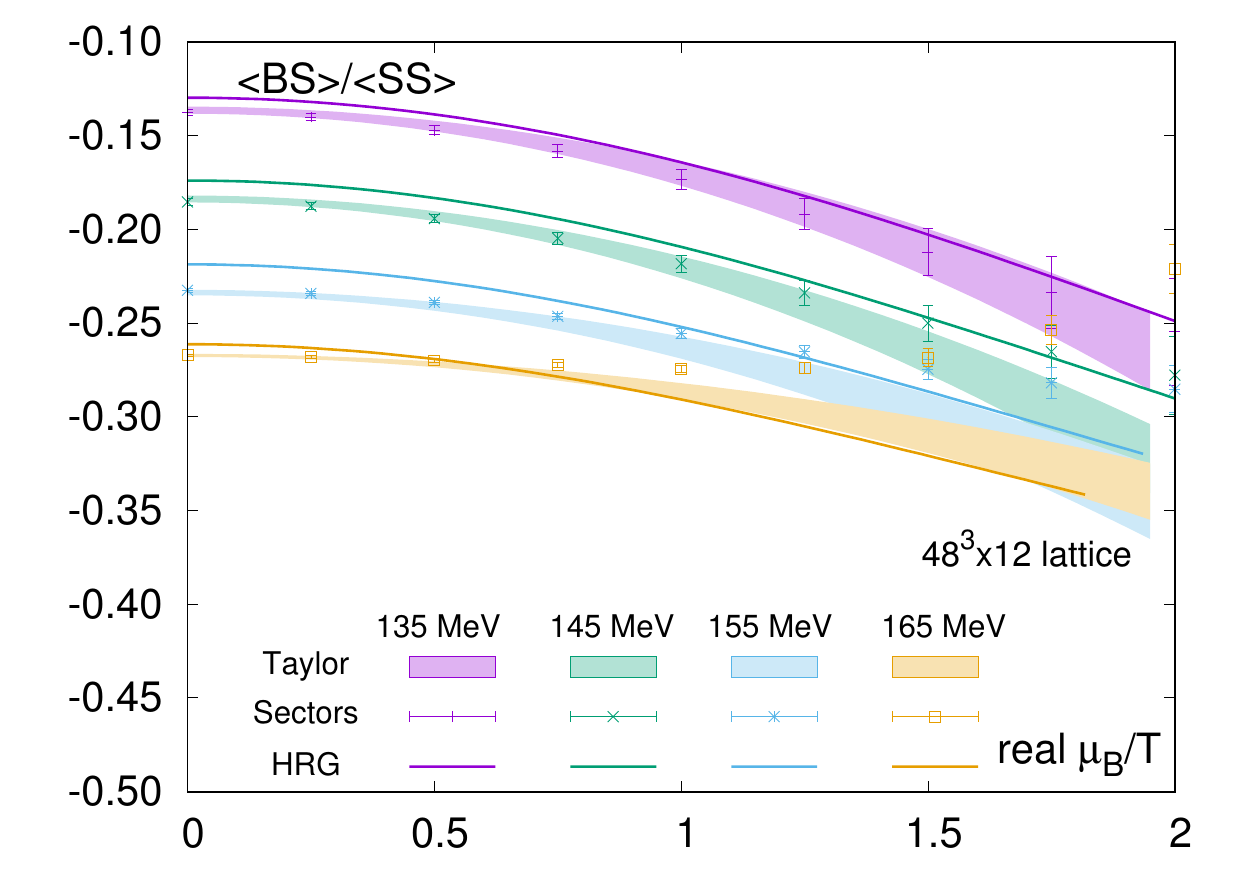}
 \caption{(Ref.~\cite{Bellwied:2019pxh})  Comparison between the Taylor and the sector method for $-\frac{\chi^{BS}_{11}}{\chi^S_2}$ on an $48^3\times12$ lattice, as well as results from the hadron resonance gas model (HRG). \label{fig:PaoloPaper2}}
\end{figure}

In general, a detailed comparison between lattice data and experiment requires the incorporation of many different aspects (Ref.~\cite{Begun:2006jf,Kitazawa:2011wh,Kitazawa:2012at,Alba:2014eba}). While on the lattice site, clearly, a continuum extrapolation is still needed, on the experimental side it is not always clear which particle species allow for good comparisons to the conserved charges measured in lattice QCD. While for the cumulants for the baryon number distribution the proton number distribution has been established as reasonable proxy, for the strangness number distribution the situation is less clear. A recent study in the hadron resonance gas (Ref.~\cite{Bellwied:2019pxh}) investigated different possibilities to find comparable observables. The comparison for different experimental proxies to the total, which would be the lattice result for $-\frac{\chi^{BS}_{11}}{\chi^S_2}$, is shown in  Fig.~\ref{fig:PaoloPaper1}. The lattice prediction for $-\frac{\chi^{BS}_{11}}{\chi^S_2}$ is  calculated on an $48^3\times12$ lattice with two different methods. One that is based on the Taylor expansion and one that is based on the sector expansion (see Ref.~\cite{Alba:2017mqu}), where the pressure is parameterized as
\begin{eqnarray}
\label{eq:pressure}
P(\hmu_B,\hmu_S) &=& P^{BS}_{00}+P^{BS}_{10}\cosh(\hmu_B)+P^{BS}_{01} \cosh(\hmu_S)
\nonumber \\
&+& P^{BS}_{11} \cosh(\hmu_B-\hmu_S)
\nonumber \\
&+& P^{BS}_{12} \cosh(\hmu_B-2\hmu_S)
\nonumber \\
&+& P^{BS}_{13} \cosh(\hmu_B-3\hmu_S)
\;.
\end{eqnarray}
As can bee seen in figure~\ref{fig:PaoloPaper2}, both expansions agree well for small chemical potentials and temperatures around 140~MeV. For higher temperatures, a deviation between the two methods becomes visible before the errorbar grows due to the increased chemical potential. 
Also the influence of different experimental cuts has been investigated.

\paragraph{The critical endpoint:}
To look for a bound on the critical endpoint in the QCD phase diagram, one can try to calculate the radius of convergence of an expansion in $\mu_B$ around $\mu_B=0$. When one wants to estimate the radius of convergence naively form the fluctuations, one defines
\begin{equation}
 r^\chi_{2n} = \sqrt{\frac{\chi_{2n}}{\chi_{2n+2}}}.\label{eqn:rChi}
\end{equation}
If $r^\chi_{2n}$ converges, in the limit of $n \to \infty$,  it is guaranteed that there
is no criticality within this radius. This has been done in Ref.~\cite{Bazavov:2017dus} for the fluctuations up to 6th order.

\begin{figure}
 \centering
 \includegraphics[width=0.7\textwidth]{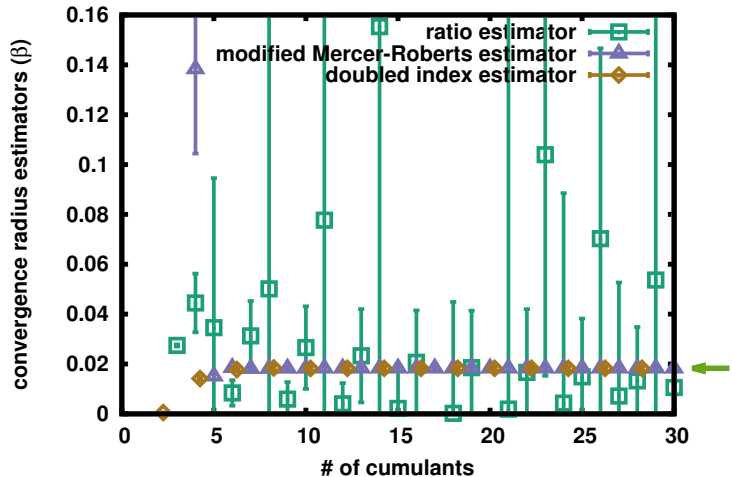}
 \caption{(Ref.~\cite{Giordano:2019slo}) The radius of convergence from three different estimators determined on an $N_t=4$ lattice. The green arrow indicates the result from reweighting.\label{fig:radius}}
\end{figure}

However, it has been shown in Ref.~\cite{Giordano:2019slo}, that  the ratio estimator as given in equation~(\ref{eqn:rChi}) is never
convergent in a finite volume. This is consistent with the fact that there is never a true phase transition in a finite volume. However, the convergence of the ratio estimator is problematic even when
using the $p_n$ extrapolated to infinite volume. It will
work if one uses infinite-volume Taylor coefficients and
the singularity determining the radius of convergence corresponds
to a real phase transition in the infinite-volume limit. In the same work, alternative estimators have been discussed. Fig.~\ref{fig:radius} shows results on an $N_t=4$ lattice for the standard estimator defined in equation~(\ref{eqn:rChi}), the modified Mercer-Roberts estimator
\begin{equation}
\label{eqn:r2}
    r^{\mathrm (MMR)}_k = \left| \frac{(k+1)(k-1)c_{k+1}c_{k-1}-k^2
        c_k^2}{(k+2)k c_{k+2}c_k - (k+1)^2 c_{k+1}^2} \right|^{\frac{1}{2}}\,,
\end{equation}
and the doubled index estimator
\begin{equation}
    r^{\mathrm (2i)}_k = \left| \frac{2}{2k c_{2k} + k^2 c_{k}^2}
    \right|^{\frac{1}{2k}}\,, \label{eqn:r3}
\end{equation}
for a Taylor series with coefficients $c_k$.

Another way to look for a critical endpoint and find the radius of convergence is the investigation of the Lee-Yang zeros (Ref.~\cite{Lee:1952ig}). These are the zeros of the partition function in the plane of complex chemical potential. The radius of convergence is the distance to the closest (leading) Lee-Yang zero in the infinite Volume limit. If this happens to be on the real axis, it is a signal for the critical endpoint. While the higher order fluctuations needed for a reliable radius of convergence estimate suffer from increasingly larger errors and are difficult to determine, Ref.~\cite{Giordano:2019slo} discusses a cancellation that allows for a calculation of the leading Lee-Yang zero. 
However, this benefit only holds only if it is possible to reweight by using a reduced matrix formulation (Ref.~\cite{Hasenfratz:1991ax,Fodor:2001pe,Fodor:2004nz,Danzer:2008xs,Alexandru:2010yb}). This is not possible for the case of rooted staggered fermions. For this case, Ref.~\cite{Giordano:2019gev} introduces a new definition of the rooted staggered determinant which allows for a numerical study of the Lee-Yang zeros. It is than tested on $N_t=4$ lattice with stout smeard staggered fermions and a Symanzik improved gauge action. Spatial lattice extends of $N_s = 8, 10$ and 12 are used to estimate the infinite volume limit. However, this approach might still suffer from the overlap problem related with reweighting.

In Ref.~\cite{Giordano:2020roi}, an algorithm called sign-reweighting is proposed, aiming to avoid uncontrolled systematics like the overlap problem. This approach separates the sign of the Dirac determinant $\det D(\mu)$ from the configuration generation which is done with a weight of $|\Re (\det D(\mu)) |$. The sign is handled separately by a discrete reweighting.

\subsection{Complex Langevin \label{sec:CL}}
One way to conduct lattice simulations, despite the sign problem that made significant progress are complex Langevin simulations. I will only give a very brief overview over the new developments. A more comprehensive recent review can be found in Ref.~\cite{Attanasio:2020spv}.  These simulations are based on the Langevin process, an evolution in a fictitious Langevin time, to generate configurations with a complex measure. This involves a complexification of all fields and, therefore, extending the $SU(3)$ gauge group to $SL(3,\mathbb{C})$, which is a non compact group. This can lead to so called runaway configurations which can cause the trajectory to converge to a wrong result (Ref.~\cite{Ambjorn:1985iw,Klauder:1985ks,RunAway,Ambjorn:1986fz}). To keep the evolution close to the unitary manifold and therefore to the correct result, gauge cooling (Ref.~\cite{Seiler:2012wz,Aarts:2013uxa}) was developed. A recent overview on this subject can be found in Ref.~\cite{Berges:2007nr}. Even if the gauge cooling as well as an adaptive step size in the numerical integration (Ref.~\cite{Aarts:2009dg}) or the addition of force to the evolution \cite{Attanasio:2018rtq} increase the stability of the Langevin simulations, it is still important to verify the correctness of the result. This can be done by different correctness criteria, which are related to the fall-off behavior of specific observables (Ref.~\cite{Nishimura:2015pba,Aarts:2017vrv,Nagata:2016vkn, Nagata:2018net, Scherzer:2018hid,Tsutsui:2019suq}).

\begin{figure}
 \begin{center}
  \includegraphics[width = 0.7\textwidth]{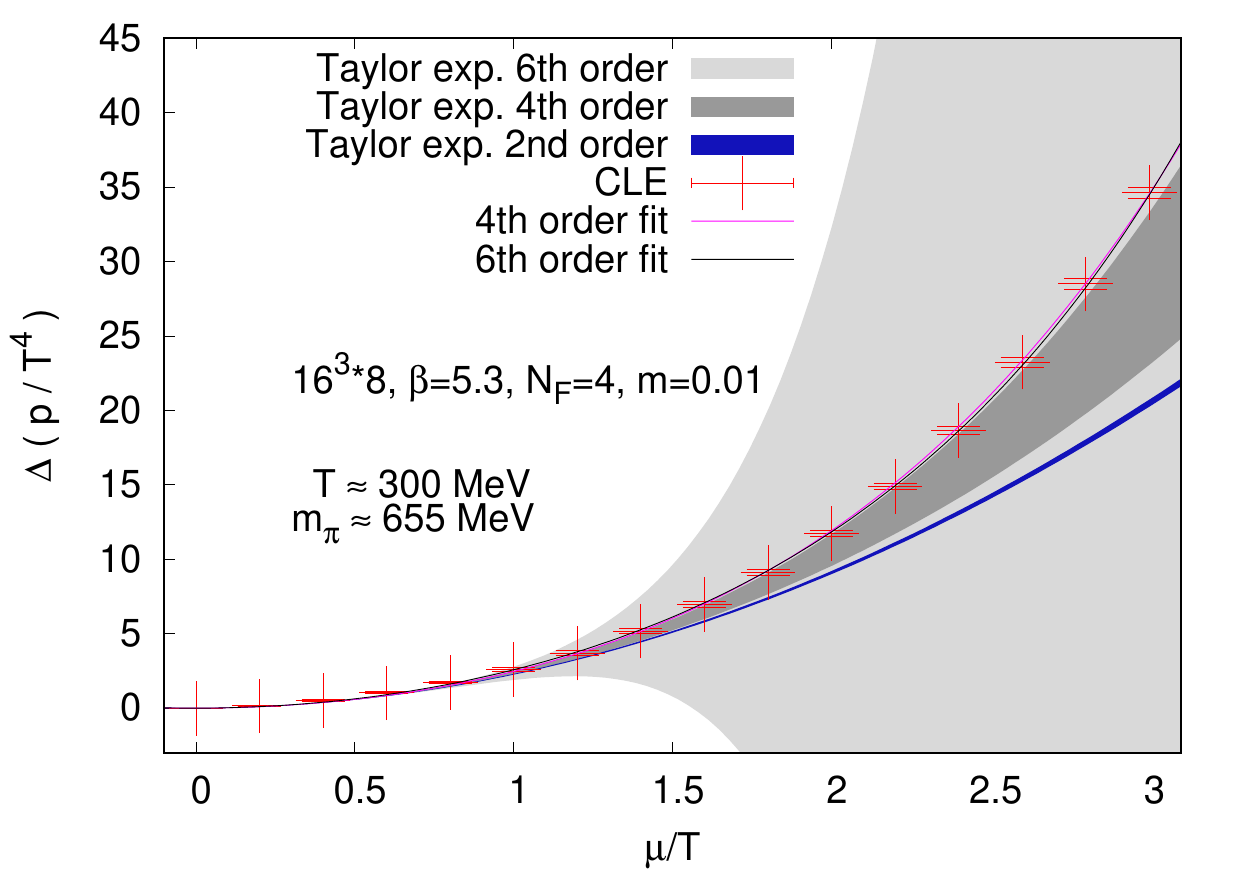}\\
  \vspace{12pt}
\includegraphics[width = 0.7\textwidth]{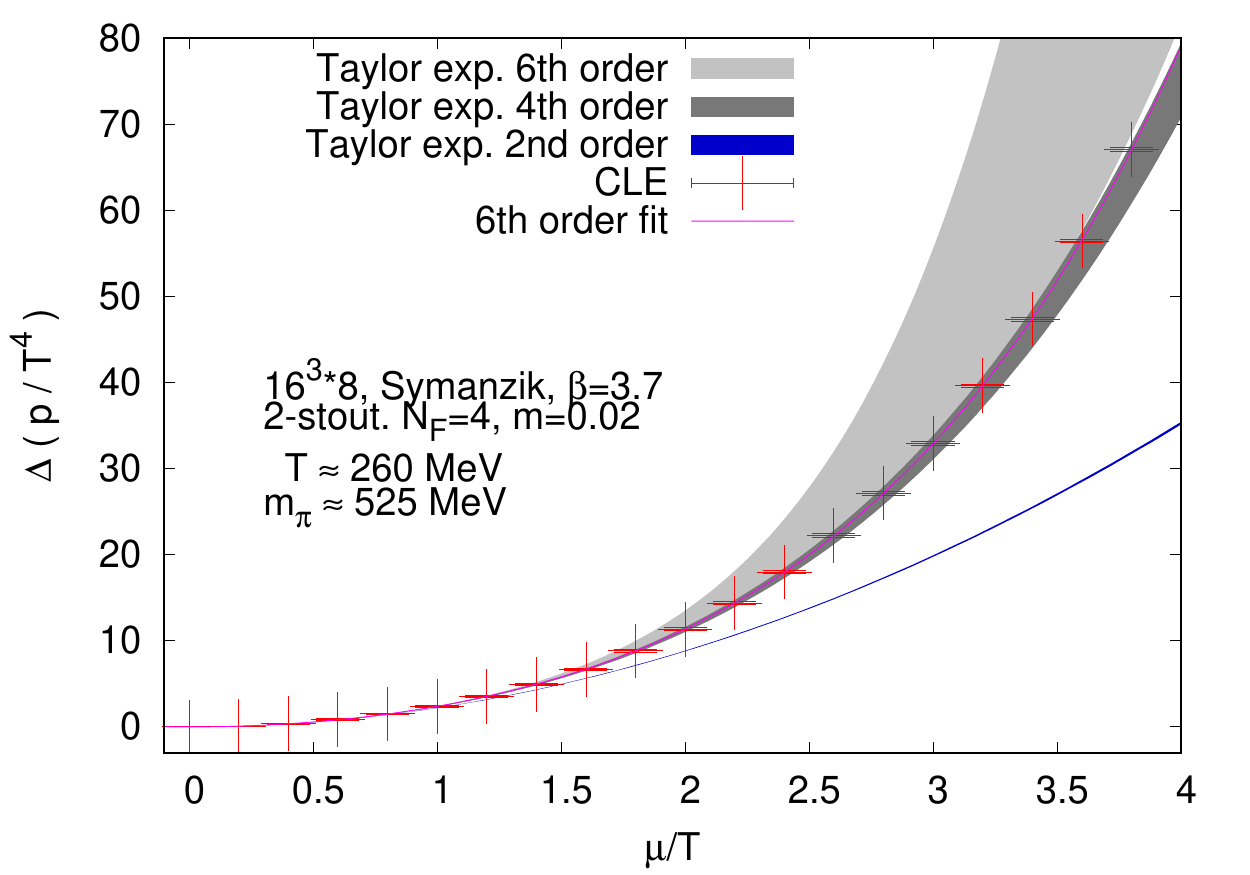}
 \end{center}
\caption{(Ref.~\cite{Sexty:2019vqx}) Top: Comparison between an extrapolation by Taylor expansion (see sec.~\ref{sec:anaCont}) and simulations with the complex Langevin equations using a naive action. Bottom: Comparison between an extrapolation by Taylor expansion and simulations with the complex Langevin equations using a Symanzik improved action.\label{abb:SextyPaper}}
\end{figure}

By now, results obtained with Complex Langevin simulations start to attacking the QCD phase diagram. In Ref.~\cite{Sexty:2019vqx}, Complex Langevin was combined with stout smearing and the result was compared with results from the Taylor expansion method (see section~\ref{sec:anaCont}). The simulations were performed with four flavours of staggered quarks on $16^3\times8$ lattices with pion masses between 500~MeV and 700~MeV. The results for two different gauge actions, one unimproved and one with the Symanzik improvement, are shown in figure~\ref{abb:SextyPaper}. The agreement between higher order Taylor and Complex Langevin results is good. These investigations were done at temperatures well above the transition temperature.

\begin{figure}[ht]
 \begin{center}
  \includegraphics[width = 0.7\textwidth]{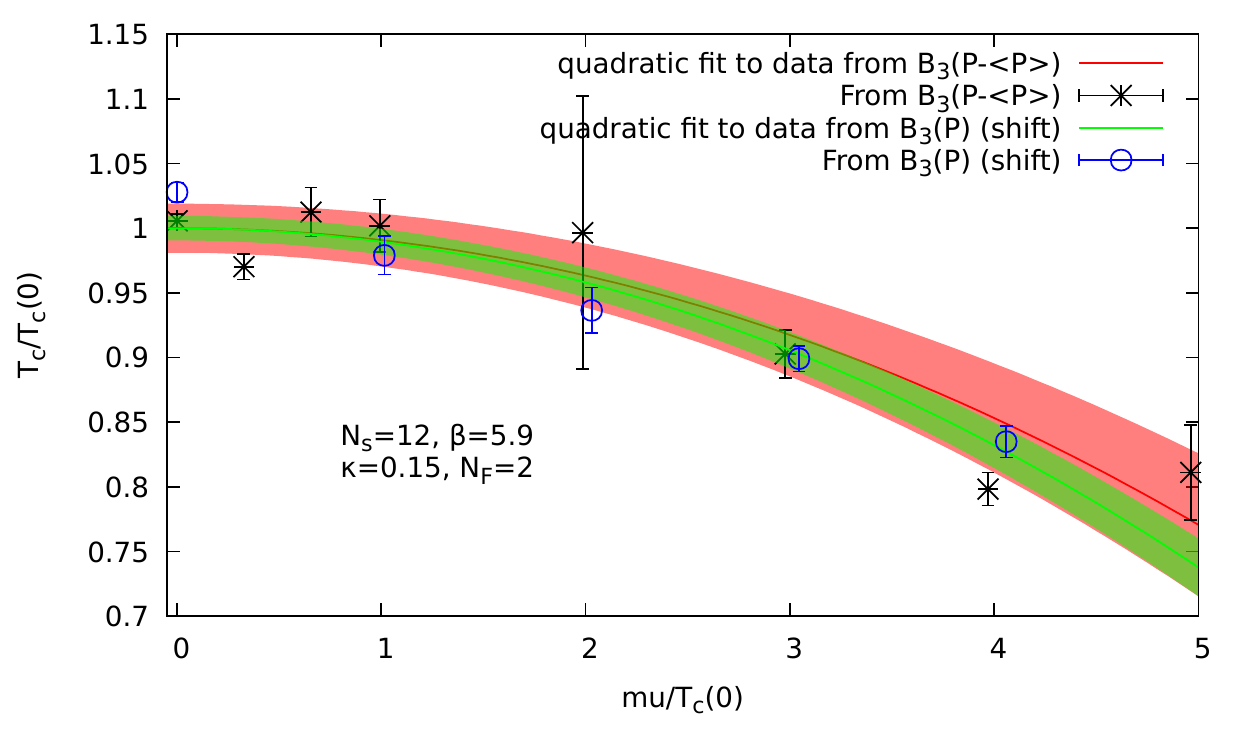}\\
 \end{center}
\caption{(Ref.~\cite{Scherzer:2020kiu}) The transition temperature from Complex Langevin Simulations at heavy pion masses ($m_\pi\approx1.3$~GeV with two flavor Wilson fermions on an $N_s=12$ lattice. The transition temperature is determined from the third order Binder cumulant $B_3$ with two different Polyakov loop related observables. The data is than fitted with a quadratic function.   \label{abb:CLTransition}}
\end{figure}

In Ref.~\cite{Scherzer:2020kiu}, the transition is studied with two flavours of naive Wilson fermions and very heavy pion masses of about 1.3~GeV. The transition temperature is determined from the third order Binder cumulant
\begin{equation}
 B_{3}(\mathcal{O})=\frac{\left<\mathcal{O}^{3}\right>}{\left<\mathcal{O}^{2}\right>^{3/2}}
\end{equation}
with two different Polyakov loop related observables $\mathcal{O}$. The data is than fitted with a quadratic function to determine the curvature of the transition line $\kappa_2$ (see equation~\ref{eq:kappa}). The transition line on an $N_s=12$ lattice from both observables is shown in Fig.~\ref{abb:CLTransition}.

In addition to finite temperature studies, Complex Lan\-gevin simulations are also used to study low or zero temperature QCD. In Ref.~\cite{Ito:2018jpo,Tsutsui:2019suq, Ito:2020mys}, the average quark number $\langle N \rangle = N_f N_c N_s$ is studied on lattices with sizes $8^3\times16$ and $16^3\times 32$. They found a plateau for $\langle N \rangle =24$ as a function of the chemical potential.  It is interpreted as the maximum number of zero momentum quarks that can exist at zero temperature. When the chemical potential is increased enough, it excites the lowest non-zero momentum states.

In Ref.~\cite{Kogut:2019qmi}, the zero temperature transition between hadronic an nuclear matter at a chemical potential of a third of the nucleon mass $\mu=\frac{m_N}{3}$ is searched for but not found. This contradiction to the physical expectation is related to possible issues with the Complex Langevin simulations, which faces difficulties at low or zero temperature.

\clearpage
\section{Effective lattice theories\label{sec:ELT}}

Full lattice QCD simulations for large baryon chemical potentials are out of reach for the moment. However, some insight can be gained from simulations of effective theories on the lattice. Here usually either the fermions and the spacial gauge links are integrated out, yielding a theory only depending on Polyakov loops, which contains the temporal link variables. Or, as the second common option, spatial and temporal gauge links are integrated out resulting in a theory with hadronic degrees of freedom. The use of effective theories is limited to a specific parameter range, where the reduction of the degrees of freedom holds true. If it is possible to find overlapping parameter ranges between different theories, one can gain inside in a wide range of phenomena.

\subsection{Heavy Quarkonium \label{sec:Quarkonium}}
 Heavy quarkonium refers to states of matter including heavy quarks, most commonly the charm and bottom quark. In heavy ion collisions, $c\overline c$ and  $b\overline b$ pairs can usually only be created in early stages of the collision due to the required large energy amount. This heavy bound states are then surrounded by the quark gluon plasma and can either meld due to the high temperatures or survive until freezout. In the later case, a signature for these particles should be visible in the detectors. Since the melting of different particles depends on the state of the quark gluon plasma, the observation of quarkonium is a useful tool to enhance our knowledge on QCD at very high temperatures. On the other hand it becomes more and more clear that not only the melting but also the possibility of recombination of quarkonium states has to be taken into account. The interaction between the heavy bound states and the surrounding medium is one of the main question fueling the work on effective theories. A sketch of the different stages a heavy quark pair goes through in a heavy ion collision is shown at the bottom for figure~\ref{fig:HIC} form Ref.~\cite{Rothkopf:2019ipj} where detailed introduction and review of recent developments for heavy quarkonium can be found.

The input from the lattice to the investigation on quarkonium can be twofold. On one hand, the calculation of quarkonium correlators and their spectral functions as for example done in Ref.~\cite{Karsch:2003wy,Aarts:2005hg}. Since these studies require zero temperature simulations, they are not part of this review.

\begin{figure}
 \centering
 \includegraphics[width=0.9\textwidth]{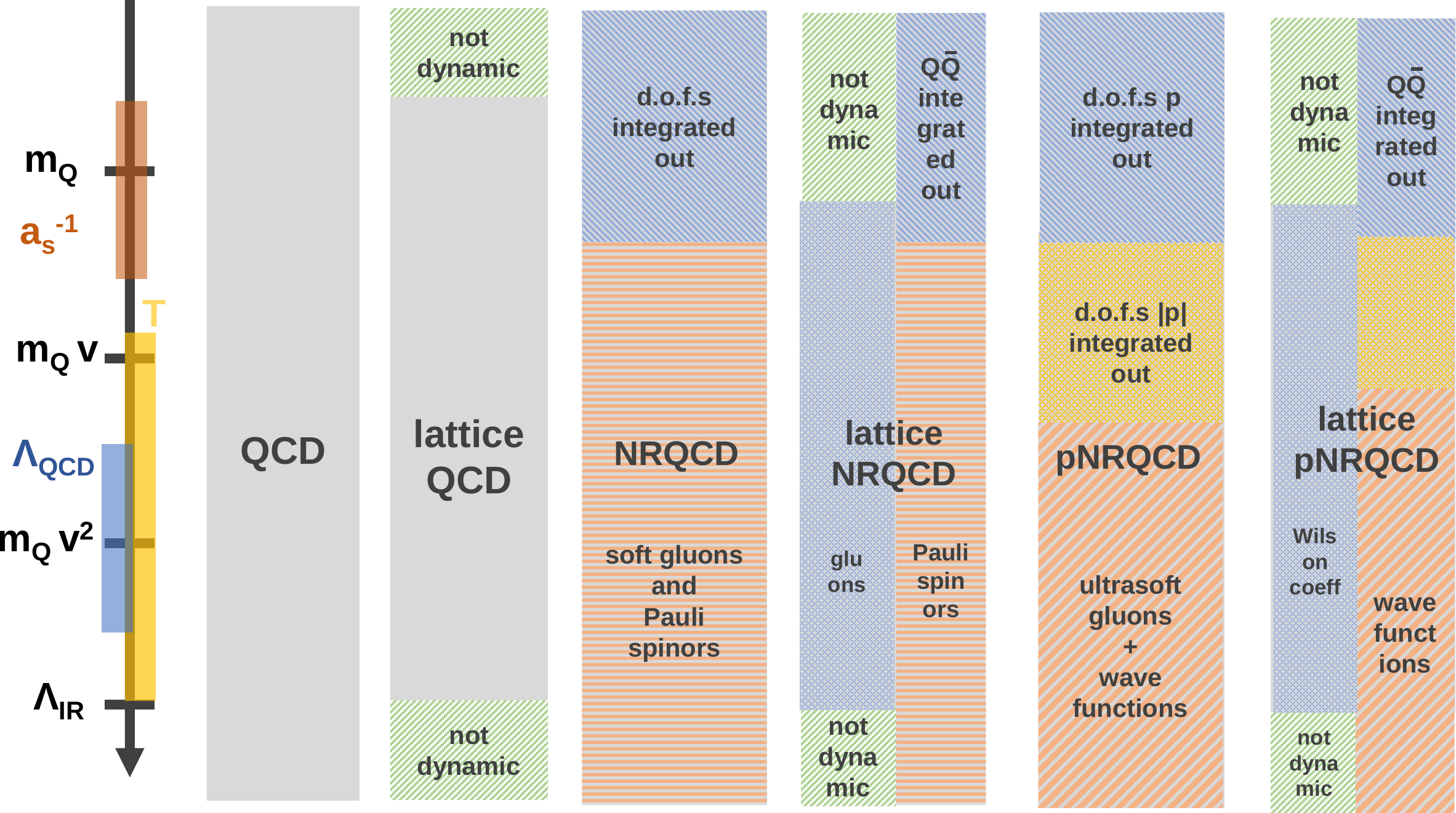}
 \caption{(Ref.~\cite{Rothkopf:2019ipj}) Overview over the different scales treated in effective field theories on the lattice. In the first step the lattice introduces UV cut off by the finite lattice volume and an infrared cut off by the finite lattice spacing. In the next step gluons and  heavy quarks are integrated out yielding a non relativistic version on QCD called NRQCD. The lattice version of NRQCD again has the respective cut offs. Instead of on the lattices NRQCD can also be treated perturbatively to integrated out further degrees of freedom down to an energy scale $m_Qv$. This leads to potential NRQCD (pNRQCD) with ultrasoft gluons and wavefunctions as remaining degrees of freedom. pNRQCD can then be studied on the lattice by the investigation of non-local Wilson coefficients. \label{fig:shematicETF}}
\end{figure}

On the other hand, the lattice simulations of effective field theories are a very helpful tool to investigate the behavior of quarkonium in a quark gluon plasma. To arrive at an effective field theory, the difference in scales within the quarkonium is exploited. The mass of a heavy quark $m_Q$ is much larger than its velocity $v$ within a bound state and, therefore,
\begin{equation}
 m_Q \gg v m_Q \gg v^2 m_Q
\end{equation}
as well as
\begin{equation}
 m_Q \gg \Lambda_{\mathrm{QCD}}.
\end{equation}
The reduction of the degrees of freedom by the integration over different scales is sketched in figure~\ref{fig:shematicETF}. While full continuum QCD is valid at all scales, the finite volume of the lattice provides an infrared cut off, while the finite lattice spacing cuts off the ultraviolet divergences. To arrive at a non relativisic version of QCD called NRQCD (Ref.~\cite{Caswell:1985ui}) the hard gluons and heavy quarks are integrated out. The relevant degrees of freedom are the Pauli spinors and soft gluons. This effective theory it self can be treated by lattice simulation. On the other hand, a perturbative treatment of NRQCD can be applied to  integrate out further degrees of freedom down to an energy scale $m_Qv$. This leads to potential NRQCD (pNRQCD, Ref.~\cite{Brambilla:1999xf}), with ultrasoft gluons and wavefunctions as remaining degrees of freedom. In this framework the in-medium real-time potential can be investigated. Again, it is possible to study pNRQCD on the lattice. To construct the respective Lagrange functions for the various effective field theories, one identifies the relevant degrees of freedom at each energy scale and constructs a general Lagrange function from symmetry considerations. The relevant prefactors of  each term, the so called Wilson coefficients, are determined by matching. The strength of effective field theories lies in the reduced number of degrees of freedom, that makes computations easier. The trade off is the reduced validity  range.

An investigation of the validity of the perturbative treatment of effective field theory was done in  Ref.~\cite{Bazavov:2018wmo}. There, lattice QCD calculations of a wide temperature range from 140~MeV up to temperatures of 5814~MeV with (2+1)-flavours of highly improved staggered fermions were performed. The evaluation of the static quark-antiquark potential led to the conclusion that effective field theories can be used between $0.3\lesssim rT \lesssim0.6$ (with $r$ beginning the separation between quark and antiquark) to describe color screening reliably.

\paragraph{NRQCD:} A large effort to use lattice NRQCD to study the behaiviour of bottemonium in a quark gluon plasma has been undertaken by the FASTSTUM collaboration in Ref.~\cite{Aarts:2014cda,Aarts:2013kaa,Aarts:2012ka,Aarts:2011sm,Aarts:2010ek}. They calculate the spectrum of bottemonium around the crossover temperature, using (2+1)-flavours of Wilson clover fermions on anisotropic lattices. The anistropy of the lattices improves the NRQCD expansion. On the other hand it also leads to heavy pion masses of $m_\pi\approx400$~MeV, if computation are to be kept afordable.

Another recent investigation of both bottemonium and charmonium using NRQCD has been done in Ref.~\cite{Kim:2014iga,Kim:2018yhk}. Here, lattice configuration of the HOTQCD collaboration with 2+1 highly imporved staggered (HISQ) quarks were used. These ensembles have a realistic pion mass of $m_\pi\approx 161$~MeV. Ref.~\cite{Kim:2018yhk} was able to sort out previous tensions between Ref.~\cite{Aarts:2014cda} and Ref.~\cite{Kim:2014iga} on the melting temperatures of quarkonium by relating them to different uncertainties in the spectral reconstruction. It was shown, that the mass of the ground state for heavy quarkonium reduces when the temperature is increased. However, in Ref.~\cite{Larsen:2019zqv,Larsen:2019bwy}, where the same configurations where analysed, the mass reduction was not observed. 

\paragraph{pNRQCD:} To investigate not only the ground state, but also the in-medium behaiviour of exited quarkonium states, one has to turn to pNRQCD. Here the spectral functions are computed by solving the Schrödinger equation with a static potential. The results show a mass reduction for rising temperatures and agree well with lattice NRQCD results, while contradicting expectation gained from perturbative pNRQCD.

\begin{figure}
 \centering
 \includegraphics[width=.7\textwidth]{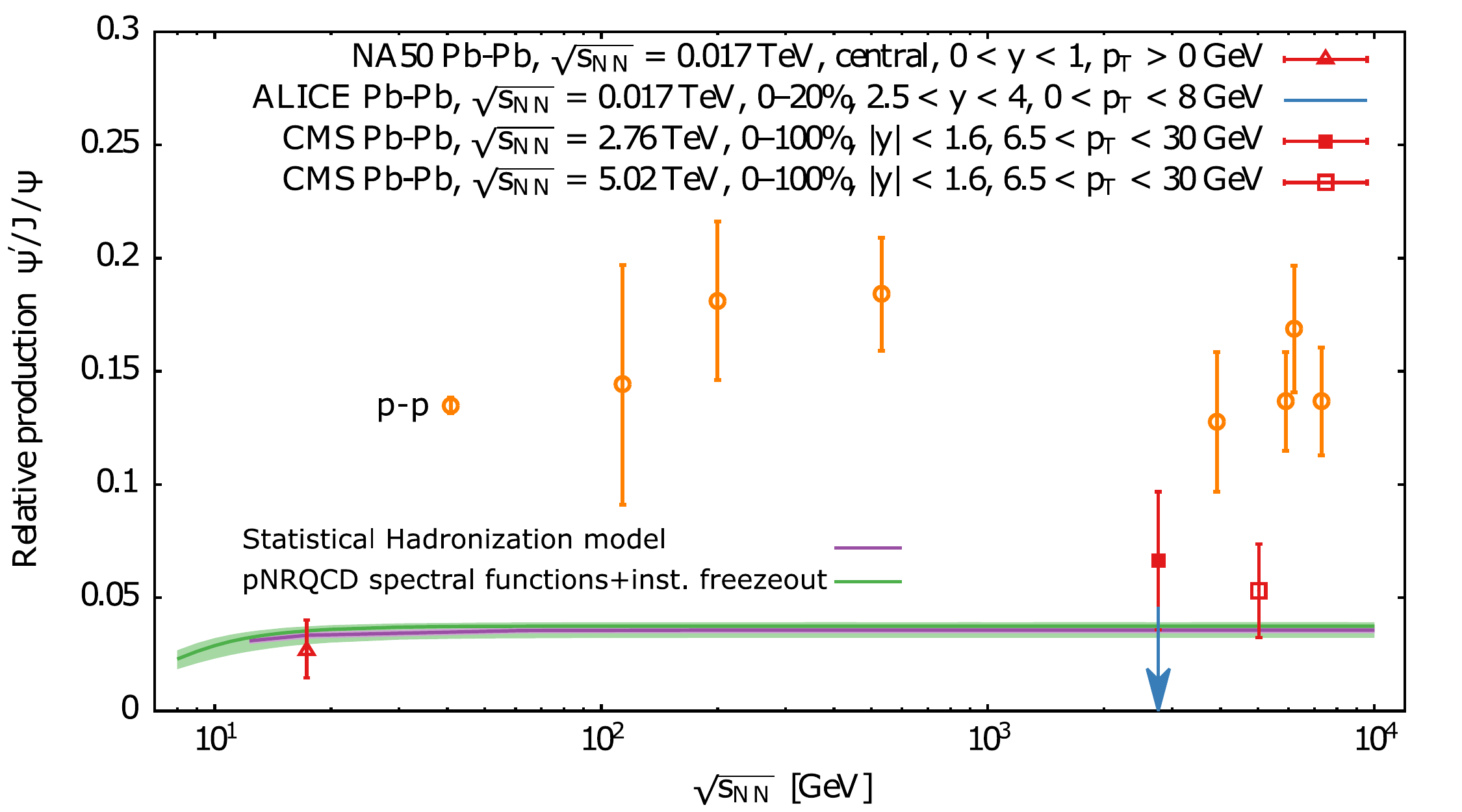}
 \caption{(Ref.~\cite{Lafferty:2019jpr,Rothkopf:2019ipj})
 The ratio between \(\psi^\prime\) and \(J/\psi\) from various heavy ion collision experiments: the NA50 (Ref.~\cite{Alessandro:2006ju}), ALICE (Ref.~\cite{Adam:2015isa}) and CMS (Ref.~\cite{Khachatryan:2014bva,Sirunyan:2016znt}). The orange points show the $pp$ baseline form Ref.~\cite{Andronic:2017pug,Drees:2017zcb}. The purple  line comes from the statistical model of hadronization from Ref.~\cite{Andronic:2017pug}. The green line is the result from a computation based on pNRQCD spectral functions combined with an instantaneous freeze-out scenario.\label{fig:QuarkoniumFreezout}}
\end{figure}

Recent progress with computations in lattice pNRQCD include the determination of the ratio between the \(\psi^\prime\) and \(J/\psi\) in Ref.~\cite{Lafferty:2019jpr}. Here a non perturbative treatmend of pNRQCD is empolyed, based on the derivation of the generalized Gauss law. The derived potential is matched to lattice QCD results from Ref.~\cite{Burnier:2015tda,Burnier:2014ssa, Bazavov:2009bb}. The calculation of inmedium spectral functions allows for a prediction on the \(\psi^\prime\) and \(J/\psi\) ratio, which is shown in figure~\ref{fig:QuarkoniumFreezout}. It is also compared to various results from heavy ion collision experiments as well as results from the statistical model of hadronization (Ref.~\cite{Andronic:2017pug}).

\clearpage
\section{Columbia plot\label{sec:Columbia}}

\begin{figure}
 \centering
 \includegraphics[width=0.5\textwidth]{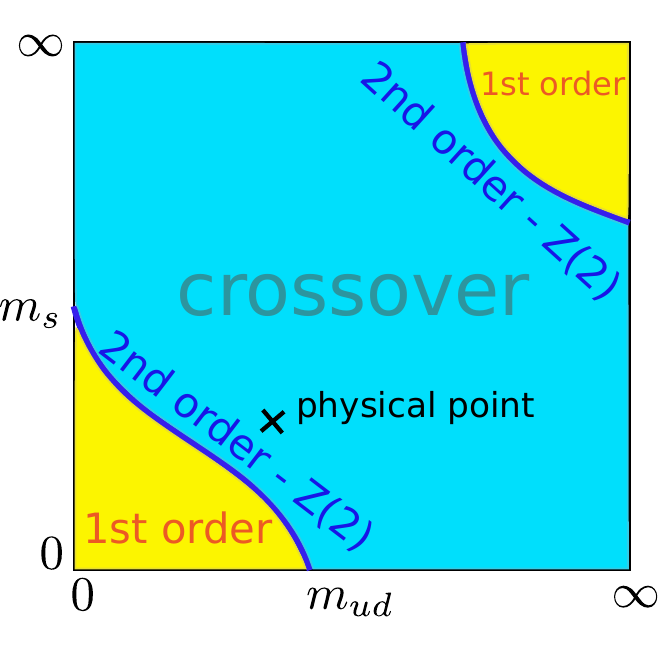}\\
 \vspace{12pt}
 \includegraphics[width=0.5\textwidth]{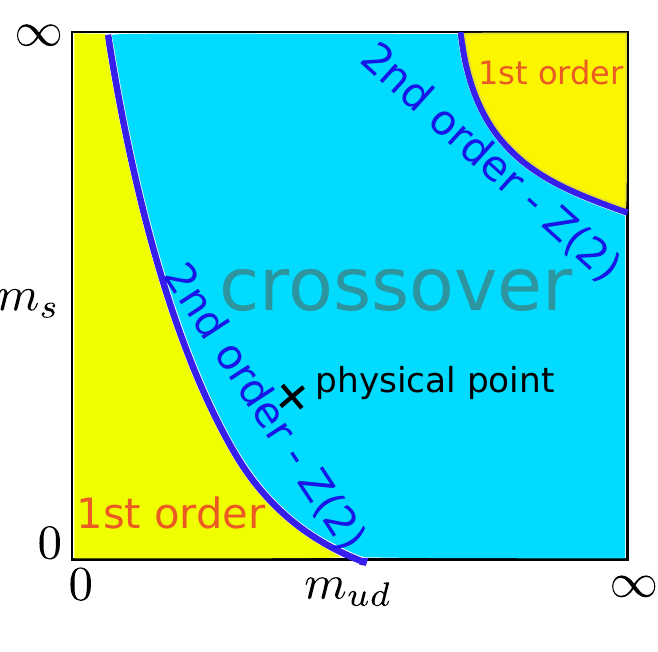}
 \caption{ Schematic of two different, possible scenarios for the Columbia plot. It shows the dependence of the transition between hadronic matter and the quark gluon plasma on the quark masses.\label{fig:Columbia}}
\end{figure}

As discussed before, for zero chemical potential and physical quark masses, the QCD transition is a crossover. However this changes when  the quark masses are varied. The type of the transition between hadronic matter and the quark gluon plasma, called the QCD transition, depends on the quark masses. This is illustrated in the Columbia plot for 2+1 quark flavours (figure~\ref{fig:Columbia}). The upper right corner, where quark masses are infinite, is the pure $SU(3)$ gauge limit with static quarks and exhibits a first order phase transition. Finite quark masses break the $Z(3)$ center symmetry of the zero mass limit explicitly, which weakens the phase transition until it becomes second order in the $Z(2)$ line bordering the upper right corner of the Columbia plot.

In the limit of vanishing quark masses, corresponding to the lower left corner of the Columbia plot, again a first order transition is expected for three flavours in the chiral limit, where $m=0$. Finite quark masses break the chiral symmetry and weaken the transition until it becomes second order. This yields another $Z(2)$ line delimiting this corner. However, the case on the left border, for vanishing light quark masses, is still under investigation. Another possible scenario is that the first order region could extent all the way up to the $N_f=2$ limit or hit the $m_{ud} =0$ line at a finite strange quark mass, as shown in Fig.~\ref{fig:Columbia}. The two scenarios are connected to the possible $U(1)_A$ symmetry reforestation at the transition temperature (Ref.~\cite{Philipsen:2019rjq, Pelissetto:2013hqa}) and are shown in figure~\ref{fig:Columbia}.

The computation of the $Z(2)$ lines, especially in the lower left corner, are numerically very challenging. Smaller quark mass, as well as critical slowing down near a phase transition increase the computational cost. At the same time, small quark masses increase the taste breaking artifacts for the computationally relatively cheap staggered quarks, making a continuum extrapolation even more difficult. Also, scans in temperature, volume and quark masses are required to determine type and position of the transition. Due to this challenges, the only continuum extrapolated point in the Columbia plot up to now is the physical point (Ref.~\cite{Aoki:2006we,Aoki:2006br,Aoki:2009sc,Borsanyi:2010bp,Bhattacharya:2014ara,Bazavov:2011nk}).

To overcome the computational  challenges faced when directly investigating the parameter space of the Columbia plot, additional extrapolation direction are used. Simulations at imaginary chemical potential can enlarge the first order region and help to determine it at $\mu_B = 0$ by limiting the search region. Similarly, simulations at variable $N_f$, even non integer ones, have been applied.

\subsection{Lower left corner\label{sec:lowerCorner}}
One common direction of investigation is along the $N_f=3$ diagonal of the Columbia plot. However, the first order region has only be found on $N_t =4$ lattices with unimproved staggered (Ref.~\cite{Karsch:2001nf,deForcrand:2003vyj}) or $O(a)$-improved Wilson \cite{Jin:2014hea} fermions. For finer lattices or with further improved actions, the phase transition remains elusive (Ref.~\cite{Bazavov:2017xul}).

A direct investigation along the $N_f=2$ border only yielded results for $N_t=4$ lattices with unimporved staggered or Wilson quarks (Ref.~\cite{Bonati:2014kpa,Cuteri:2017gci,Philipsen:2016hkv}).

\begin{figure}
 \begin{center}
  \includegraphics[width = 0.7\textwidth]{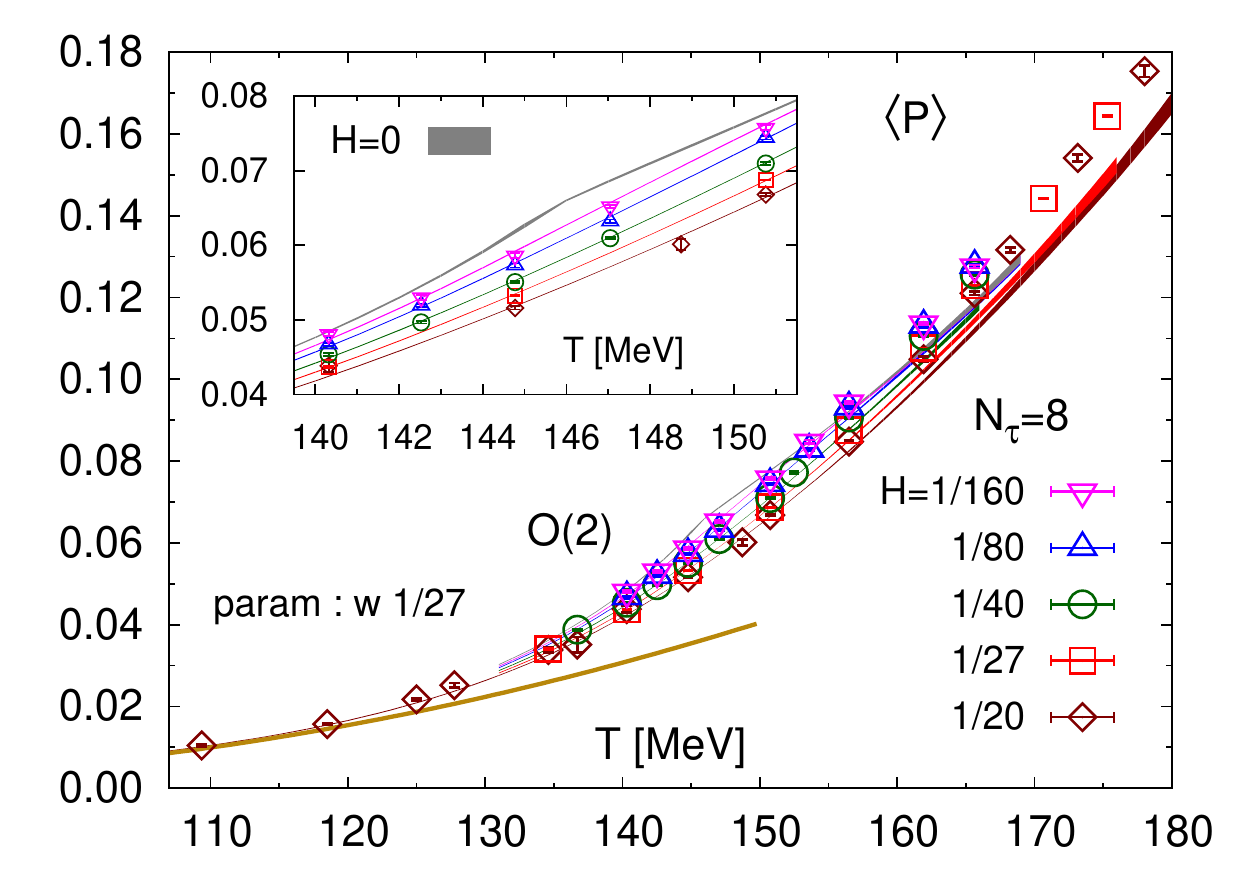}\\
  \vspace{12pt}
\includegraphics[width = 0.7\textwidth]{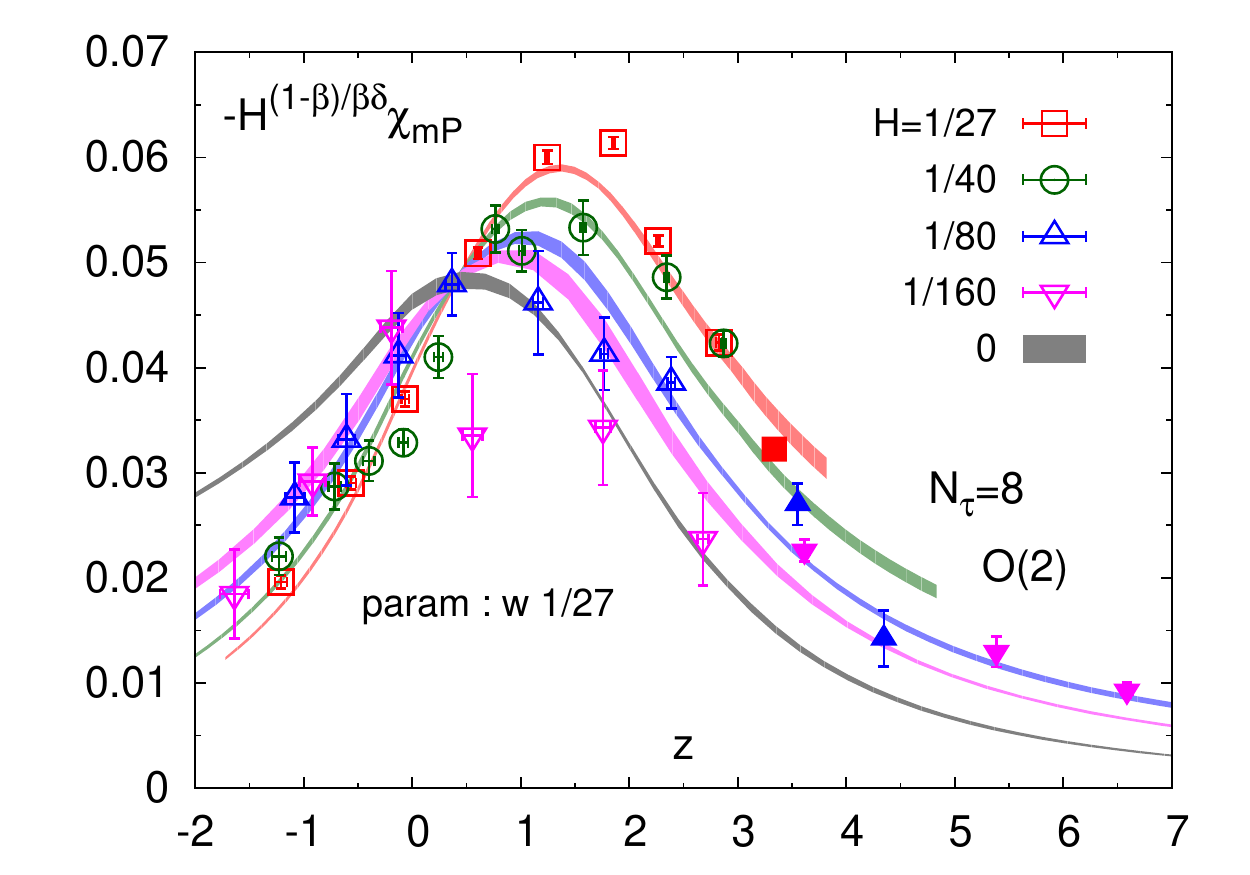}
 \end{center}
 \caption{(Ref.~\cite{Clarke:2020htu}) Top: The temperature dependence of the Polyakov loop $\langle P \rangle$. Bottom: The derivative of the Polyakov loop  as a function of the scaling varibale $z=z_0 t  H^{-1/\beta\delta}$ with $t=(T-T_c)/T_c$ and $z_0,\ T_c$ being  non-universal constants. \label{fig:KarschColumbia}}
\end{figure}

A recent investigation of $2+1$-flavor QCD with small quark masses is reported in Ref.~\cite{Clarke:2020htu}. It studies the the Polyakov loop expectation value $\langle P \rangle$ and the heavy quark free energy
\begin{equation}
 F_q(T,H)  = -T\ln \langle P \rangle 
= -\frac{T}{2} \lim_{\left|\vec{x}-\vec{y} \right|\rightarrow \infty}
\ln \langle P^{\phantom\dagger}_{\vec{x}} P^\dagger_{\vec{y}} \rangle 
\end{equation}
with $H=m_l/m_s$ on $N_t=8$ lattices. The analysis is done for several volumes to reach the limit $H\longrightarrow0$. Result for the Polyakov loop and its derivative are shown in Fig.~\ref{fig:KarschColumbia}. It is found that the scaling behavior is consistent with the scaling behavior for energy-like observables in the 3-d, $O(N )$ universality classes. The authors observe a singular behavior of the quark mass derivatives of $\langle P \rangle$ and $F_q/T$. In the chiral limit the divergence is consistent with the 3-d, $O(2)$ universality class.

\begin{figure}
 \begin{center}
  \includegraphics[width = 0.7\textwidth]{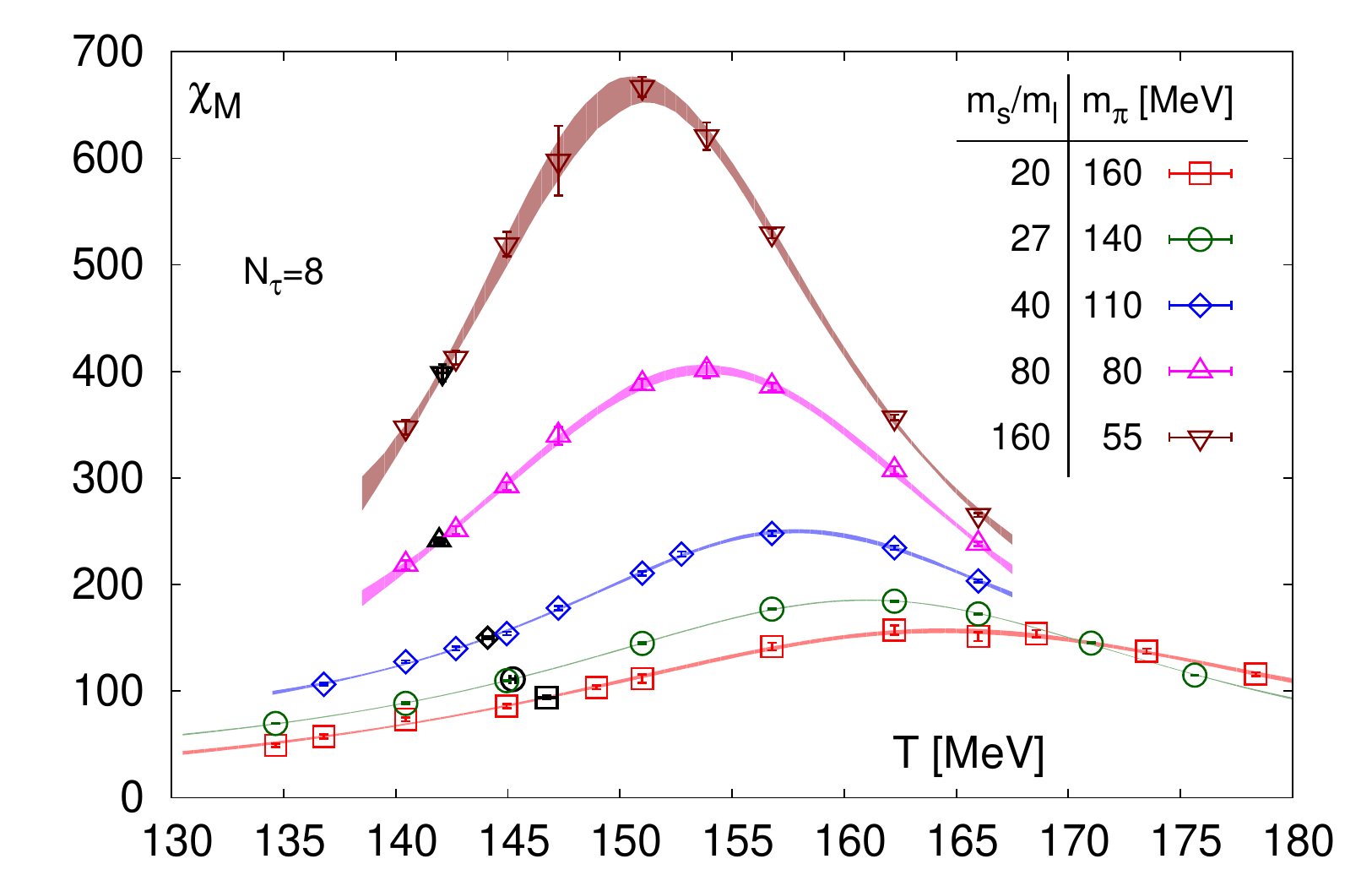}\\
  \vspace{12pt}
\includegraphics[width = 0.7\textwidth]{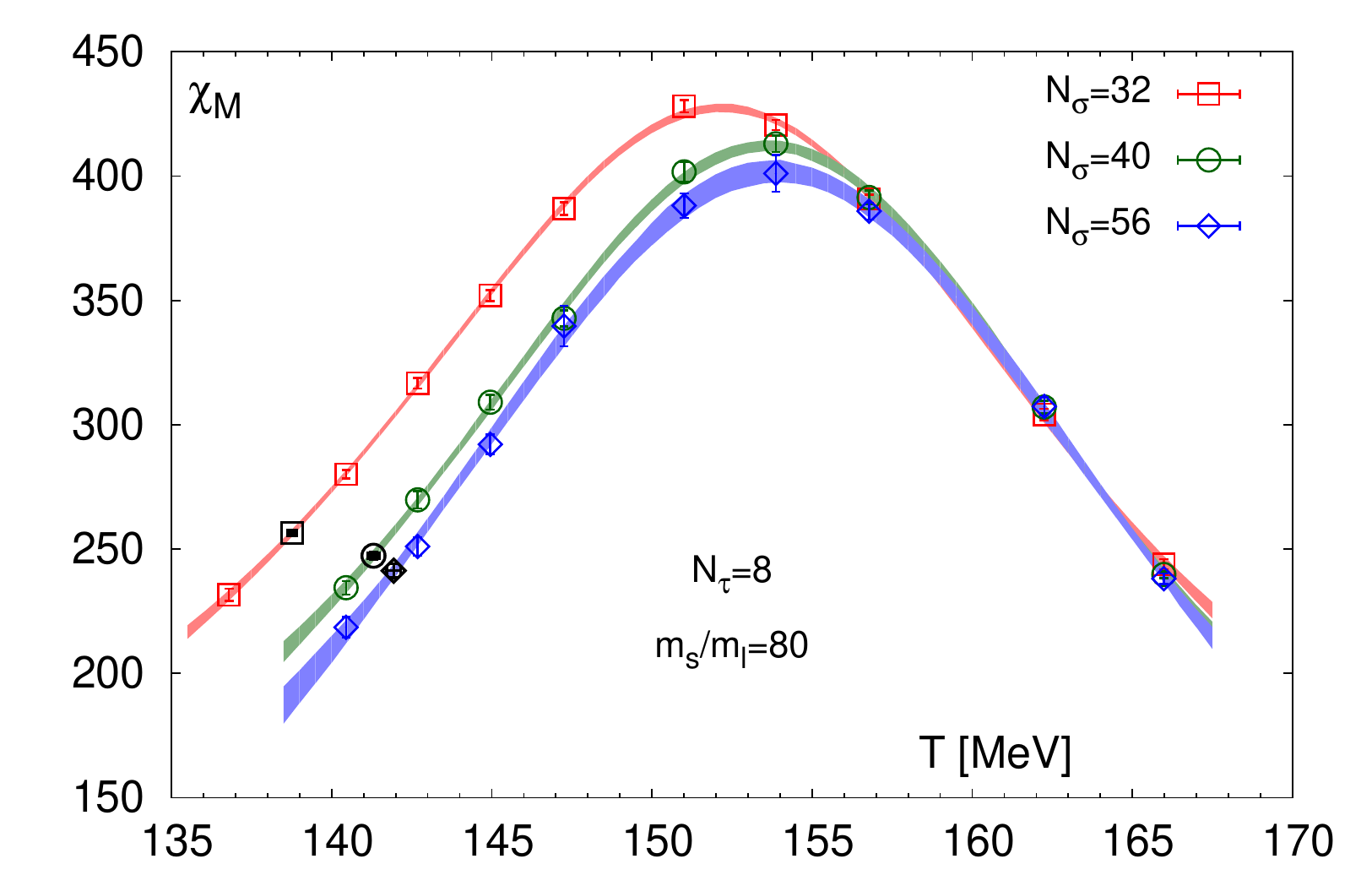}
 \end{center}
\caption{(Ref.~\cite{Ding:2019prx}) Top: The quark mass dependence of the chiral susceptibility as defined in equation~(\ref{eqn:chiM}).  Bottom: The volume dependence of the chiral susceptibility as defined in equation~(\ref{eqn:chiM}).\label{fig:Tc0}}
\end{figure}

Previously, scaling studies have been carried out in Ref.~\cite{Ding:2019fzc,Ding:2019prx}. Here the transition temperature in the chiral limit $T_c^0$ and the magnetization
\begin{equation}
 M=2 \frac{m_s\langle\bar{\psi}\psi\rangle_l-m_l\langle \bar{\psi}\psi\rangle_s}{f_K^4}
\end{equation}
as well as its derivative, the chiral susceptibility
\begin{eqnarray}
 \chi_M &=& \left. 
	m_s (\partial_{m_u}+\partial_{m_d}) M \right|_{m_u=m_d}
\nonumber \\
&=&m_s \frac{ m_s \chi_l - 2 \langle \bar\psi \psi\rangle_s -4 m_l \chi_{su}
}{f_K^4}\; \label{eqn:chiM}
\end{eqnarray}
were determined from lattices with $N_t = 6,8$ and 12 using the HISQ action. The pion mass was lowered down to $m_\pi \approx 55$~MeV and an $O(4)$ scaling ansatz was used to extrapolate to the chiral limit. The final result for the transition temperature in the chiral limit determined from different analyses is given as
\begin{equation}
 T_c^0 = 132^{+3}_{-6} \ \mathrm{MeV}.
\end{equation}
The dependence of the chiral susceptibility on the quark masses is shown in the top Fig.~\ref{fig:Tc0}. It was determined on an $N_t = 8$ lattice. The spacial extent of the lattice is increased with decreasing light quark mass $m_l$. It is $N_s = 32$ for $\frac{1}{H} = \frac{m_s}{m_l} = 20$ or 27, $N_s = 40$ for $\frac{1}{H} = 40$ and $N_s = 56$ for $\frac{1}{H} = 80$ or 160. The dependence on the volume for fixed $\frac{1}{H} = 80$ is shown on the bottom for Fig.~\ref{fig:Tc0}. 

\begin{figure}
  \begin{center}
  \includegraphics[width = 0.7\textwidth]{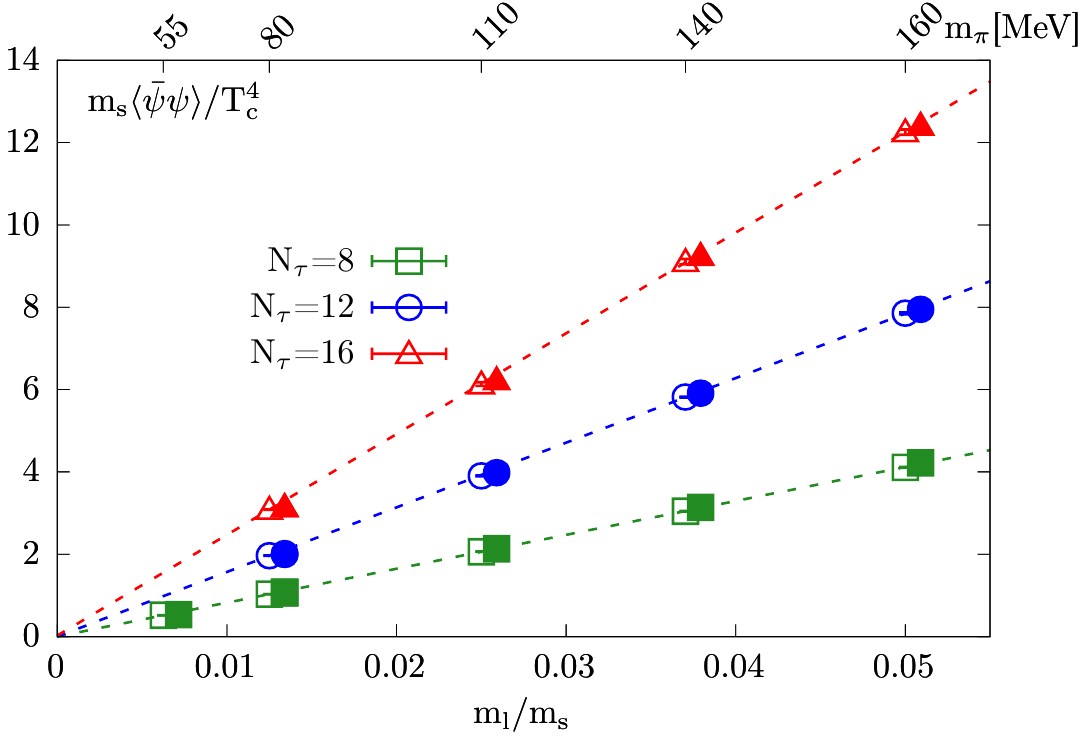}\\
  \vspace{12pt}
\includegraphics[width = 0.7\textwidth]{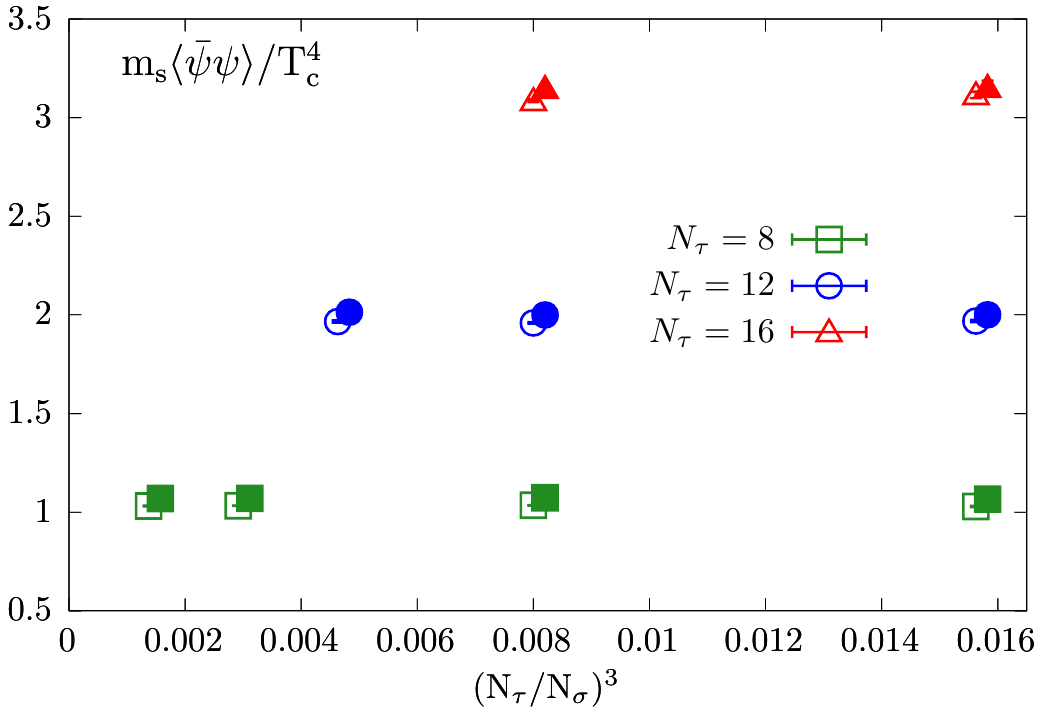}
 \end{center}
 \caption{(Ref.\cite{Ding:2020xlj}) Top: The quark mass dependence of the chiral condensate for different lattice spacing. Bottom: The volume dependence of the chiral condensate for different lattice spacings at $m_\pi=80$~MeV. \label{fig:DingNew}}
\end{figure}

A further study on the chiral limit is done in Ref.~\cite{Ding:2020xlj}. Here again pion masses down to $\mu_\pi = 55$~MeV are used. Lattice spacing of $a = 0.12$~fm~0.08~fm and 0.06~fm corresponding to $N_t = 8, 10$ and 16. The authors invesgate the dependence of the Dirac eigenvalue spectrum and the axial anomaly on the quark masses. For the chiral condensate they find that while there is a clear dependence on the quark masses and lattice spacing, the volume dependence is small. This is illustated in Fig.~\ref{fig:DingNew}.

\begin{figure}
 \centering
 \includegraphics[width=0.7\textwidth]{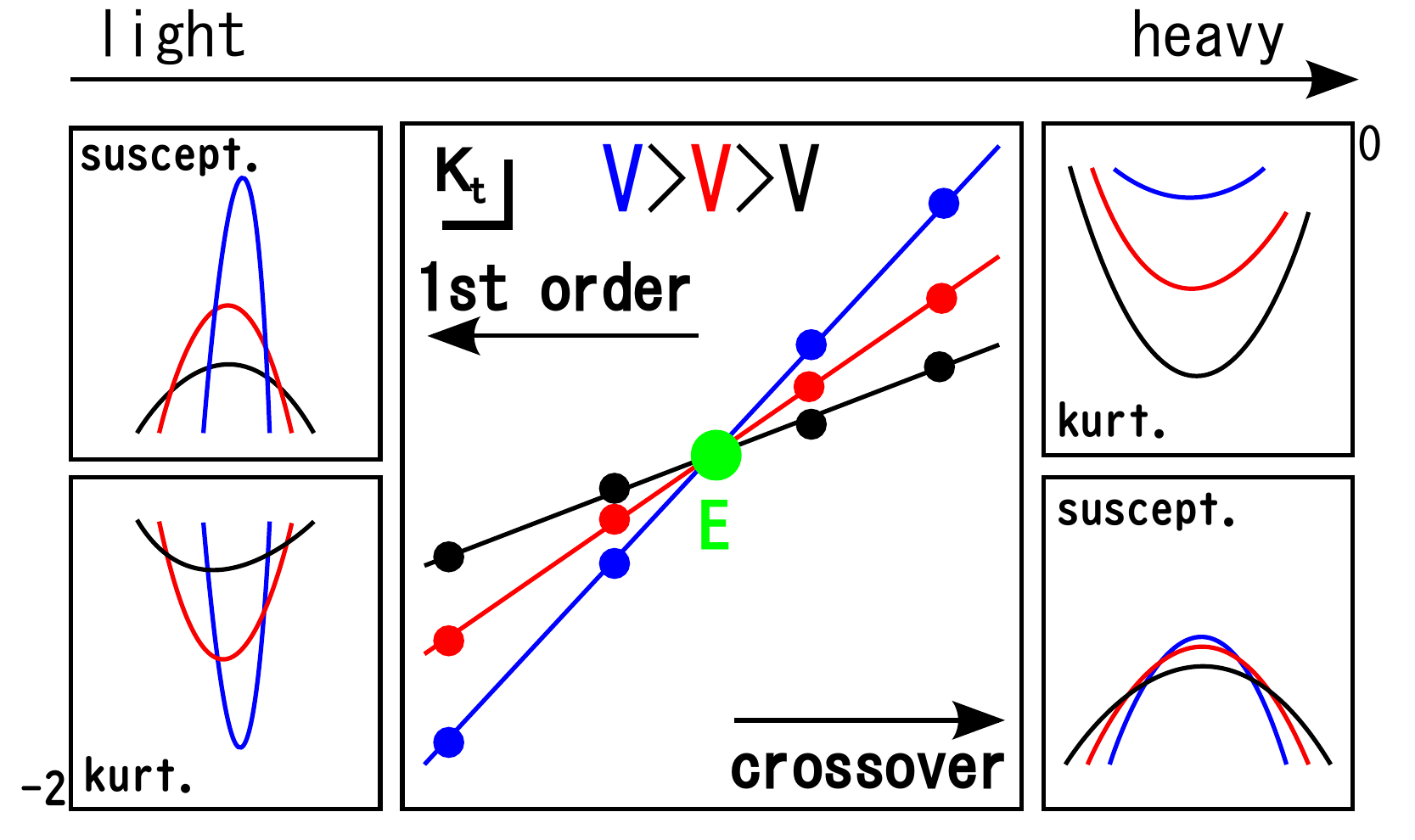}
 \caption{(Ref.~\cite{Kuramashi:2020meg}) Skatch of the kurtosis instersection analysis. The minimum of the kurtosis is determined from multiensemble reweighting and the then fittet for different volumes. The intersection is taken to be the value for the critical endpoint, as here there should be no volume scaling. \label{fig:cartoonNt12}}
\end{figure}

\begin{figure}
 \centering
 \includegraphics[width=0.7\textwidth]{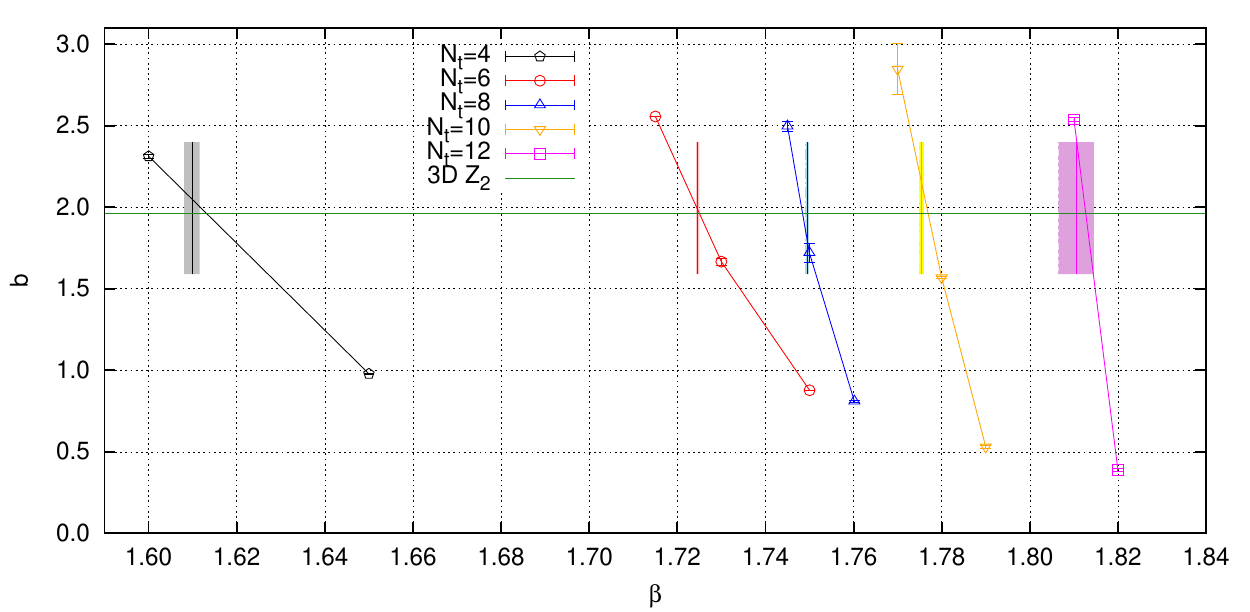}
 \caption{(Ref.~\cite{Kuramashi:2020meg}) The scaling exponent of the maximum susceptibility with the volumen: $\chi_{\mathrm{max}}\propto (N_s)^b$. The results are shown for $N_t=4,6,8,10$ and 12 with connecting lines. The green horizontal line shows the expectation for $Z(2)$-scaling. The shaded areas show the results for the critical $\beta$ from the kurtosis intersection analysis.  \label{fig:resultNt12}}
\end{figure}
Ref.~\cite{Kuramashi:2020meg} investigates the Columbia plot for the $N_f=3$ case with improved Wilson quarks and the Iwasaki gaugue action up to $N_t=12$. It uses multiensemble reweighting to determine the minimum of the kurtosis. The change in this minimum as a function of $\beta$ is then studied for different volumes. The point, where it agrees for several volumes defines the critical $\beta$ value. A sketch of this so called kurtosis intersection analysis is shown in figure~\ref{fig:cartoonNt12}. The results for this analysis for five different values of $N_t$ as well as the scaling of the maximal chiral sucsptibilty  $\chi_{\mathrm{max}}\propto (N_s)^b$ is shown in figure~\ref{fig:resultNt12}. The authors conclude that the $Z(2)$ line in the columbia plot for the $N_f=3$ case (along the diagonal) is located at a pion mass $m_\pi\lesssim110$~MeV.

\subsection{Imaginary chemical potential\label{sec:immuColumbia}}

\begin{figure}
 \centering
 \includegraphics[width=0.5\textwidth]{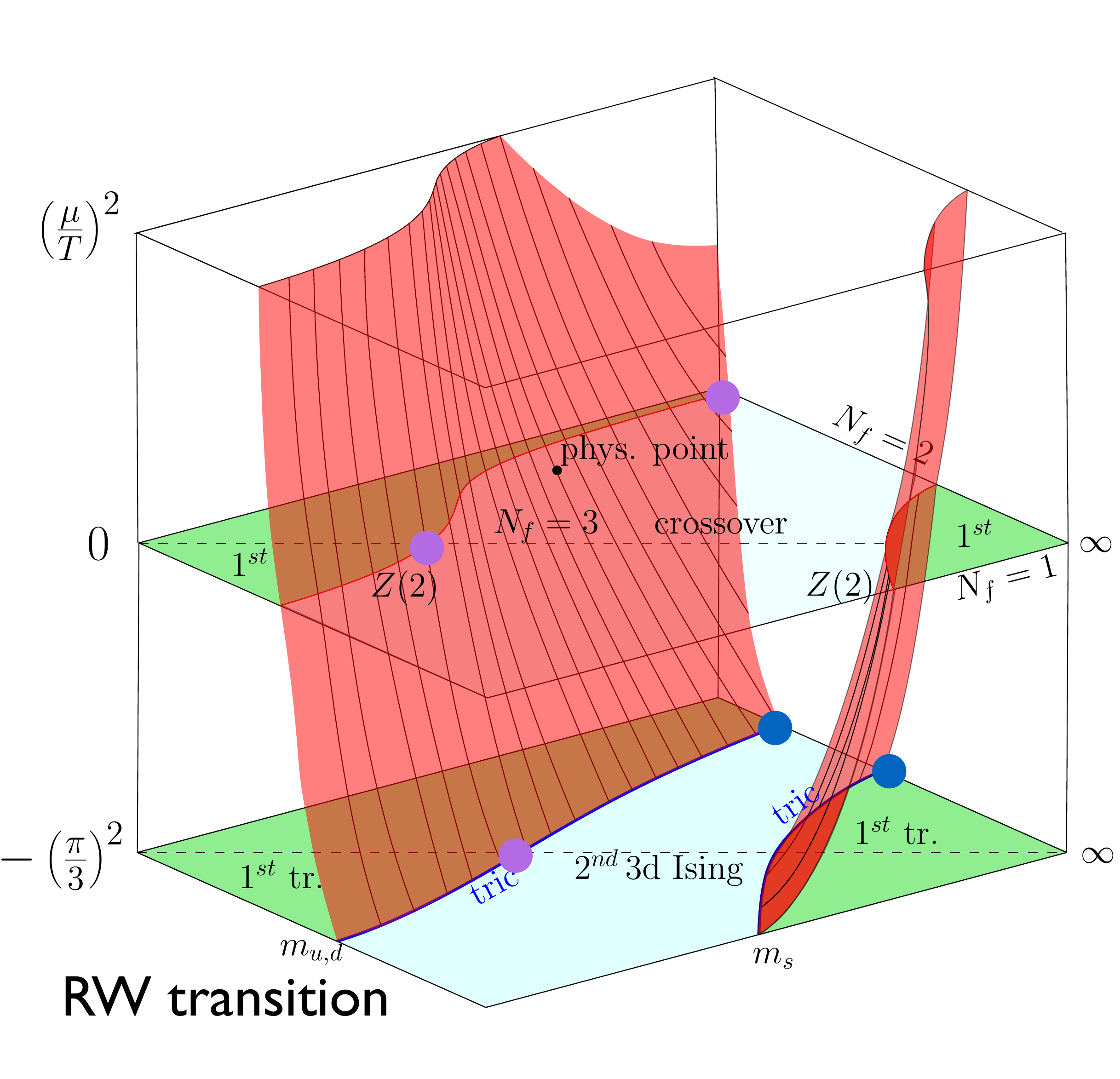}
 \caption{(Ref.~\cite{Philipsen:2019rjq}) Skatch of the 3d-Columbia plot with added chemical potential.\label{fig:3d}}
\end{figure}
One way to enlarge the first order region of the Columbia plot, at least on corse lattices, is the introduction of an imaginary chemical potential. As discussed in section~\ref{sec:anaCont} there is no sign problem in QCD the case of an purley imaginary chemical potential. A sketch of this 3d-Columbia plot is shown in Fig.~\ref{fig:3d}.  When adding the chemical potential axis to the Columbia plot, its curvature at $\mu=0$ is negative. This increases, therefore, the first order region for imaginary chemical potential and decreases it for real $\mu$. This investigations have been done both with unimporved Wilson (Ref.~\cite{Philipsen:2014rpa}) and staggered  (Ref.~\cite{deForcrand:2010he,Bonati:2010gi}) fermions on $N_t=4$ lattices.

\subsection{Variable $N_f$\label{sec:Nf}} 

\begin{figure}
 \centering
 \includegraphics[width=0.7\textwidth]{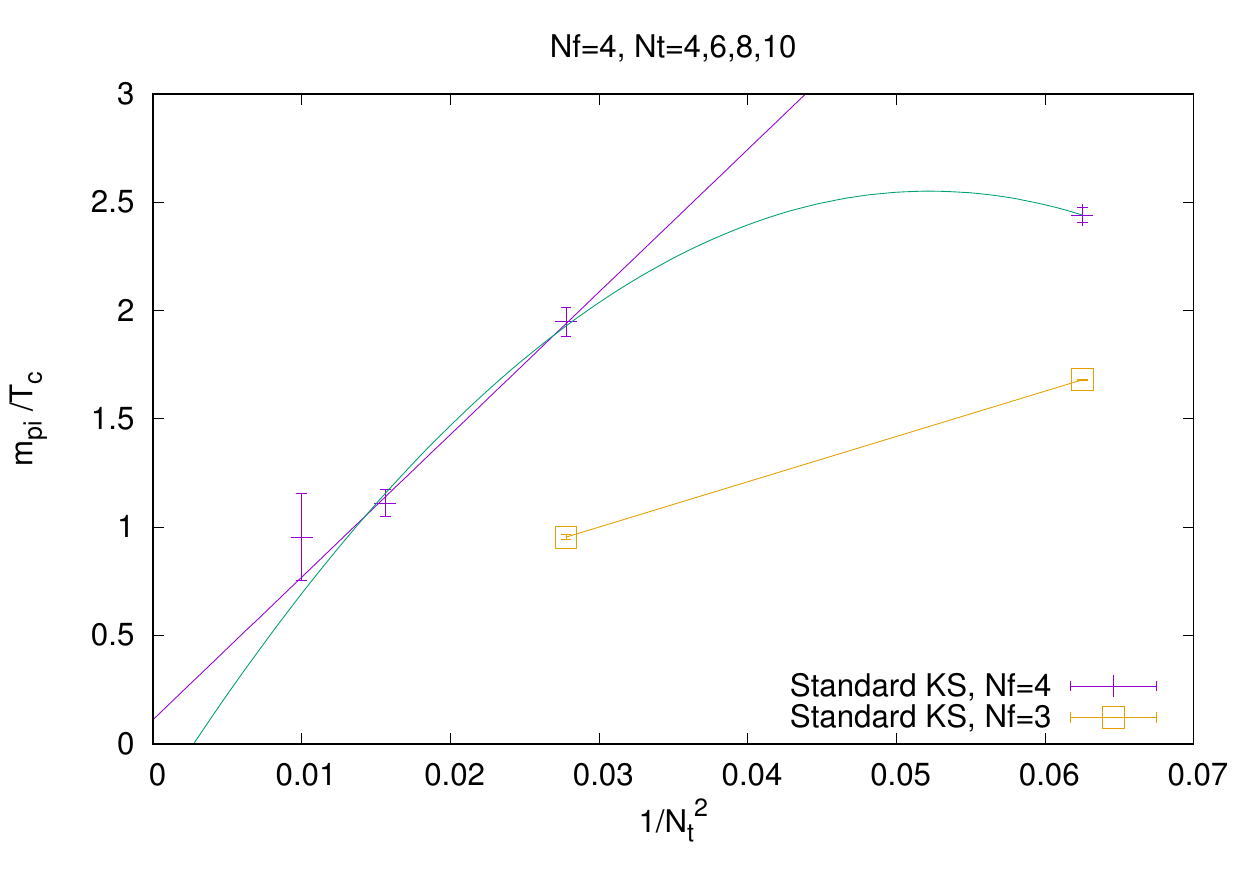}
 \caption{(Ref.~\cite{deForcrand:2017cgb}) The value of the critical pion mass for four or three staggered quarks, for different values of $N_t$. For four flavours the critical mass is larger, allowing to to to finer lattices.\label{fig:Nf4Coulumbia}}
\end{figure}

Another way to increase the strength of the phase transition, and therefor helping with the investigation of the Columbia plot, is to increase the number of falvours $N_f$. Ref.~\cite{deForcrand:2017cgb} studies the Columbia plot with four staggered quarks and the Wilson plaquette action. The choice of $N_f=4$ has the added benefit that no rooting is required. Simulations were performed for $N_t=4$, 6, 8 and 10 for several spacial lattice extends. Figure~\ref{fig:Nf4Coulumbia} shows the difference between the critical pion mass for four or three staggered quarks. The increased number of flavours increases the pion mass, so that simulations on lattices with larger temporal extend are possible.

\begin{figure}
 \centering
 \includegraphics[width=0.7\textwidth]{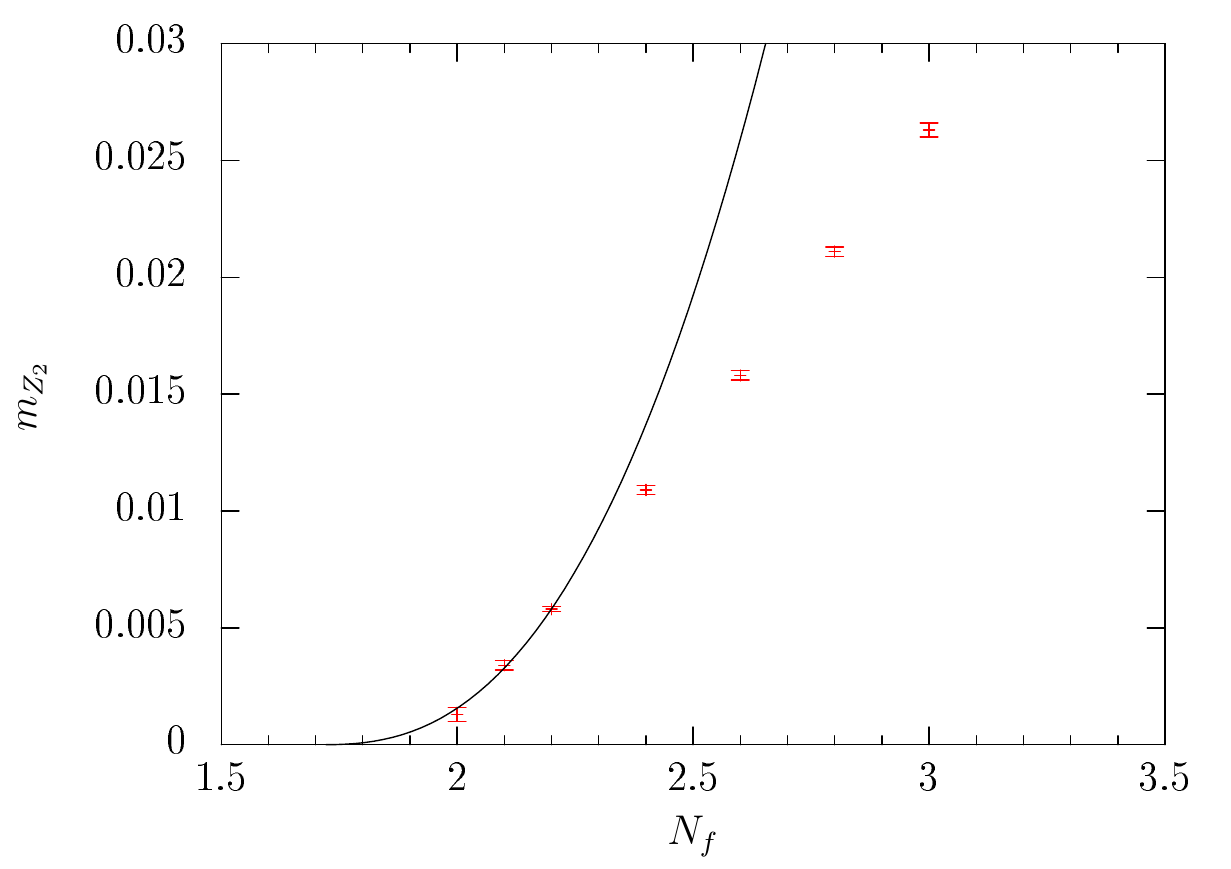}
\hfill
 \includegraphics[width=0.7\textwidth]{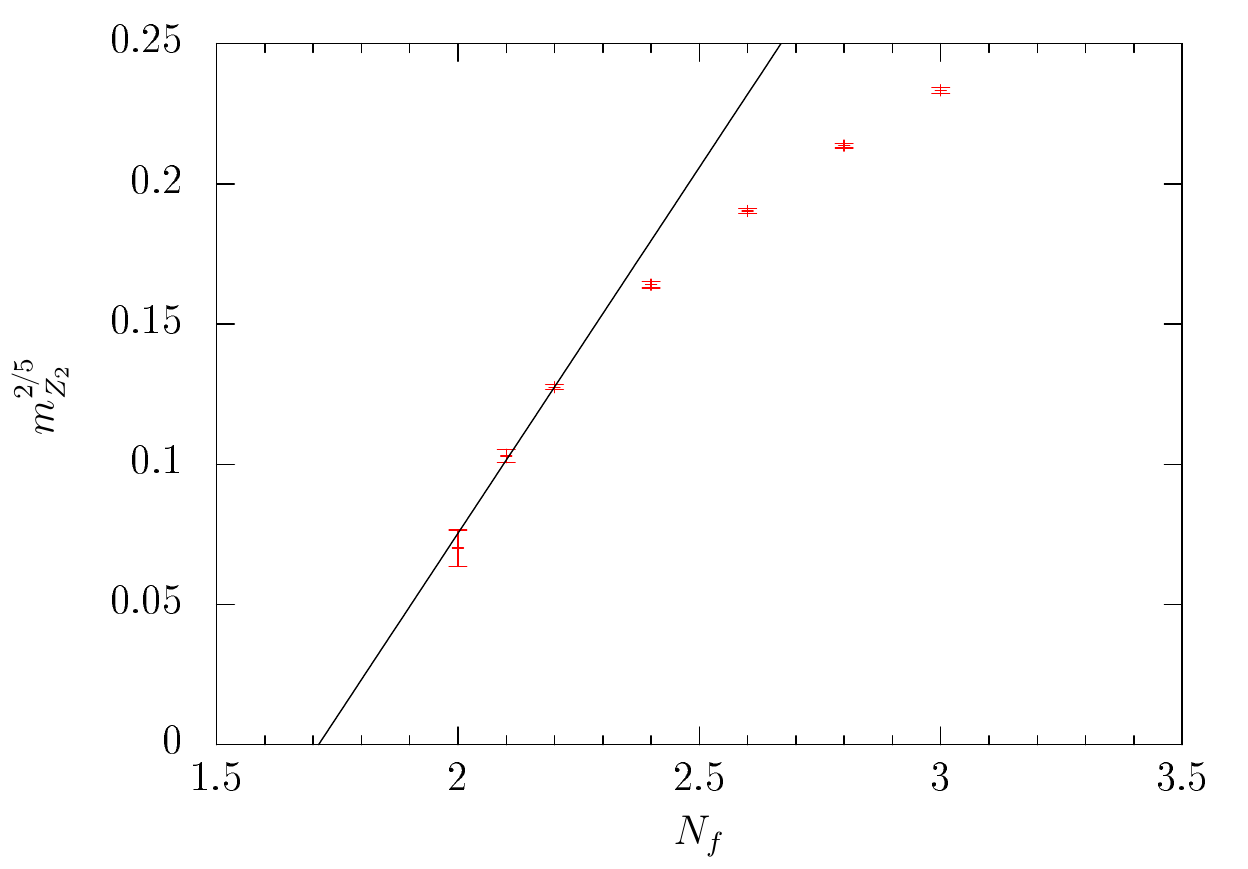}
 \caption{(Ref.~\cite{Cuteri:2017gci}) Top: The critical mass $m_{Z_2}$ as a function of the number of flavour. The black line is the rescaled version of the fit done to the resacled mass below. Bottom: The rescaled critical mass $m_{Z_2}$ as a function of the number of flavour. As well as a linear fit to the data. \label{fig:NfColumbia}}
\end{figure}
To get a smooth transition between the  integer number of flavours, it is even possible to consider non-integer powers of the quark determinant (Ref.~\cite{Cuteri:2017gci}). This allows for a smooth interpolation in $N_f$. Since in the ciral limit the transition is second order for $N_f=2$ and first order for $N_f=3$ there has to be a tricritical scaling area, with a critical point at $N_f^{\mathrm{tric}}$, in between. Ref.~\cite{Cuteri:2017gci} presents results on $N_t=4$ lattices for $N_f=2.8$, $2.6$, 2.4, 2.2 and 2.1. The dependence of the critical mass $m_{Z_2}$ for a $Z(2)$ scaling is expected to follow
\begin{equation}
 m_{Z_2}^{2/5}(N_f) = C\left( N_f-N_f^{\mathrm{tric}} \right).
\end{equation}
The result are shown in the top of figure~\ref{fig:NfColumbia}. The rescaled mass is shown in the bottom of figure~\ref{fig:NfColumbia}. A linear function was fitted to the rescaled mass to determine $C$ and $N_f^{\mathrm{tric}}$.

\subsection{Overview - lower left corner\label{sec:overview}}
\begin{figure}
 \centering
 \includegraphics[width=0.6\textwidth]{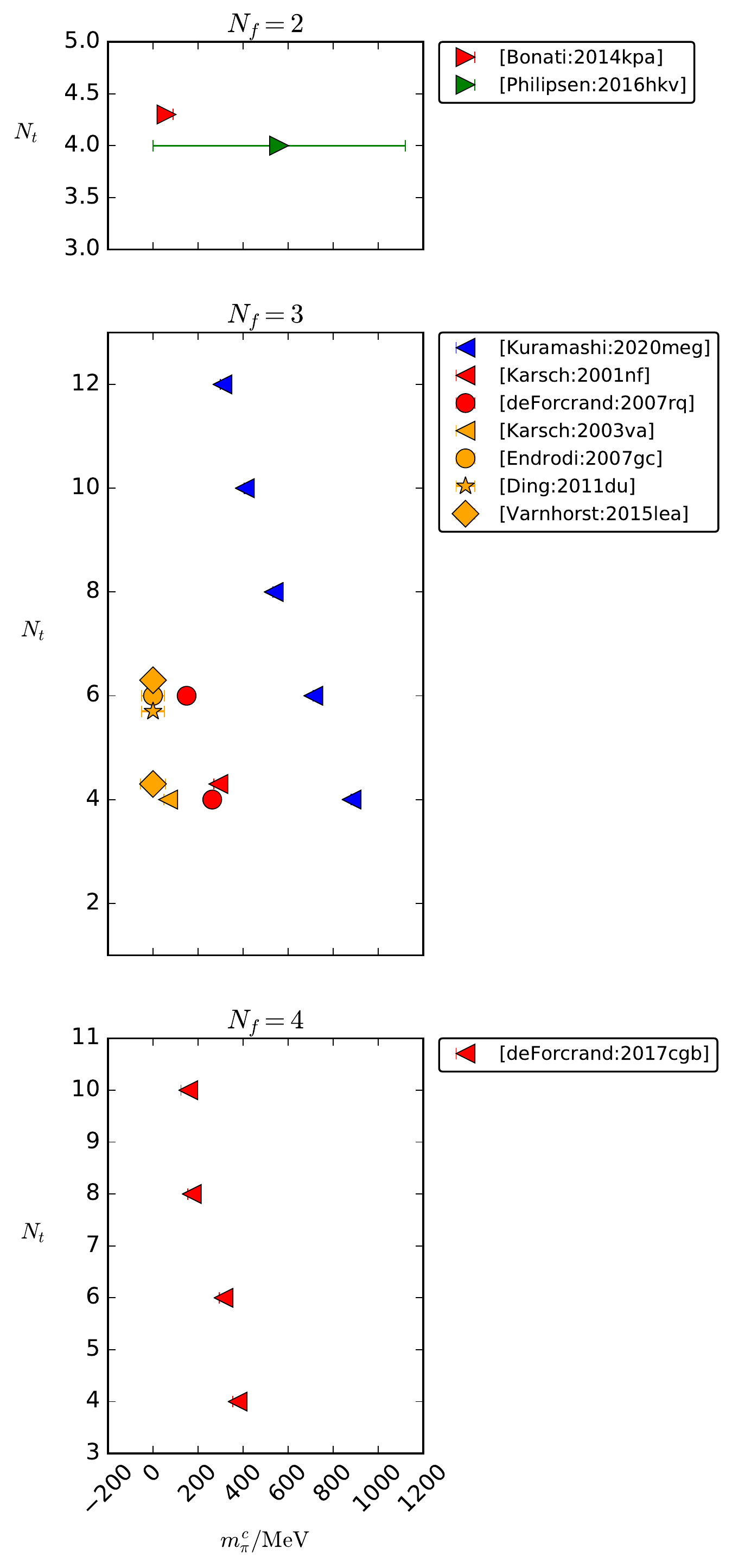}
 \caption{Overview over different available results for the critical pion mass in the Columbia plot for $N_f=2$, 3 and 4. Values are from Ref.~\cite{Varnhorst:2015lea,deForcrand:2017cgb,Kuramashi:2020meg,Karsch:2001nf,Karsch:2000kv,deForcrand:2007rq,Karsch:2003va,Endrodi:2007gc,Ding:2011du,deForcrand:2003vyj,Jin:2014hea}. The color of the points denotes the fermion type used in the simulations. Red stands for unimproved staggered, orange for imporved staggered, green for unimporved Wilson and blue for imporved Wilson quarks. Upper bounds are denoted by zero with the errorbar showning the bound. Values given without any error, are denoted with 50\% relative error. Some values were converted to MeV using the value for the critical temperature from Ref.~\cite{Karsch:2000kv} $T_c = (156\pm8)$~MeV. Similar values for the same $N_t$ have been shifted lightly for more clarity. \label{fig:columbiaOverview}}
\end{figure}

The investigation of the Columbia plot remains a challenge for lattice QCD. For the determination of the second order line in the lower left corner (see figure~\ref{fig:Columbia}) several results are available on finite lattices however there is no clear solution for the continuum limit. Figure~\ref{fig:columbiaOverview} shows an overview over the results for different fermion types. For this plot, some conversions were necessary. Some values were converted to MeV using the value for the critical temperature from Ref.~\cite{Karsch:2000kv} $T_c = (156\pm8)$~MeV for coarse lattices and staggered quarks. Upper bounds are denoted by zero with the errorbar showing the bound. Values given without any error, are assumed to have 50\% relative error. The color of the points denotes the fermion type used in the simulations. Red stands for unimproved staggered, orange for improved staggered, green for unimproved Wilson and blue for improved Wilson quarks. The largest number of results is available for the three flavour case. Here both Wilson and staggered quarks show a shrinking value when the lattices become finer. The same trend can be observed for staggered quarks with $N_f=4$. For both two and three flavors, staggered quarks show smaller values than Wilson quarks. Also for staggered quarks an improvement of the quark formulation reduces the value for the critical pion mass. Assuming that all quark formulations agree in the continuum, which they should, it seems that still a large amount of computational power is needed to settle the shape of the lower left corner of the Columbia plot. Investigating this region becomes especially challenging since both staggered and Wilson quarks suffer from large lattice artifacts when the quark masses become small (Ref.~\cite{Iwasaki:1996zt}).

\subsection{Upper right corner\label{sec:upperCorner}}

\begin{figure}
 \centering
 \includegraphics[width=0.7\textwidth]{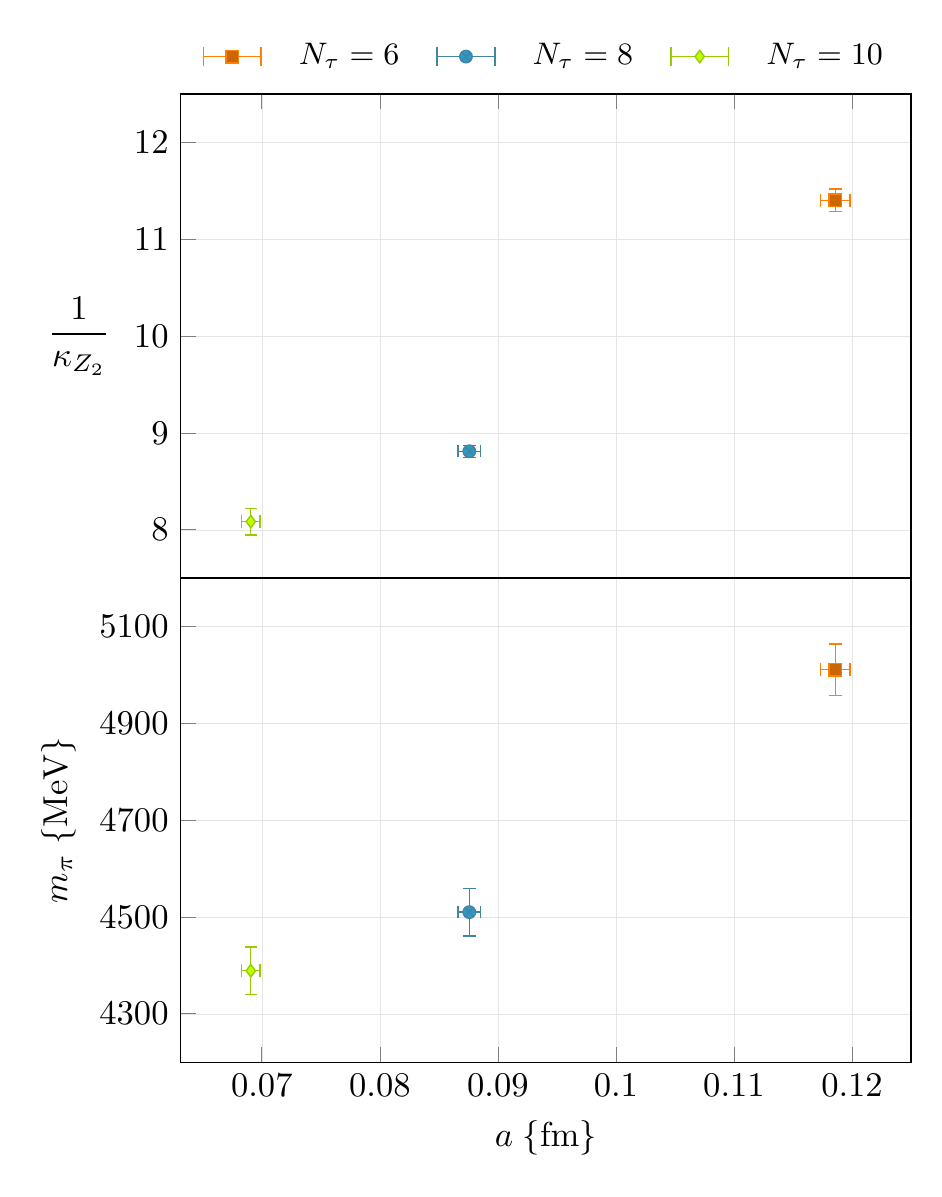}
 \caption{(Ref.~\cite{Cuteri:2020yke}) Top: The critical value for the hopping parameter $\kappa$ (see equations (\ref{eqn:kappa}) and (\ref{eqn:kappaWilson})) as a function of the lattice spacing. Bottom: Value of the critical pion mass as a function of the lattice spacing.\label{fig:columbiaUpperRight}}
\end{figure}

Recent progress on the determination of the upper right corner of the Columbia plot (see figure~\ref{fig:Columbia}) was reported in Ref.~\cite{Cuteri:2020yke}. The authors determine the critical pion mass for the $N_f=2$ flavour case. This corresponds to the upper border of the Columbia plot. The lattice setup consisted of two dynamical, degenerate Wilson fermions on lattices with $N_t=6,$~8~and~10. These $N_t$ values correspond to lattice spacings between 0.07~fm and 0.12~fm. The results for the critical pion masses $m^c_\pi$ for all three lattices are shown in the bottom of figure~\ref{fig:columbiaUpperRight}. While a precise continuum limit is not possible from these three lattices alone the authors estimate a conutinuum value of $m^c_\pi\approx 4$~GeV with an error of about 20\%.
In addition to the ciritical pion mass, also the critical hopping parameter $\kappa_{Z_3}$ has been determined. It appears in fermion matrix for the Wilson quark action as
\begin{equation}\label{eqn:kappaWilson}
    D(n_1|n_2) = \delta_{n_1,n_2} - \kappa \sum_{\mu=\pm 1}^{\pm 4}\left[ (\mathbb{1}-\gamma_\mu)U_{\mu}(n_1) \delta_{j+\hat\mu,n_2} \right] 
\end{equation}
and controlles the bare quark mass $m_b$ as
\begin{equation}\label{eqn:kappa}
    \kappa = \frac{1}{2(am_b+4)}.
\end{equation}

These results do not agree with those previously computed in Ref.~\cite{Ejiri:2019csa}. Here  reweighting from quenched QCD has been used to investigate the end-line in 2+1 flavor QCD. For the $N_f=2$ and $N_t=6$ case, the results differ by about 50\%. Since the same lattice action has been used in both cases, the discrepancy is most likely related to the different methods and not to the cut off effects. Ref.~\cite{Ejiri:2019csa} describes finite volume effects as well as effects related to the hopping parameter expansion which seems to shift the results in the direction of  Ref.~\cite{Cuteri:2020yke}. 

\clearpage
\section{Magnetic fields\label{sec:magneticFields}}
When dealing with heavy ion collision experiments, in addition to a finite chemical potential, also the effects of magnetic fields have to be considered (Ref.~\cite{Kharzeev:2007jp,Skokov:2009qp,Deng:2012pc}). Form zero temperature lattice QCD studies with staggered fermions (Ref.~\cite{DElia:2010abb,Shovkovy:2012zn,Ding:2020hxw}) the so-called magnetic catalysis is found. It describes that the chiral condensate, which is an order parameter of the QCD chiral transition, increases with the strength of the magnetic field. From this it was derived that also the transition temperature increases with the magnetic field strength.

\begin{figure}
 \centering
 \includegraphics[width=0.7\textwidth]{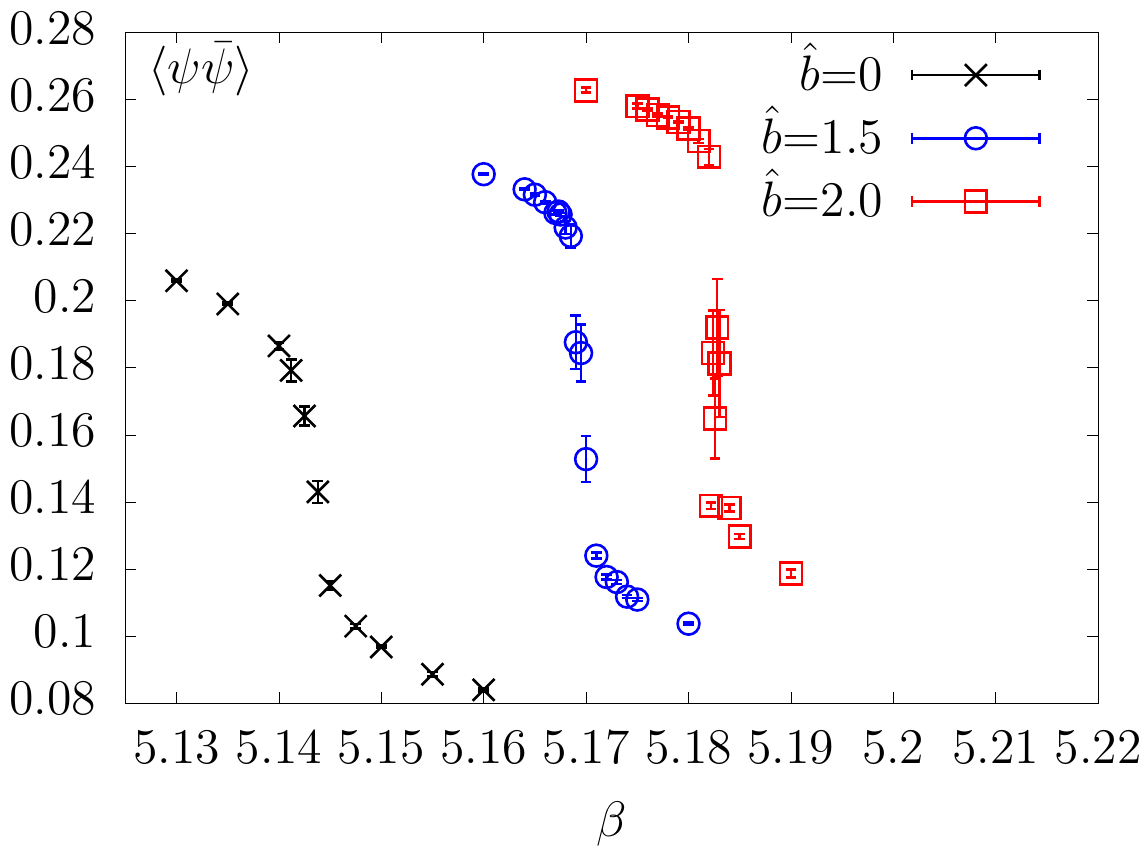}\\
 \vspace{12pt}
 \includegraphics[width=0.7\textwidth]{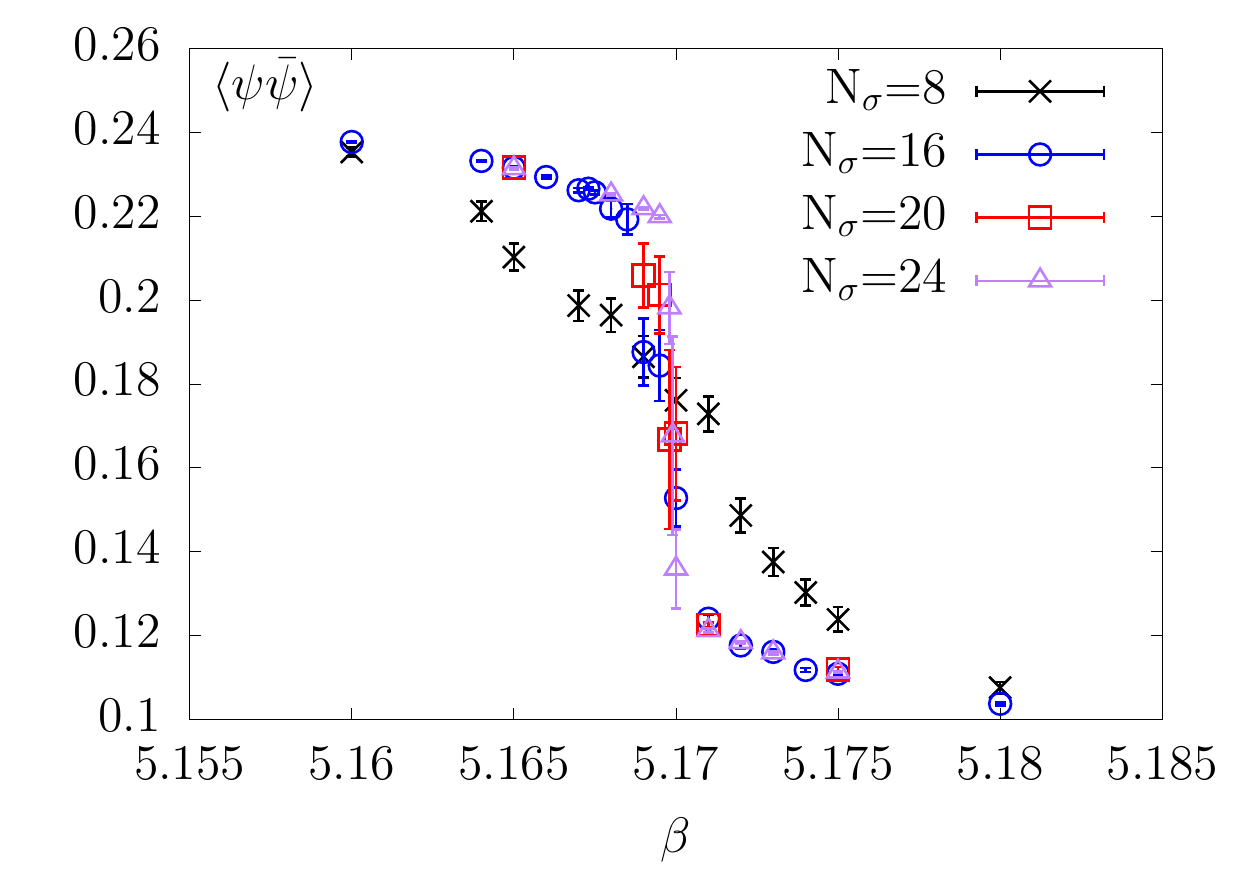}
 \caption{(Ref.~\cite{Ding:2020inp}) Top: The chiral condensate on an $16^3\times4$ lattice for different magnetic field strengths $\hat b = a\sqrt{eB}$. Bottom: The chiral condensate for different volumes on lattices with temporal extend $N_t =4$ at a magnetic field strength of $\hat b = a\sqrt{eB}=1.5$. \label{fig:DingMagnetic}}
\end{figure}

A recent study (Ref.\cite{Ding:2020inp}) investigated the chiral phase structure of three flavour QCD with a magnetic field on $N_t =4$ lattices with four different volumes between $N_s =8$ and $N_s=24$. It uses a Wilson plaquette action with staggered fermions and pion masses of $m_\pi\approx 280$ MeV and two values of finite magnetic field strength. They find a strengthening in the transition, which turns to first order, when a magnetic field is added. The results for the chiral condensate for different magnetic field strengths and lattice volumes are shown in Fig.~\ref{fig:DingMagnetic}. The first order nature is derived both from the volume scaling as well as from meta stable states of the chiral condensate in the simulation stream. They also see an increase in the transition temperature with the magnetic field strength.

On the other hand, Ref.~\cite{Bali:2011qj,Bali:2012zg,Ilgenfritz:2013ara,Bornyakov:2013eya,Bali:2014kia,Tomiya:2019nym} find a decreasing transition temperature with growing magnetic fields. Here continuum extrapolations or improved actions were employed. It is assumed that the discrepancy is related to discretization artifacts (Ref.~\cite{Ding:2020inp}). The decreasing of the transition temperature as well as the chiral condensate is called inverse magnetic catalysis. Most recently, Ref.~\cite{Tomiya:2019nym} uses the HISQ quark action to study the effects of magnetic fields on a $16^3\times6$ lattice. They can observe the inverse magnetic catalysis even on a finite lattice.

 To get a qualitative understanding of the processes leading to a magnetic catalysis or inverse magnetic catalysis, many model studies have been performed (Ref.~\cite{DElia:2011koc,Shovkovy:2012zn,Andersen:2014xxa,Kojo:2012js,Bruckmann:2013oba,Fukushima:2012kc,Ferreira:2014kpa,Yu:2014sla,Feng:2015qpi,Li:2019nzj,Mao:2016lsr,Gursoy:2016ofp,Xu:2020yag}). Lattice studies with heavy pion masses (Ref.~\cite{DElia:2018xwo,Endrodi:2019zrl}) hint towards an explanation for the inverse magnetic catalysis unrelated to the decreasing of the chiral condensate, but related to the quark masses.

  \begin{figure}
 \centering
 \includegraphics[width=0.7\textwidth]{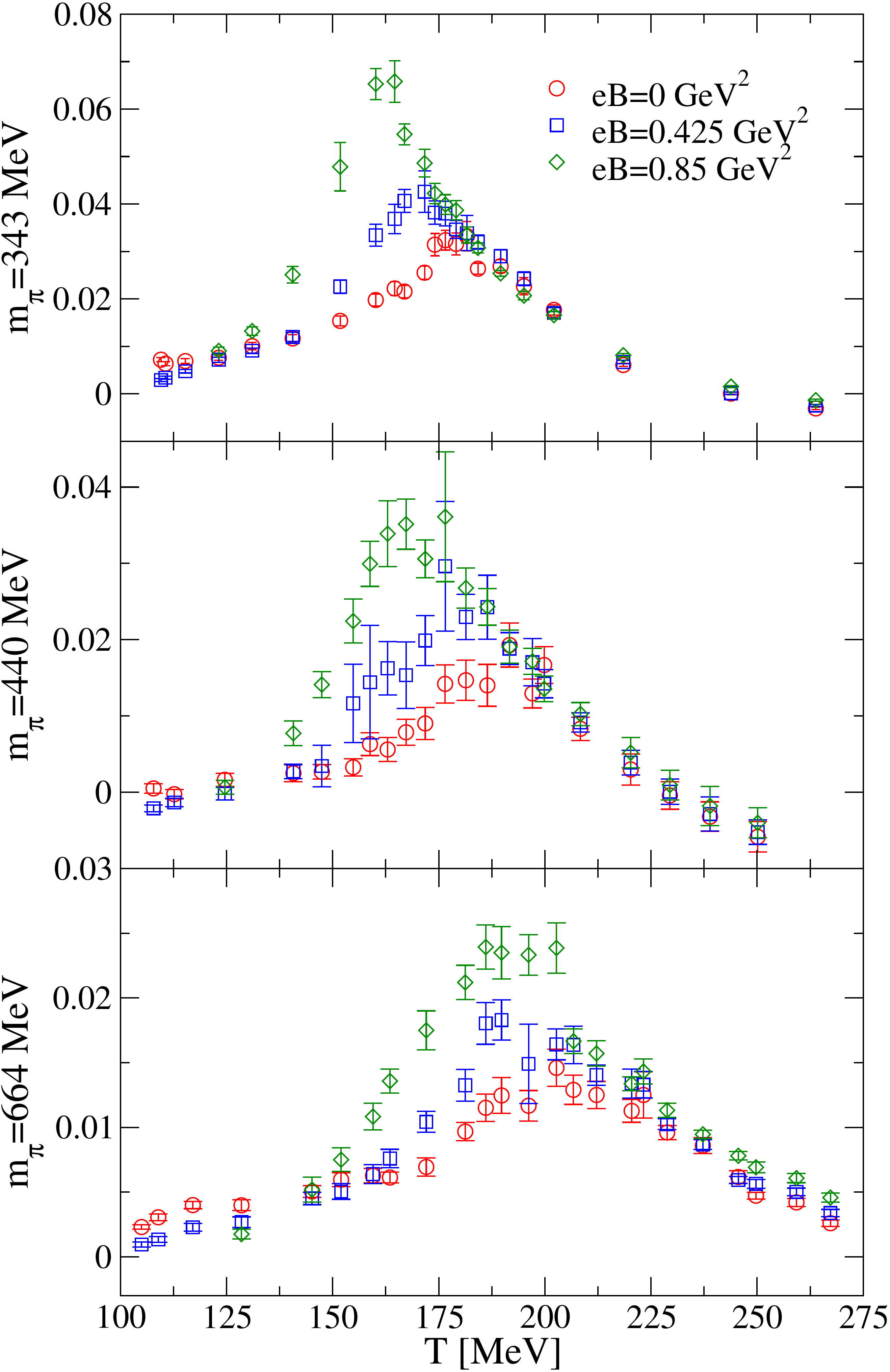}
 \caption{(Ref.~\cite{DElia:2018xwo}) The chiral susceptibility for different strengths of the magnetic field and different pion masses. Simulations were done with $N_f=2+1$ stout smeard staggered fermions on $N_t=6$ lattices. One can observe a decreasing peak position and therefore transition temperature for all pion masses. \label{fig:DEliaMagnetic}}
\end{figure}

\begin{figure}
 \centering
 \includegraphics[width=0.7\textwidth]{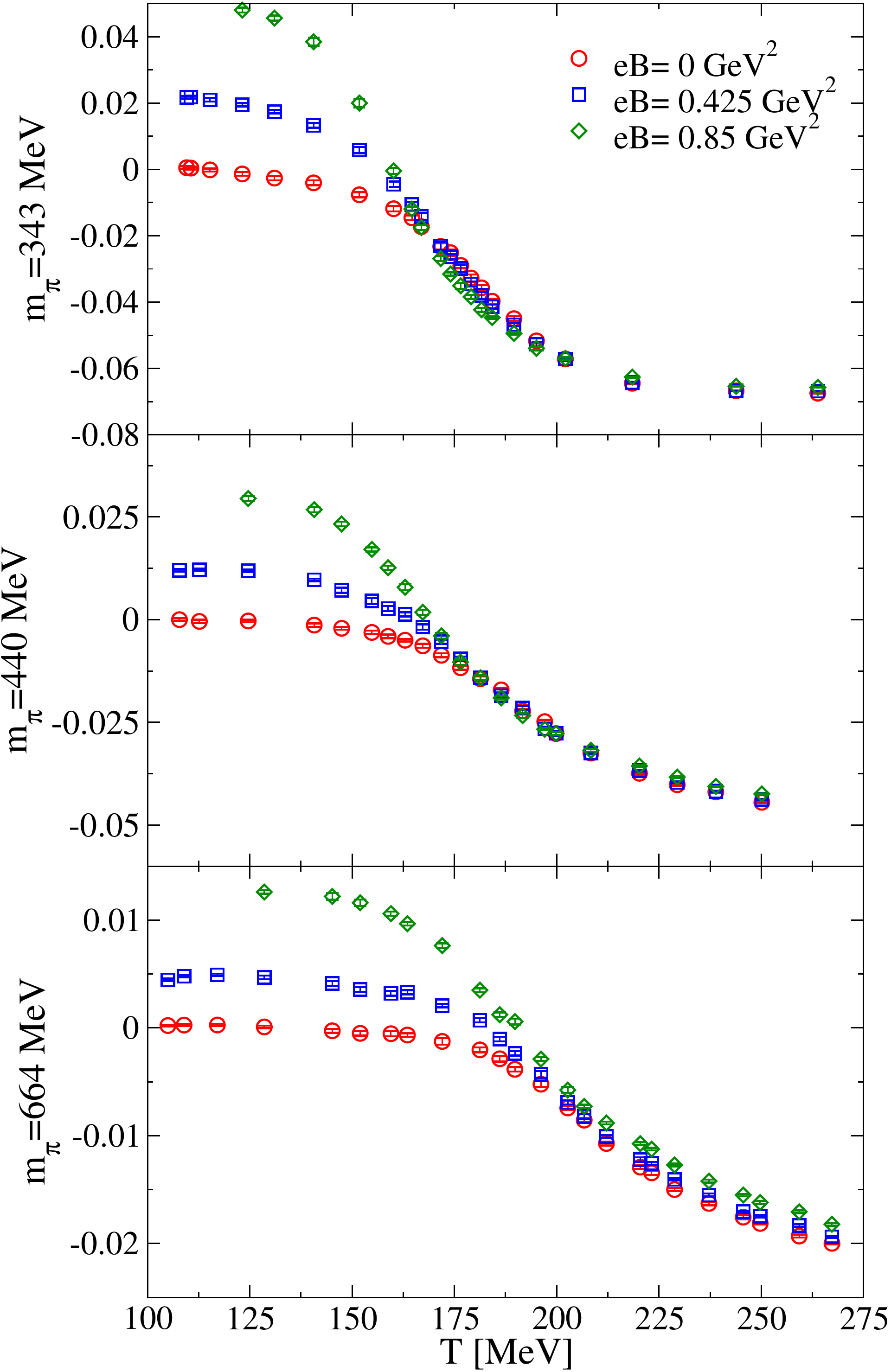}
 \caption{(Ref.~\cite{DElia:2018xwo}) The chiral condensate for different strengths of the magnetic field and different pion masses. Simulations were done with $N_f=2+1$ stout smeard staggered fermions on $N_t=6$ lattices. One can observe that while for the lightest pion mass the condensate decreases for a stronger magnetic field, it increases for the highest pion mass. \label{fig:DEliaMagnetic2}}
\end{figure}

Ref.~\cite{DElia:2018xwo} uses 2+1 flavours of stout smeared staggered quarks on $N_t=6$ lattices. They use pion masses up to $m_\pi\approx600$~MeV and investigate the transition temperature as a function of the magnetic field strength. The transition temperature is determined both from the chiral condensate and the Polyakov loop leading to similar results. The chiral condensate for different strengths of the magnetic field and different pion masses is shown in figure~\ref{fig:DEliaMagnetic} and figure~\ref{fig:DEliaMagnetic2}. They find that, while the decrease of the transition temperature is present for all pion masses, the decrease of the chiral condensate, however, is not observable for the highest pion masses. This strengthens the idea that the inverse magnetic catalysis, the decrease of the chiral condensate, is not directly related to the transition temperature. The authors suggest that the decrease of the transition temperature might be a deconfinement catalysis (Ref.~\cite{Bonati:2016kxj}).

\begin{figure}
 \centering
 \includegraphics[width=0.7\textwidth]{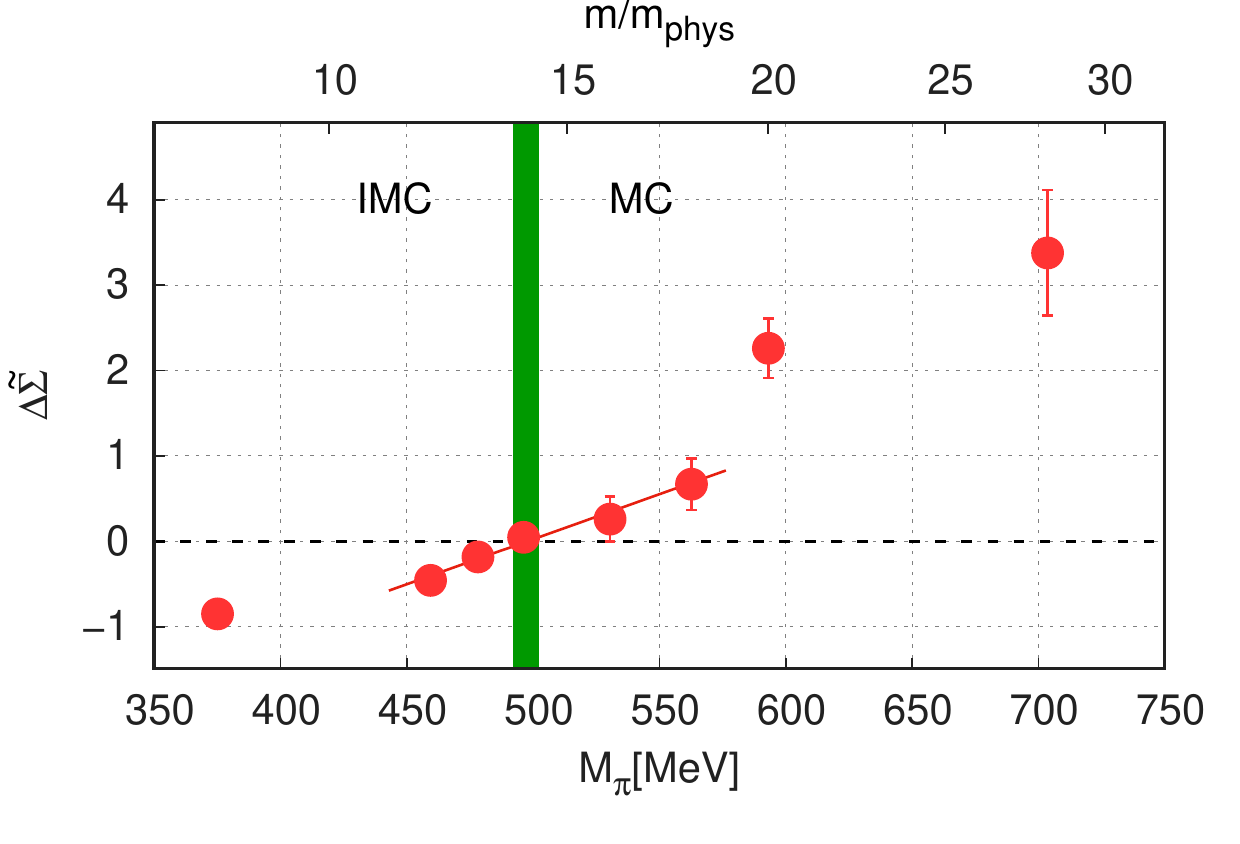}\\
 \vspace{12pt}
 \includegraphics[width=0.7\textwidth]{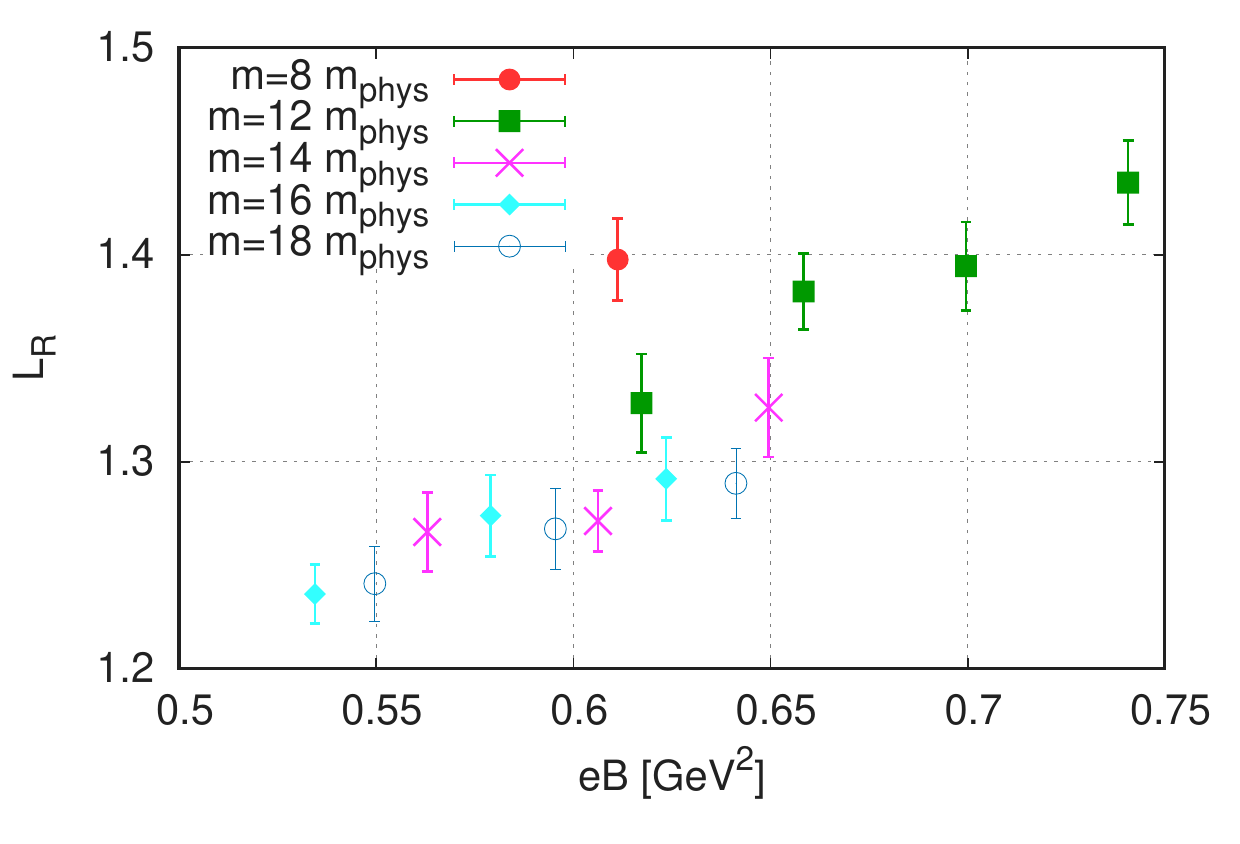}
 \caption{(Ref.~\cite{Endrodi:2019zrl}) Top: The dependence of the chiral condensate $\Delta \tilde \Sigma$ defined in equation~(\ref{eq:DS}) as a function of the pion mass in a magnetic field of $eB_0=0.6$~GeV. The green line separates the area of the inverse magnetic catalysis (IMC) from the area of the magnetic catalysis (MC) Bottom: The Polyakov loop ratio as defined in equation~(\ref{eqn:LR}) as a function of the magnetic field close to the turning point between inverse magnetic catalysis and magnetic catalysis. \label{fig:EndrodiHeavyQuarks}}
\end{figure}

Ref.~\cite{Endrodi:2019zrl} determines the transition point in the quark mass between magnetic catalysis and inverse catalysis further. They use the three stout smeared staggered fermions with physical strange and variable light quark masses. The finite temperature and magnetic field simulations are performed on a $24^3\times6$ lattices. In addition four different zero temperature lattice simulations without a magnetic field were needed. To determine the mass $\tilde m$ on the border between magnetic catalysis and inverse magnetic catalysis they look at the difference between the chiral condensate introduced by a magnetic field defined as
\begin{eqnarray}
&&\Delta \Sigma(B,T,m) =
\Sigma(B,T,m)-\Sigma(0,T,m)\\&=&  \frac{2m_{\mathrm{phys}}}{M_\pi^2F^2}\left[
\langle\bar\psi\psi\rangle{B,T,m} -\langle\bar\psi\psi\rangle_{0,T,m}
\right],
\end{eqnarray}
for $eB_0=0.6 \ \mathrm{GeV}^2$ at the pseudo critical transition temperature $T_c$:
\begin{equation}
\Delta \tilde \Sigma(m)=\Delta\Sigma(B_0,T_c(m,B=0),m)\,.
\label{eq:DS}
\end{equation}
A sign change in $\Delta \tilde \Sigma$ than determines the value of $\tilde m$ since for $\Delta \tilde \Sigma>0$ the chiral condensate increases compared to the case without magnetic field while it decreases when $\Delta \tilde \Sigma<0$. $\Delta \tilde \Sigma$ is shown in the top of Fig.~\ref{fig:EndrodiHeavyQuarks}, indicating $\frac{\tilde m}{m_{\mathrm{phys}}} = 14.07\pm0.55$. This corresponds to a pion mass of $\tilde m_\pi=(497\pm4)$~MeV. In addition also the Polyakov loop

\begin{equation}
P = \frac{1}{V} \left\langle \sum_{\mathbf x} \Re
  \textmd{Tr}\prod_{t=0}^{N_t-1} U_4(\mathbf{x},t) \right\rangle
\end{equation}
from which the ratio
\begin{equation}
L_R = P(B,T,m) \big/ P(0,T,m)
\label{eqn:LR}
\end{equation}
is defined, is investigated. Its behavior around $\tilde m$ is shown in the top of Fig.~\ref{fig:EndrodiHeavyQuarks}.

\begin{figure}
	\centering
	\includegraphics[width=0.7\textwidth]{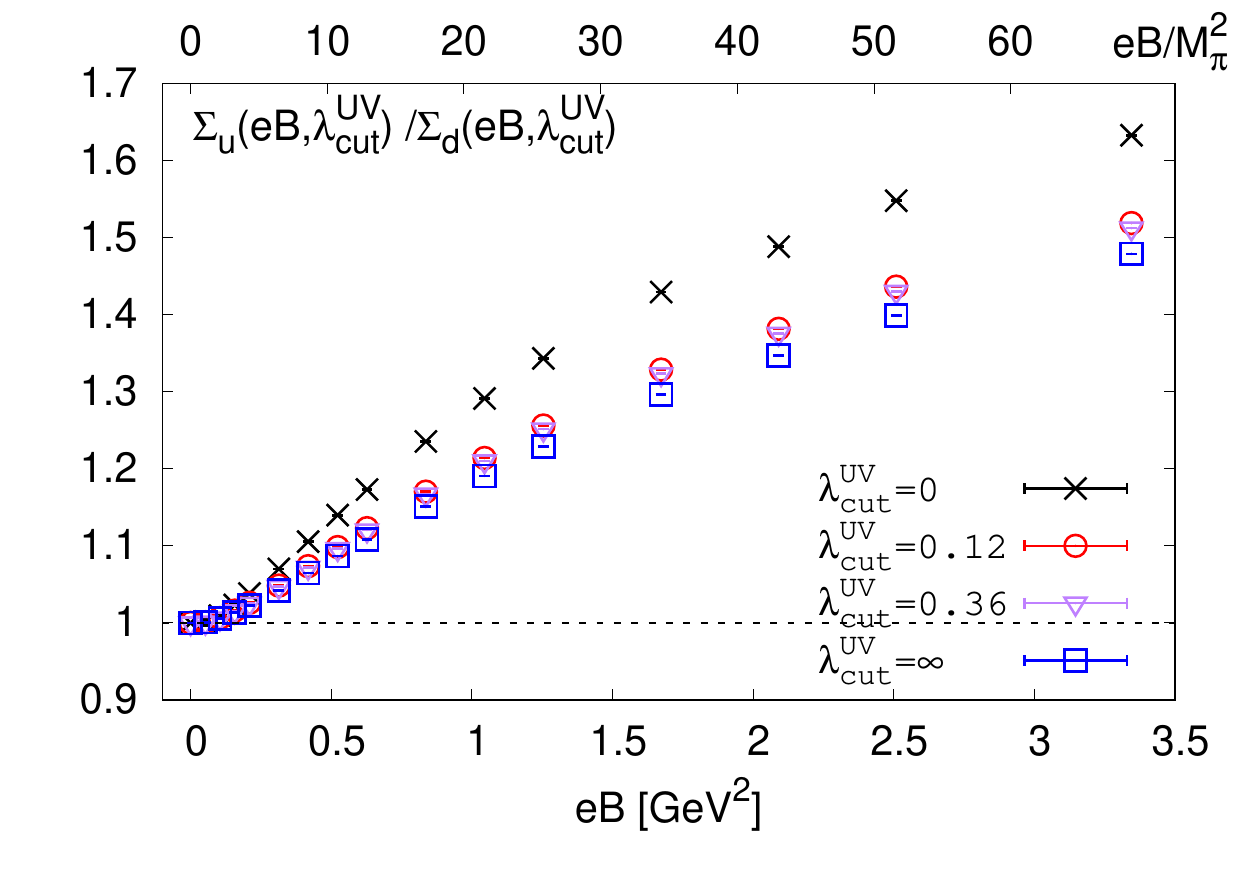}\\
	\vspace{12pt}
	\includegraphics[width=0.7\textwidth]{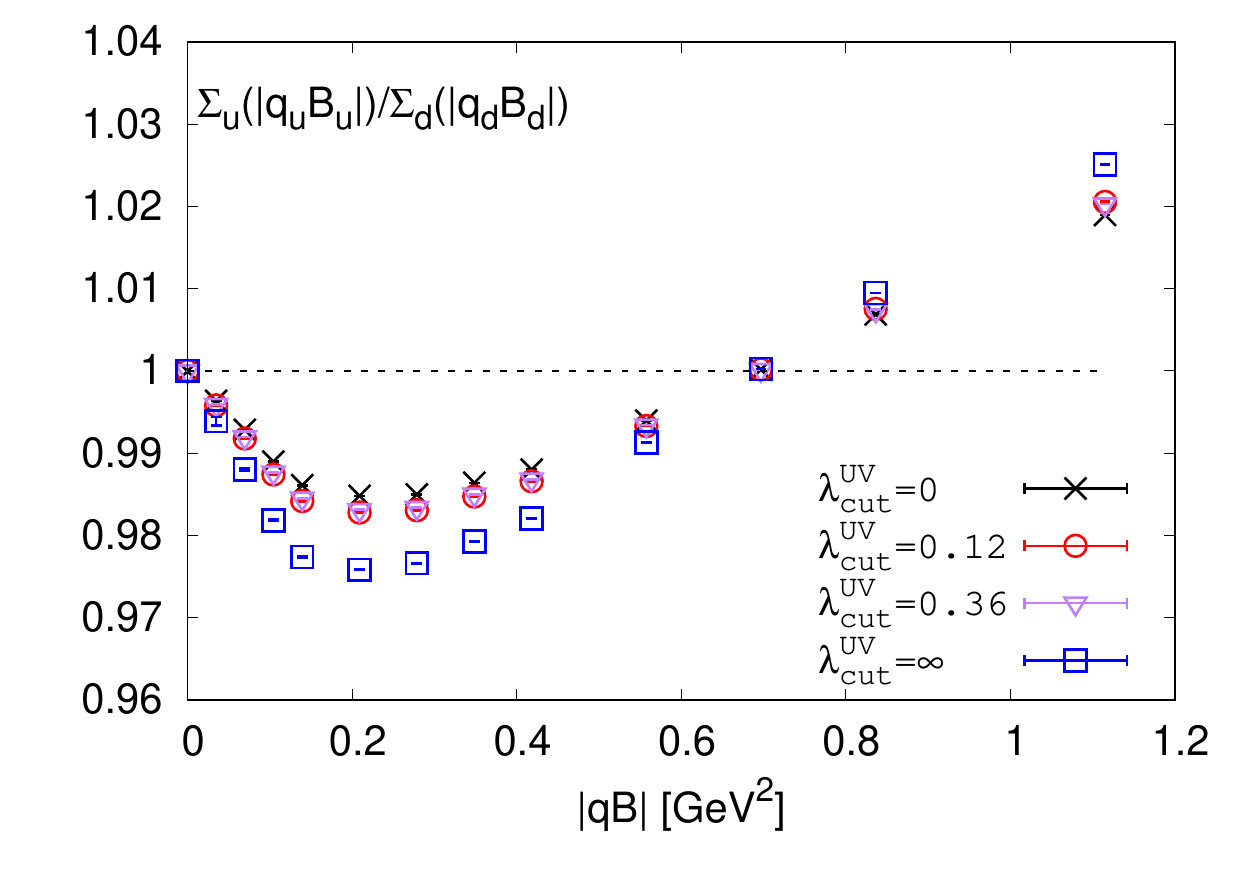}
	\caption{(Ref.~\cite{Ding:2020hxw}) Top: Ratio between the renormalized up quark and the down quark condensate $\Sigma_u(eB)/\Sigma_d(eB)$ as a function of $eB$. Bottom: Ratio between the renormalized up quark and the down quark condensate $\Sigma_u(eB)/\Sigma_d(eB)$ as a function of $|qB|=|q_uB_u|=|q_dB_d|$. \label{fig:zeroTMagnetic}}
\end{figure}

The connections between the transition temperature, the quark masses and the chiral condensate in a magnetic field are further investigated in Ref.~\cite{Ding:2020hxw} for zero temperature.  They connect the chiral condensate and the pion mass by the Gell-Mann-Oakes-Renner relation (Ref.~\cite{GellMann:1968rz}) for two flavours
\begin{equation}
(m_u + m_d) ~\left(\pbp_u + \pbp_d\right)=2 f_\pi^2 M_\pi^2\, (1-\delta_\pi) ,
\label{eq:GMOR}
\end{equation}
and its extension to the 3-flavor case (Ref.~\cite{Gasser:1984gg}) with an additional strange quark
\begin{equation}
(m_s + m_d) ~\left(\pbp_s + \pbp_d\right) =  2 f_K^2 M_K^2 \,(1-\delta_K).
\label{eq:GMOR_K}
\end{equation}
They use a $2+1$-flavour HISQ quark action on a $32^3\times96$ lattice. The results for the light quark condensate defined as
\begin{equation}
\Sigma_{l} (B) = \frac{2 m_l}{M_\pi^2 f_\pi^2}\left (\pbp_l(B\ne0) - \pbp_l(B=0)\right ) +1 ,
\label{eq:SigmaB}
\end{equation}
are shown in Fig.~\ref{fig:zeroTMagnetic}. They find that the Gell-Mann-Oakes-Renner relation for two flavours (equation~(\ref{eq:GMOR})) has only corrections of about 6\% when a magnetic field up to $eB=3.35\ \mathrm{GeV}^2$ is introduced. This illuminates the connection between the reduction of the transition temperature and the pion mass. For the three flavour case the corrections to the Gell-Mann-Oakes-Renner relation (equation~(\ref{eq:GMOR_K})) are much larger namely up to 56\%. Another recent investigation of the magnetic catalysis and the inverse magnetic catalysis can be found in Ref.\cite{Ahmad:2020jzn}, that study uses Dyson-Schwinger-equations instead of lattice QCD.

If one wants to match the conditions in heavy ion collisions in addition to the magnetic field a chemical potintal should be considered. This causes the above discussed sign problem to reappear. Ref.~\cite{Braguta:2019yci} that studies the effect of a magnetic field and a finite baryon density with analytic continuation from imaginary $\mu_B$ finds that the inverse magnetic catalysis becomes slightly stronger with increased chemical potential.

\section{Conclusion\label{sec:conclusion}}

The investigation of the QCD phase diagram, especially in relation to heavy ion collision experiments, remains a lively topic for lattice QCD calculations. With newly published results from LHC and RHIC as well as upcoming data from facilities like NICA, CBM and PARC-HI, a better understanding of the QCD phase diagram including various different influences that duplicate the situation in the colliders as well as possible, is needed.

After the introduction (section~\ref{sec:Introduction}) and the brief description of  the stages of a heavy ion collision experiment (section~\ref{sec:HIC}), this review started with the discussion of lattice QCD results at low, finite density (section~\ref{sec:lowDensity}). Accessing the phase diagram with a finite chemical potential is hindered by the infamous sign problem (section~\ref{sec:signProblem}). It is the reason why, for now, continuum extrapolated, physical results, are only available as extrapolations (section~\ref{sec:anaCont}). Because the QCD phase transition is an analytic crossover at $\mu=0$, one can describe observables with an analytic function that is extrapolated to $\mu>0$. There are two techniques for this kind of extrapolation which are commonly used. One can gain information on the $\mu$ dependence of the system either form calculating expansion coefficients from lattice simulations at zero chemical potential  (Taylor method) or by describing simulation results at imaginary chemical potential with various functions. Both methods are by now agreeing well on various results. New publication include higher order coefficients on the Taylor expansion of the transition temperature, higher order cumulants of the baryon number distribution, which is used to compare to heavy ion collision experiments, as well as higher order cumulants of the baryon number distribution, which are used to compare to heavy ion collision experiments. In addition the possibility to find the QCD critical endpoint are under further investigation. There the development of new observables and techniques is necessary.

One method that is getting relatively close to direct lattice simulations at finite density are Complex Langevin simulations (section~\ref{sec:CL}). These simulations are based on an evolution in a fictitious Langevin time to generate configurations with a complex measure. Here results are now available for different actions (both fermion and gauge). First comparisons to results from the Taylor expansion method are promising. However, for now results are only available with heavier than physical quark masses and on relatively small lattices.

Another way to accesses finite density QCD are effective field theories (section~\ref{sec:ELT}), which also can be simulated on the lattice. This review focused  on the results for  heavy quarkonium (section~\ref{sec:Quarkonium}). There the separation of scales in heavy quark bound states is used to integrate out various degrees of freedom. Lattice simulation can then employ NRQCD or pNRQCD to gain information of heavy quarkonium immersed in the quark gluon plasma. 

Leaving the topic of finite density also  zero density simulations show fascinating progress. By now there are a lot of results on the investigation of the Columbia plot (section~\ref{sec:Columbia}). The Columbia plot summarizes the dependence of the type of the QCD transition (crossover, first or second order phase transition) on the lighter than physical  quark masses. The lower left corner (section~\ref{sec:lowerCorner}) poses a special computational challenge due to the light quark masses. Therefore, various techniques are employed to gain accesses to the first order region located there and its critical boundary (section~\ref{sec:immuColumbia} and section~\ref{sec:Nf}). For the case of three degenerate light quarks results are now available on up to $N_t=12$ lattices with improved Wilson quarks. However, the continuum extrapolation cannot be carried out yet. This review is therefore restricted to an overview (section~\ref{sec:overview}) of the available results for different discretizations. Also the upper right corner of the Columbia plot has been investigated recently (section~\ref{sec:upperCorner}). Here the current estimate for the critical pion mass in the continuum is $m_\pi^c\approx4$~GeV with an error of about 20\%.

Finally this review discusses recent results on the influence of a magnetic field (section~\ref{sec:magneticFields}). Like the inclusion of a chemical potential, these effects are relevant for heavy ion collision experiments. An important topic is the distinction between the magnetic catalysis and inverse catalysis which has been related to  the light quark masses. Here the lattice results are not in agreement with the common expectation from perturbation theory.

Investigating the QCD phase diagram is a fascinating topic from a purely theoretical point of view. However, many recent results in finite temperature lattice QCD are linked to heavy ion collision experiments. The experimental progress on the quark gluon plasma allows on the one hand to confirm the theoretical predictions made by QCD. On the other hand, the computation from lattice QCD are needed by experiments to understand the observations made by the detectors and link them to different stages of collision. This can be done either directly by lattice QCD or through different models or effective theories. In any case the continuation of the experimental heavy ion collision programs will provide interesting checks and challenges for finite temperature lattice QCD calculations.

\section*{Acknowledgments}
 The author thanks Lukas Varnhorst for proofreading and discussion.
The project leading to this publication has received funding from Excellence Initiative of Aix-Marseille University - A*MIDEX, a French “Investissements d’Avenir” programme, AMX-18-ACE-005.

\clearpage

\bibliography{finiteT}{}

\begin{thebibliography}{100}

\bibitem{Aoki:2006we}
Y.~Aoki, G.~Endrodi, Z.~Fodor, S.~D. Katz, and K.~K. Szabo.
\newblock {The Order of the quantum chromodynamics transition predicted by the
  standard model of particle physics}.
\newblock {\em Nature}, 443:675--678, 2006.

\bibitem{Aoki:2006br}
Y.~Aoki, Z.~Fodor, S.D. Katz, and K.K. Szabo.
\newblock {The QCD transition temperature: Results with physical masses in the
  continuum limit}.
\newblock {\em Phys. Lett. B}, 643:46--54, 2006.

\bibitem{Aoki:2009sc}
Y.~Aoki, Szabolcs Borsanyi, Stephan Durr, Zoltan Fodor, Sandor~D. Katz, Stefan
  Krieg, and Kalman~K. Szabo.
\newblock {The QCD transition temperature: results with physical masses in the
  continuum limit II.}
\newblock {\em JHEP}, 06:088, 2009.

\bibitem{Borsanyi:2010bp}
Szabolcs Borsanyi, Zoltan Fodor, Christian Hoelbling, Sandor~D Katz, Stefan
  Krieg, Claudia Ratti, and Kalman~K. Szabo.
\newblock {Is there still any $T_c$ mystery in lattice QCD? Results with
  physical masses in the continuum limit III}.
\newblock {\em JHEP}, 09:073, 2010.

\bibitem{Bhattacharya:2014ara}
Tanmoy Bhattacharya et~al.
\newblock {QCD Phase Transition with Chiral Quarks and Physical Quark Masses}.
\newblock {\em Phys. Rev. Lett.}, 113(8):082001, 2014.

\bibitem{Bazavov:2011nk}
A.~Bazavov et~al.
\newblock {The chiral and deconfinement aspects of the QCD transition}.
\newblock {\em Phys. Rev.}, D85:054503, 2012.

\bibitem{Barbour:1997ej}
Ian~M. Barbour, Susan~E. Morrison, Elyakum~G. Klepfish, John~B. Kogut, and
  Maria-Paola Lombardo.
\newblock {Results on finite density QCD}.
\newblock {\em Nucl. Phys. B Proc. Suppl.}, 60:220--234, 1998.

\bibitem{Fodor:2001au}
Z.~Fodor and S.~D. Katz.
\newblock {A New method to study lattice QCD at finite temperature and chemical
  potential}.
\newblock {\em Phys. Lett.}, B534:87--92, 2002.

\bibitem{Fodor:2001pe}
Z.~Fodor and S.~D. Katz.
\newblock {Lattice determination of the critical point of QCD at finite T and
  mu}.
\newblock {\em JHEP}, 03:014, 2002.

\bibitem{Csikor:2002ic}
F.~Csikor, G.I. Egri, Z.~Fodor, S.D. Katz, K.K. Szabo, and A.I. Toth.
\newblock {The QCD equation of state at finite T and mu}.
\newblock {\em Nucl. Phys. B Proc. Suppl.}, 119:547--549, 2003.

\bibitem{Fodor:2007vv}
Zoltan Fodor, Sandor~D. Katz, and Christian Schmidt.
\newblock {The Density of states method at non-zero chemical potential}.
\newblock {\em JHEP}, 03:121, 2007.

\bibitem{Alexandru:2014hga}
Andrei Alexandru, C.~Gattringer, H.~P. Schadler, K.~Splittorff, and J.J.M.
  Verbaarschot.
\newblock {Distribution of Canonical Determinants in QCD}.
\newblock {\em Phys. Rev. D}, 91(7):074501, 2015.

\bibitem{Alexandru:2005ix}
Andrei Alexandru, Manfried Faber, Ivan Horvath, and Keh-Fei Liu.
\newblock {Lattice QCD at finite density via a new canonical approach}.
\newblock {\em Phys. Rev. D}, 72:114513, 2005.

\bibitem{Kratochvila:2005mk}
Slavo Kratochvila and Philippe de~Forcrand.
\newblock {The Canonical approach to finite density QCD}.
\newblock {\em PoS}, LAT2005:167, 2006.

\bibitem{Ejiri:2008xt}
Shinji Ejiri.
\newblock {Canonical partition function and finite density phase transition in
  lattice QCD}.
\newblock {\em Phys. Rev. D}, 78:074507, 2008.

\bibitem{Gattringer:2014nxa}
Christof Gattringer.
\newblock {New developments for dual methods in lattice field theory at
  non-zero density}.
\newblock {\em PoS}, LATTICE2013:002, 2014.

\bibitem{Scorzato:2015qts}
Luigi Scorzato.
\newblock {The Lefschetz thimble and the sign problem}.
\newblock {\em PoS}, LATTICE2015:016, 2016.

\bibitem{Alexandru:2015xva}
Andrei Alexandru, G\"ok\c{c}e Basar, and Paulo Bedaque.
\newblock {Monte Carlo algorithm for simulating fermions on Lefschetz
  thimbles}.
\newblock {\em Phys. Rev. D}, 93(1):014504, 2016.

\bibitem{Rothkopf:2019ipj}
Alexander Rothkopf.
\newblock {Heavy Quarkonium in Extreme Conditions}.
\newblock {\em Phys. Rept.}, 858:1--117, 2020.

\bibitem{Gelis:2010nm}
Francois Gelis, Edmond Iancu, Jamal Jalilian-Marian, and Raju Venugopalan.
\newblock {The Color Glass Condensate}.
\newblock {\em Ann. Rev. Nucl. Part. Sci.}, 60:463--489, 2010.

\bibitem{Gelis:2012ri}
F.~Gelis.
\newblock {Color Glass Condensate and Glasma}.
\newblock {\em Int. J. Mod. Phys. A}, 28:1330001, 2013.

\bibitem{Rischke:2003mt}
Dirk~H. Rischke.
\newblock {The Quark gluon plasma in equilibrium}.
\newblock {\em Prog. Part. Nucl. Phys.}, 52:197--296, 2004.

\bibitem{Hasenfratz:1983ba}
P.~Hasenfratz and F.~Karsch.
\newblock {Chemical Potential on the Lattice}.
\newblock {\em Phys. Lett. B}, 125:308--310, 1983.

\bibitem{Allton:2002zi}
C.~R. Allton, S.~Ejiri, S.~J. Hands, O.~Kaczmarek, F.~Karsch, E.~Laermann,
  C.~Schmidt, and L.~Scorzato.
\newblock {The QCD thermal phase transition in the presence of a small chemical
  potential}.
\newblock {\em Phys. Rev.}, D66:074507, 2002.

\bibitem{Allton:2005gk}
C.~R. Allton, M.~Doring, S.~Ejiri, S.~J. Hands, O.~Kaczmarek, F.~Karsch,
  E.~Laermann, and K.~Redlich.
\newblock {Thermodynamics of two flavor QCD to sixth order in quark chemical
  potential}.
\newblock {\em Phys. Rev.}, D71:054508, 2005.

\bibitem{Gavai:2008zr}
R.~V. Gavai and Sourendu Gupta.
\newblock {QCD at finite chemical potential with six time slices}.
\newblock {\em Phys. Rev.}, D78:114503, 2008.

\bibitem{Basak:2009uv}
S.~Basak et~al.
\newblock {QCD equation of state at non-zero chemical potential}.
\newblock {\em PoS}, LATTICE2008:171, 2008.

\bibitem{Kaczmarek:2011zz}
O.~Kaczmarek, F.~Karsch, E.~Laermann, C.~Miao, S.~Mukherjee, P.~Petreczky,
  C.~Schmidt, W.~Soeldner, and W.~Unger.
\newblock {Phase boundary for the chiral transition in (2+1) -flavor QCD at
  small values of the chemical potential}.
\newblock {\em Phys. Rev.}, D83:014504, 2011.

\bibitem{Seiler:2012wz}
Erhard Seiler, Denes Sexty, and Ion-Olimpiu Stamatescu.
\newblock {Gauge cooling in complex Langevin for QCD with heavy quarks}.
\newblock {\em Phys. Lett. B}, 723:213--216, 2013.

\bibitem{Sexty:2013ica}
D\'enes Sexty.
\newblock {Simulating full QCD at nonzero density using the complex Langevin
  equation}.
\newblock {\em Phys. Lett. B}, 729:108--111, 2014.

\bibitem{Borsanyi:2011sw}
Szabolcs Borsanyi, Zoltan Fodor, Sandor~D. Katz, Stefan Krieg, Claudia Ratti,
  and Kalman Szabo.
\newblock {Fluctuations of conserved charges at finite temperature from lattice
  QCD}.
\newblock {\em JHEP}, 01:138, 2012.

\bibitem{Borsanyi:2012cr}
Sz. Borsanyi, G.~Endrodi, Z.~Fodor, S.D. Katz, S.~Krieg, C.~Ratti, and K.K.
  Szabo.
\newblock {QCD equation of state at nonzero chemical potential: continuum
  results with physical quark masses at order $mu^2$}.
\newblock {\em JHEP}, 08:053, 2012.

\bibitem{Bellwied:2015lba}
R.~Bellwied, S.~Borsanyi, Z.~Fodor, S.~D. Katz, A.~Pasztor, C.~Ratti, and K.~K.
  Szabo.
\newblock {Fluctuations and correlations in high temperature QCD}.
\newblock {\em Phys. Rev.}, D92(11):114505, 2015.

\bibitem{Ding:2015fca}
H.~T. Ding, Swagato Mukherjee, H.~Ohno, P.~Petreczky, and H.~P. Schadler.
\newblock {Diagonal and off-diagonal quark number susceptibilities at high
  temperatures}.
\newblock {\em Phys. Rev. D}, 92(7):074043, 2015.

\bibitem{Bazavov:2017dus}
A.~Bazavov et~al.
\newblock {The QCD Equation of State to $\mathcal{O}(\mu_B^6)$ from Lattice
  QCD}.
\newblock {\em Phys. Rev.}, D95(5):054504, 2017.

\bibitem{Bazavov:2018mes}
A.~Bazavov et~al.
\newblock {Chiral crossover in QCD at zero and non-zero chemical potentials}.
\newblock {\em Phys. Lett.}, B795:15--21, 2019.

\bibitem{Bazavov:2020bjn}
A.~Bazavov et~al.
\newblock {Skewness, kurtosis, and the fifth and sixth order cumulants of net
  baryon-number distributions from lattice QCD confront high-statistics STAR
  data}.
\newblock {\em Phys. Rev. D}, 101(7):074502, 2020.

\bibitem{Bonati:2018nut}
Claudio Bonati, Massimo D'Elia, Francesco Negro, Francesco Sanfilippo, and
  Kevin Zambello.
\newblock {Curvature of the pseudocritical line in QCD: Taylor expansion
  matches analytic continuation}.
\newblock {\em Phys. Rev. D}, 98(5):054510, 2018.

\bibitem{Endrodi:2011gv}
G.~Endrodi, Z.~Fodor, S.~D. Katz, and K.~K. Szabo.
\newblock {The QCD phase diagram at nonzero quark density}.
\newblock {\em JHEP}, 04:001, 2011.

\bibitem{deForcrand:2002hgr}
Philippe de~Forcrand and Owe Philipsen.
\newblock {The QCD phase diagram for small densities from imaginary chemical
  potential}.
\newblock {\em Nucl. Phys.}, B642:290--306, 2002.

\bibitem{DElia:2002tig}
Massimo D'Elia and Maria-Paola Lombardo.
\newblock {Finite density QCD via imaginary chemical potential}.
\newblock {\em Phys. Rev.}, D67:014505, 2003.

\bibitem{DElia:2009pdy}
Massimo D'Elia and Francesco Sanfilippo.
\newblock {Thermodynamics of two flavor QCD from imaginary chemical
  potentials}.
\newblock {\em Phys. Rev. D}, 80:014502, 2009.

\bibitem{Cea:2014xva}
Paolo Cea, Leonardo Cosmai, and Alessandro Papa.
\newblock {Critical line of 2+1 flavor QCD}.
\newblock {\em Phys. Rev. D}, 89(7):074512, 2014.

\bibitem{Bonati:2014kpa}
Claudio Bonati, Philippe de~Forcrand, Massimo D'Elia, Owe Philipsen, and
  Francesco Sanfilippo.
\newblock {Chiral phase transition in two-flavor QCD from an imaginary chemical
  potential}.
\newblock {\em Phys. Rev. D}, 90(7):074030, 2014.

\bibitem{Cea:2015cya}
Paolo Cea, Leonardo Cosmai, and Alessandro Papa.
\newblock {Critical line of 2+1 flavor QCD: Toward the continuum limit}.
\newblock {\em Phys. Rev.}, D93(1):014507, 2016.

\bibitem{Bonati:2015bha}
Claudio Bonati, Massimo D'Elia, Marco Mariti, Michele Mesiti, Francesco Negro,
  and Francesco Sanfilippo.
\newblock {Curvature of the chiral pseudocritical line in QCD: Continuum
  extrapolated results}.
\newblock {\em Phys. Rev.}, D92(5):054503, 2015.

\bibitem{Bellwied:2015rza}
R.~Bellwied, S.~Borsanyi, Z.~Fodor, J.~Guenther, S.~D. Katz, C.~Ratti, and
  K.~K. Szabo.
\newblock {The QCD phase diagram from analytic continuation}.
\newblock {\em Phys. Lett.}, B751:559--564, 2015.

\bibitem{DElia:2016jqh}
Massimo D'Elia, Giuseppe Gagliardi, and Francesco Sanfilippo.
\newblock {Higher order quark number fluctuations via imaginary chemical
  potentials in $N_f=2+1$ QCD}.
\newblock {\em Phys. Rev.}, D95(9):094503, 2017.

\bibitem{Gunther:2016vcp}
J.~N. Guenther, R.~Bellwied, S.~Borsanyi, Z.~Fodor, S.~D. Katz, A.~Pasztor,
  C.~Ratti, and K.~K. Szabó.
\newblock {The QCD equation of state at finite density from analytical
  continuation}.
\newblock {\em Nucl. Phys.}, A967:720--723, 2017.

\bibitem{Alba:2017mqu}
Paolo Alba et~al.
\newblock {Constraining the hadronic spectrum through QCD thermodynamics on the
  lattice}.
\newblock {\em Phys. Rev. D}, 96(3):034517, 2017.

\bibitem{Vovchenko:2017xad}
Volodymyr Vovchenko, Attila Pasztor, Zoltan Fodor, Sandor~D. Katz, and Horst
  Stoecker.
\newblock {Repulsive baryonic interactions and lattice QCD observables at
  imaginary chemical potential}.
\newblock {\em Phys. Lett. B}, 775:71--78, 2017.

\bibitem{Borsanyi:2018grb}
Szabolcs Borsanyi, Zoltan Fodor, Jana~N. Guenther, Sandor~K. Katz, Kalman~K.
  Szabo, Attila Pasztor, Israel Portillo, and Claudia Ratti.
\newblock {Higher order fluctuations and correlations of conserved charges from
  lattice QCD}.
\newblock 2018.

\bibitem{Borsanyi:2020fev}
Szabolcs Borsanyi, Zoltan Fodor, Jana~N. Guenther, Ruben Kara, Sandor~D. Katz,
  Paolo Parotto, Attila Pasztor, Claudia Ratti, and Kalman~K. Szabo.
\newblock {The QCD crossover at finite chemical potential from lattice
  simulations}.
\newblock 2020.

\bibitem{Pasztor:2016iqd}
Ren\'e Bellwied, Szabolcs Bors\'anyi, Zolt\'an Fodor, Jana G\"unther,
  S\'andor~D. Katz, K\'alm\'an~K. Szab\'o, Claudia Ratti, and Attila Pasztor.
\newblock {Fluctuations and correlations in finite temperature QCD}.
\newblock {\em PoS}, ICHEP2016:369, 2016.

\bibitem{Pasztor:2020dur}
Attila P\'asztor, Zsolt Sz\'ep, and Gergely Mark\'o.
\newblock {Apparent convergence of Pad\textbackslash{}'e approximants for the
  crossover line in finite density QCD}.
\newblock 10 2020.

\bibitem{Steinbrecher:2018phh}
Patrick Steinbrecher.
\newblock {The QCD crossover at zero and non-zero baryon densities from Lattice
  QCD}.
\newblock {\em Nucl. Phys. A}, 982:847--850, 2019.

\bibitem{Isserstedt:2019pgx}
Philipp Isserstedt, Michael Buballa, Christian~S. Fischer, and Pascal~J.
  Gunkel.
\newblock {Baryon number fluctuations in the QCD phase diagram from
  Dyson-Schwinger equations}.
\newblock {\em Phys. Rev.}, D100(7):074011, 2019.

\bibitem{Andronic:2005yp}
A.~Andronic, P.~Braun-Munzinger, and J.~Stachel.
\newblock {Hadron production in central nucleus-nucleus collisions at chemical
  freeze-out}.
\newblock {\em Nucl. Phys.}, A772:167--199, 2006.

\bibitem{Becattini:2012xb}
Francesco Becattini, Marcus Bleicher, Thorsten Kollegger, Tim Schuster, Jan
  Steinheimer, and Reinhard Stock.
\newblock {Hadron Formation in Relativistic Nuclear Collisions and the QCD
  Phase Diagram}.
\newblock {\em Phys. Rev. Lett.}, 111:082302, 2013.

\bibitem{Alba:2014eba}
Paolo Alba, Wanda Alberico, Rene Bellwied, Marcus Bluhm, Valentina
  Mantovani~Sarti, Marlene Nahrgang, and Claudia Ratti.
\newblock {Freeze-out conditions from net-proton and net-charge fluctuations at
  RHIC}.
\newblock {\em Phys. Lett.}, B738:305--310, 2014.

\bibitem{Vovchenko:2015idt}
V.~Vovchenko, V.~V. Begun, and M.~I. Gorenstein.
\newblock {Hadron multiplicities and chemical freeze-out conditions in
  proton-proton and nucleus-nucleus collisions}.
\newblock {\em Phys. Rev.}, C93(6):064906, 2016.

\bibitem{Adamczyk:2017iwn}
L.~Adamczyk et~al.
\newblock {Bulk Properties of the Medium Produced in Relativistic Heavy-Ion
  Collisions from the Beam Energy Scan Program}.
\newblock {\em Phys. Rev.}, C96(4):044904, 2017.

\bibitem{Hatta:2003wn}
Y.~Hatta and M.A. Stephanov.
\newblock {Proton number fluctuation as a signal of the QCD critical endpoint}.
\newblock {\em Phys. Rev. Lett.}, 91:102003, 2003.
\newblock [Erratum: Phys.Rev.Lett. 91, 129901 (2003)].

\bibitem{Stephanov:2008qz}
M.A. Stephanov.
\newblock {Non-Gaussian fluctuations near the QCD critical point}.
\newblock {\em Phys. Rev. Lett.}, 102:032301, 2009.

\bibitem{Friman:2011pf}
B.~Friman, F.~Karsch, K.~Redlich, and V.~Skokov.
\newblock {Fluctuations as probe of the QCD phase transition and freeze-out in
  heavy ion collisions at LHC and RHIC}.
\newblock {\em Eur. Phys. J. C}, 71:1694, 2011.

\bibitem{Halasz:1998qr}
Adam~Miklos Halasz, A.D. Jackson, R.E. Shrock, Misha~A. Stephanov, and J.J.M.
  Verbaarschot.
\newblock {On the phase diagram of QCD}.
\newblock {\em Phys. Rev. D}, 58:096007, 1998.

\bibitem{Stephanov:1999zu}
Misha~A. Stephanov, K.~Rajagopal, and Edward~V. Shuryak.
\newblock {Event-by-event fluctuations in heavy ion collisions and the QCD
  critical point}.
\newblock {\em Phys. Rev.}, D60:114028, 1999.

\bibitem{Cheng:2007jq}
M.~Cheng et~al.
\newblock {The QCD equation of state with almost physical quark masses}.
\newblock {\em Phys. Rev.}, D77:014511, 2008.

\bibitem{Bazavov:2017tot}
A.~Bazavov et~al.
\newblock {Skewness and kurtosis of net baryon-number distributions at small
  values of the baryon chemical potential}.
\newblock 2017.

\bibitem{Karsch:2012wm}
Frithjof Karsch.
\newblock {Determination of Freeze-out Conditions from Lattice QCD
  Calculations}.
\newblock {\em Central Eur. J. Phys.}, 10:1234--1237, 2012.

\bibitem{Bazavov:2012vg}
A.~Bazavov et~al.
\newblock {Freeze-out Conditions in Heavy Ion Collisions from QCD
  Thermodynamics}.
\newblock {\em Phys. Rev. Lett.}, 109:192302, 2012.

\bibitem{Borsanyi:2013hza}
S.~Borsanyi, Z.~Fodor, S.~D. Katz, S.~Krieg, C.~Ratti, and K.~K. Szabo.
\newblock {Freeze-out parameters: lattice meets experiment}.
\newblock {\em Phys. Rev. Lett.}, 111:062005, 2013.

\bibitem{Borsanyi:2014ewa}
S.~Borsanyi, Z.~Fodor, S.~D. Katz, S.~Krieg, C.~Ratti, and K.~K. Szabo.
\newblock {Freeze-out parameters from electric charge and baryon number
  fluctuations: is there consistency?}
\newblock {\em Phys. Rev. Lett.}, 113:052301, 2014.

\bibitem{Ratti:2018ksb}
Claudia Ratti.
\newblock {Lattice QCD and heavy ion collisions: a review of recent progress}.
\newblock {\em Rept. Prog. Phys.}, 81(8):084301, 2018.

\bibitem{Adam:2020unf}
J.~Adam et~al.
\newblock {Net-proton number fluctuations and the Quantum Chromodynamics
  critical point}.
\newblock 1 2020.

\bibitem{Nonaka:2020crv}
Toshihiro Nonaka.
\newblock {Measurement of the Sixth-Order Cumulant of Net-Proton Distributions
  in Au+Au Collisions from the STAR Experiment}.
\newblock In {\em {28th International Conference on Ultrarelativistic
  Nucleus-Nucleus Collisions}}, 2 2020.

\bibitem{Bellwied:2019pxh}
Rene Bellwied, Szabolcs Borsanyi, Zoltan Fodor, Jana~N. Guenther, Jacquelyn
  Noronha-Hostler, Paolo Parotto, Attila Pasztor, Claudia Ratti, and Jamie~M.
  Stafford.
\newblock {Off-diagonal correlators of conserved charges from lattice QCD and
  how to relate them to experiment}.
\newblock {\em Phys. Rev. D}, 101(3):034506, 2020.

\bibitem{Begun:2006jf}
V.V. Begun, Mark~I. Gorenstein, M.~Hauer, V.P. Konchakovski, and O.S. Zozulya.
\newblock {Multiplicity Fluctuations in Hadron-Resonance Gas}.
\newblock {\em Phys. Rev. C}, 74:044903, 2006.

\bibitem{Kitazawa:2011wh}
Masakiyo Kitazawa and Masayuki Asakawa.
\newblock {Revealing baryon number fluctuations from proton number fluctuations
  in relativistic heavy ion collisions}.
\newblock {\em Phys. Rev. C}, 85:021901, 2012.

\bibitem{Kitazawa:2012at}
Masakiyo Kitazawa and Masayuki Asakawa.
\newblock {Relation between baryon number fluctuations and experimentally
  observed proton number fluctuations in relativistic heavy ion collisions}.
\newblock {\em Phys. Rev. C}, 86:024904, 2012.
\newblock [Erratum: Phys.Rev.C 86, 069902 (2012)].

\bibitem{Giordano:2019slo}
Matteo Giordano and Attila P\'asztor.
\newblock {Reliable estimation of the radius of convergence in finite density
  QCD}.
\newblock {\em Phys. Rev. D}, 99(11):114510, 2019.

\bibitem{Lee:1952ig}
T.D. Lee and Chen-Ning Yang.
\newblock {Statistical theory of equations of state and phase transitions. 2.
  Lattice gas and Ising model}.
\newblock {\em Phys. Rev.}, 87:410--419, 1952.

\bibitem{Hasenfratz:1991ax}
A.~Hasenfratz and D.~Toussaint.
\newblock {Canonical ensembles and nonzero density quantum chromodynamics}.
\newblock {\em Nucl. Phys. B}, 371:539--549, 1992.

\bibitem{Fodor:2004nz}
Z.~Fodor and S.~D. Katz.
\newblock {Critical point of QCD at finite T and mu, lattice results for
  physical quark masses}.
\newblock {\em JHEP}, 04:050, 2004.

\bibitem{Danzer:2008xs}
Julia Danzer and Christof Gattringer.
\newblock {Winding expansion techniques for lattice QCD with chemical
  potential}.
\newblock {\em Phys. Rev. D}, 78:114506, 2008.

\bibitem{Alexandru:2010yb}
Andrei Alexandru and Urs Wenger.
\newblock {QCD at non-zero density and canonical partition functions with
  Wilson fermions}.
\newblock {\em Phys. Rev. D}, 83:034502, 2011.

\bibitem{Giordano:2019gev}
Matteo Giordano, Kornel Kapas, Sandor~D. Katz, Daniel Nogradi, and Attila
  Pasztor.
\newblock {Radius of convergence in lattice QCD at finite $\mu_B$ with rooted
  staggered fermions}.
\newblock {\em Phys. Rev. D}, 101(7):074511, 2020.

\bibitem{Giordano:2020roi}
Matteo Giordano, Kornel Kapas, Sandor~D. Katz, Daniel Nogradi, and Attila
  Pasztor.
\newblock {New approach to lattice QCD at finite density; results for the
  critical end point on coarse lattices}.
\newblock {\em JHEP}, 05:088, 2020.

\bibitem{Attanasio:2020spv}
Felipe Attanasio, Benjamin Jäger, and Felix~P.G. Ziegler.
\newblock {Complex Langevin and the QCD phase diagram: Recent developments}.
\newblock 5 2020.

\bibitem{Ambjorn:1985iw}
Jan Ambjorn and S.K. Yang.
\newblock {Numerical Problems in Applying the Langevin Equation to Complex
  Effective Actions}.
\newblock {\em Phys. Lett. B}, 165:140, 1985.

\bibitem{Klauder:1985ks}
John~R. Klauder and Wesley~P. Petersen.
\newblock {SPECTRUM OF CERTAIN NONSELFADJOINT OPERATORS AND SOLUTIONS OF
  LANGEVIN EQUATIONS WITH COMPLEX DRIFT}.
\newblock {\em J. Stat. Phys.}, 39:53--72, 1985.

\bibitem{RunAway}
H.~Q. Lin and J.~E. Hirsch.
\newblock Monte carlo versus langevin methods for nonpositive definite weights.
\newblock {\em Phys. Rev. B}, 34:1964--1967, Aug 1986.

\bibitem{Ambjorn:1986fz}
Jan Ambjorn, M.~Flensburg, and C.~Peterson.
\newblock {The Complex Langevin Equation and Monte Carlo Simulations of Actions
  With Static Charges}.
\newblock {\em Nucl. Phys. B}, 275:375--397, 1986.

\bibitem{Aarts:2013uxa}
Gert Aarts, Lorenzo Bongiovanni, Erhard Seiler, Denes Sexty, and Ion-Olimpiu
  Stamatescu.
\newblock {Controlling complex Langevin dynamics at finite density}.
\newblock {\em Eur. Phys. J. A}, 49:89, 2013.

\bibitem{Berges:2007nr}
Juergen Berges and Denes Sexty.
\newblock {Real-time gauge theory simulations from stochastic quantization with
  optimized updating}.
\newblock {\em Nucl. Phys. B}, 799:306--329, 2008.

\bibitem{Aarts:2009dg}
Gert Aarts, Frank~A. James, Erhard Seiler, and Ion-Olimpiu Stamatescu.
\newblock {Adaptive stepsize and instabilities in complex Langevin dynamics}.
\newblock {\em Phys. Lett. B}, 687:154--159, 2010.

\bibitem{Attanasio:2018rtq}
Felipe Attanasio and Benjamin Jäger.
\newblock {Dynamical stabilisation of complex Langevin simulations of QCD}.
\newblock {\em Eur. Phys. J. C}, 79(1):16, 2019.

\bibitem{Nishimura:2015pba}
Jun Nishimura and Shinji Shimasaki.
\newblock {New Insights into the Problem with a Singular Drift Term in the
  Complex Langevin Method}.
\newblock {\em Phys. Rev. D}, 92(1):011501, 2015.

\bibitem{Aarts:2017vrv}
Gert Aarts, Erhard Seiler, Denes Sexty, and Ion-Olimpiu Stamatescu.
\newblock {Complex Langevin dynamics and zeroes of the fermion determinant}.
\newblock {\em JHEP}, 05:044, 2017.
\newblock [Erratum: JHEP 01, 128 (2018)].

\bibitem{Nagata:2016vkn}
Keitaro Nagata, Jun Nishimura, and Shinji Shimasaki.
\newblock {Argument for justification of the complex Langevin method and the
  condition for correct convergence}.
\newblock {\em Phys. Rev. D}, 94(11):114515, 2016.

\bibitem{Nagata:2018net}
Keitaro Nagata, Jun Nishimura, and Shinji Shimasaki.
\newblock {Testing the criterion for correct convergence in the complex
  Langevin method}.
\newblock {\em JHEP}, 05:004, 2018.

\bibitem{Scherzer:2018hid}
Manuel Scherzer, Erhard Seiler, Dénes Sexty, and Ion-Olimpiu Stamatescu.
\newblock {Complex Langevin and boundary terms}.
\newblock {\em Phys. Rev. D}, 99(1):014512, 2019.

\bibitem{Tsutsui:2019suq}
Shoichiro Tsutsui, Yuta Ito, Hideo Matsufuru, Jun Nishimura, Shinji Shimasaki,
  and Asato Tsuchiya.
\newblock {Exploring the QCD phase diagram at finite density by the complex
  Langevin method on a $16^3\times 32$ lattice}.
\newblock {\em PoS}, LATTICE2019:151, 2019.

\bibitem{Sexty:2019vqx}
Dénes Sexty.
\newblock {Calculating the equation of state of dense quark-gluon plasma using
  the complex Langevin equation}.
\newblock {\em Phys. Rev. D}, 100(7):074503, 2019.

\bibitem{Scherzer:2020kiu}
M.~Scherzer, D.~Sexty, and I.-O. Stamatescu.
\newblock {Deconfinement transition line with the complex Langevin equation up
  to $\mu /T \sim 5$}.
\newblock {\em Phys. Rev. D}, 102(1):014515, 2020.

\bibitem{Ito:2018jpo}
Yuta Ito, Hideo Matsufuru, Jun Nishimura, Shinji Shimasaki, Asato Tsuchiya, and
  Shoichiro Tsutsui.
\newblock {Exploring the phase diagram of finite density QCD at low temperature
  by the complex Langevin method}.
\newblock {\em PoS}, LATTICE2018:146, 2018.

\bibitem{Ito:2020mys}
Yuta Ito, Hideo Matsufuru, Yusuke Namekawa, Jun Nishimura, Shinji Shimasaki,
  Asato Tsuchiya, and Shoichiro Tsutsui.
\newblock {Complex Langevin calculations in QCD at finite density}.
\newblock {\em JHEP}, 10:144, 2020.

\bibitem{Kogut:2019qmi}
J.B. Kogut and D.K. Sinclair.
\newblock {Applying Complex Langevin Simulations to Lattice QCD at Finite
  Density}.
\newblock {\em Phys. Rev. D}, 100(5):054512, 2019.

\bibitem{Karsch:2003wy}
F.~Karsch, E.~Laermann, P.~Petreczky, and S.~Stickan.
\newblock {Infinite temperature limit of meson spectral functions calculated on
  the lattice}.
\newblock {\em Phys. Rev. D}, 68:014504, 2003.

\bibitem{Aarts:2005hg}
Gert Aarts and Jose~M. Martinez~Resco.
\newblock {Continuum and lattice meson spectral functions at nonzero momentum
  and high temperature}.
\newblock {\em Nucl. Phys. B}, 726:93--108, 2005.

\bibitem{Caswell:1985ui}
W.E. Caswell and G.P. Lepage.
\newblock {Effective Lagrangians for Bound State Problems in QED, QCD, and
  Other Field Theories}.
\newblock {\em Phys. Lett. B}, 167:437--442, 1986.

\bibitem{Brambilla:1999xf}
Nora Brambilla, Antonio Pineda, Joan Soto, and Antonio Vairo.
\newblock {Potential NRQCD: An Effective theory for heavy quarkonium}.
\newblock {\em Nucl. Phys. B}, 566:275, 2000.

\bibitem{Bazavov:2018wmo}
Alexei Bazavov, Nora Brambilla, Peter Petreczky, Antonio Vairo, and
  Johannes~Heinrich Weber.
\newblock {Color screening in (2+1)-flavor QCD}.
\newblock {\em Phys. Rev. D}, 98(5):054511, 2018.

\bibitem{Aarts:2014cda}
Gert Aarts, Chris Allton, Tim Harris, Seyong Kim, Maria~Paola Lombardo,
  Sin\'ead~M. Ryan, and Jon-Ivar Skullerud.
\newblock {The bottomonium spectrum at finite temperature from N$_{f}$ = 2 + 1
  lattice QCD}.
\newblock {\em JHEP}, 07:097, 2014.

\bibitem{Aarts:2013kaa}
G.~Aarts, C.~Allton, S.~Kim, M.P. Lombardo, S.M. Ryan, and J.-I. Skullerud.
\newblock {Melting of P wave bottomonium states in the quark-gluon plasma from
  lattice NRQCD}.
\newblock {\em JHEP}, 12:064, 2013.

\bibitem{Aarts:2012ka}
Gert Aarts, Chris Allton, Seyong Kim, Maria~Paola Lombardo, Mehmet~B. Oktay,
  Sinead~M. Ryan, D.K. Sinclair, and Jon-Ivar Skullerud.
\newblock {S wave bottomonium states moving in a quark-gluon plasma from
  lattice NRQCD}.
\newblock {\em JHEP}, 03:084, 2013.

\bibitem{Aarts:2011sm}
G.~Aarts, C.~Allton, S.~Kim, M.P. Lombardo, M.B. Oktay, S.M. Ryan, D.K.
  Sinclair, and J.I. Skullerud.
\newblock {What happens to the Upsilon and eta\_b in the quark-gluon plasma?
  Bottomonium spectral functions from lattice QCD}.
\newblock {\em JHEP}, 11:103, 2011.

\bibitem{Aarts:2010ek}
G.~Aarts, S.~Kim, M.P. Lombardo, M.B. Oktay, S.M. Ryan, D.K. Sinclair, and
  J.-I. Skullerud.
\newblock {Bottomonium above deconfinement in lattice nonrelativistic QCD}.
\newblock {\em Phys. Rev. Lett.}, 106:061602, 2011.

\bibitem{Kim:2014iga}
Seyong Kim, Peter Petreczky, and Alexander Rothkopf.
\newblock {Lattice NRQCD study of S- and P-wave bottomonium states in a thermal
  medium with $N_f=2+1$ light flavors}.
\newblock {\em Phys. Rev. D}, 91:054511, 2015.

\bibitem{Kim:2018yhk}
Seyong Kim, Peter Petreczky, and Alexander Rothkopf.
\newblock {Quarkonium in-medium properties from realistic lattice NRQCD}.
\newblock {\em JHEP}, 11:088, 2018.

\bibitem{Larsen:2019zqv}
Rasmus Larsen, Stefan Meinel, Swagato Mukherjee, and Peter Petreczky.
\newblock {Excited bottomonia in quark-gluon plasma from lattice QCD}.
\newblock {\em Phys. Lett. B}, 800:135119, 2020.

\bibitem{Larsen:2019bwy}
Rasmus Larsen, Stefan Meinel, Swagato Mukherjee, and Peter Petreczky.
\newblock {Thermal broadening of bottomonia: Lattice nonrelativistic QCD with
  extended operators}.
\newblock {\em Phys. Rev. D}, 100(7):074506, 2019.

\bibitem{Lafferty:2019jpr}
David Lafferty and Alexander Rothkopf.
\newblock {Improved Gauss law model and in-medium heavy quarkonium at finite
  density and velocity}.
\newblock {\em Phys. Rev. D}, 101(5):056010, 2020.

\bibitem{Alessandro:2006ju}
B.~Alessandro et~al.
\newblock {psi-prime production in Pb-Pb collisions at 158-GeV/nucleon}.
\newblock {\em Eur. Phys. J. C}, 49:559--567, 2007.

\bibitem{Adam:2015isa}
Jaroslav Adam et~al.
\newblock {Differential studies of inclusive J/\ensuremath{\psi} and
  \ensuremath{\psi}(2S) production at forward rapidity in Pb-Pb collisions at $
  \sqrt{s_{\mathrm{NN}}}=2.76 $ TeV}.
\newblock {\em JHEP}, 05:179, 2016.

\bibitem{Khachatryan:2014bva}
Vardan Khachatryan et~al.
\newblock {Measurement of Prompt $\psi(2S) \to J/\psi$ Yield Ratios in Pb-Pb
  and $p-p$ Collisions at $\sqrt {s_{NN}}=$ 2.76 TeV}.
\newblock {\em Phys. Rev. Lett.}, 113(26):262301, 2014.

\bibitem{Sirunyan:2016znt}
Albert~M Sirunyan et~al.
\newblock {Relative Modification of Prompt \ensuremath{\psi}(2S) and
  J/\ensuremath{\psi} Yields from pp to PbPb Collisions at $\sqrt{s_{NN}}=5.02$
  TeV}.
\newblock {\em Phys. Rev. Lett.}, 118(16):162301, 2017.

\bibitem{Andronic:2017pug}
Anton Andronic, Peter Braun-Munzinger, Krzysztof Redlich, and Johanna Stachel.
\newblock {Decoding the phase structure of QCD via particle production at high
  energy}.
\newblock {\em Nature}, 561(7723):321--330, 2018.

\bibitem{Drees:2017zcb}
Axel Drees.
\newblock {Relative Yields and Nuclear Modification of $\psi$' to J /$\psi$
  mesons in p+p, p($^3$He)+A Collisions at $\sqrt {s_{NN}}$ = 200 GeV ,
  measured in PHENIX}.
\newblock {\em Nucl. Part. Phys. Proc.}, 289-290:417--420, 2017.

\bibitem{Burnier:2015tda}
Yannis Burnier, Olaf Kaczmarek, and Alexander Rothkopf.
\newblock {Quarkonium at finite temperature: Towards realistic phenomenology
  from first principles}.
\newblock {\em JHEP}, 12:101, 2015.

\bibitem{Burnier:2014ssa}
Yannis Burnier, Olaf Kaczmarek, and Alexander Rothkopf.
\newblock {Static quark-antiquark potential in the quark-gluon plasma from
  lattice QCD}.
\newblock {\em Phys. Rev. Lett.}, 114(8):082001, 2015.

\bibitem{Bazavov:2009bb}
A.~Bazavov et~al.
\newblock {Nonperturbative QCD Simulations with 2+1 Flavors of Improved
  Staggered Quarks}.
\newblock {\em Rev. Mod. Phys.}, 82:1349--1417, 2010.

\bibitem{Philipsen:2019rjq}
Owe Philipsen.
\newblock {Constraining the QCD phase diagram at finite temperature and
  density}.
\newblock {\em PoS}, LATTICE2019:273, 2019.

\bibitem{Pelissetto:2013hqa}
Andrea Pelissetto and Ettore Vicari.
\newblock {Relevance of the axial anomaly at the finite-temperature chiral
  transition in QCD}.
\newblock {\em Phys. Rev. D}, 88(10):105018, 2013.

\bibitem{Karsch:2001nf}
F~Karsch, E~Laermann, and Ch~Schmidt.
\newblock {The Chiral critical point in three-flavor QCD}.
\newblock {\em Phys. Lett. B}, 520:41--49, 2001.

\bibitem{deForcrand:2003vyj}
Philippe de~Forcrand and Owe Philipsen.
\newblock {The QCD phase diagram for three degenerate flavors and small baryon
  density}.
\newblock {\em Nucl. Phys. B}, 673:170--186, 2003.

\bibitem{Jin:2014hea}
Xiao-Yong Jin, Yoshinobu Kuramashi, Yoshifumi Nakamura, Shinji Takeda, and
  Akira Ukawa.
\newblock {Critical endpoint of the finite temperature phase transition for
  three flavor QCD}.
\newblock {\em Phys. Rev. D}, 91(1):014508, 2015.

\bibitem{Bazavov:2017xul}
A.~Bazavov, H.~T. Ding, P.~Hegde, F.~Karsch, E.~Laermann, Swagato Mukherjee,
  P.~Petreczky, and C.~Schmidt.
\newblock {Chiral phase structure of three flavor QCD at vanishing baryon
  number density}.
\newblock {\em Phys. Rev. D}, 95(7):074505, 2017.

\bibitem{Cuteri:2017gci}
Francesca Cuteri, Owe Philipsen, and Alessandro Sciarra.
\newblock {QCD chiral phase transition from noninteger numbers of flavors}.
\newblock {\em Phys. Rev. D}, 97(11):114511, 2018.

\bibitem{Philipsen:2016hkv}
Owe Philipsen and Christopher Pinke.
\newblock {The $N_f=2$ QCD chiral phase transition with Wilson fermions at zero
  and imaginary chemical potential}.
\newblock {\em Phys. Rev. D}, 93(11):114507, 2016.

\bibitem{Clarke:2020htu}
David~Anthony Clarke, Olaf Kaczmarek, Frithjof Karsch, Anirban Lahiri, and
  Mugdha Sarkar.
\newblock {Sensitivity of the Polyakov loop and related observables to chiral
  symmetry restoration}.
\newblock 8 2020.

\bibitem{Ding:2019prx}
H.T. Ding et~al.
\newblock {Chiral Phase Transition Temperature in ( 2+1 )-Flavor QCD}.
\newblock {\em Phys. Rev. Lett.}, 123(6):062002, 2019.

\bibitem{Ding:2019fzc}
Heng-Tong Ding, Prasad Hegde, Olaf Kaczmarek, Frithjof Karsch, Anirban Lahiri,
  Sheng-Tai Li, Swagato Mukherjee, and Peter Petreczky.
\newblock {Chiral phase transition in (2 + 1)-flavor QCD}.
\newblock {\em PoS}, LATTICE2018:171, 2019.

\bibitem{Ding:2020xlj}
H.-T. Ding, S.-T. Li, Swagato Mukherjee, A.~Tomiya, X.-D. Wang, and Y.~Zhang.
\newblock {Correlated Dirac eigenvalues and axial anomaly in chiral symmetric
  QCD}.
\newblock 10 2020.

\bibitem{Kuramashi:2020meg}
Yoshinobu Kuramashi, Yoshifumi Nakamura, Hiroshi Ohno, and Shinji Takeda.
\newblock {Nature of the phase transition for finite temperature $N_{\mathrm
  f}=3$ QCD with nonperturbatively O($a$) improved Wilson fermions at
  $N_{\mathrm t}=12$}.
\newblock {\em Phys. Rev. D}, 101(5):054509, 2020.

\bibitem{Philipsen:2014rpa}
Owe Philipsen and Christopher Pinke.
\newblock {Nature of the Roberge-Weiss transition in $N_f=2$ QCD with Wilson
  fermions}.
\newblock {\em Phys. Rev. D}, 89(9):094504, 2014.

\bibitem{deForcrand:2010he}
Philippe de~Forcrand and Owe Philipsen.
\newblock {Constraining the QCD phase diagram by tricritical lines at imaginary
  chemical potential}.
\newblock {\em Phys. Rev. Lett.}, 105:152001, 2010.

\bibitem{Bonati:2010gi}
Claudio Bonati, Guido Cossu, Massimo D'Elia, and Francesco Sanfilippo.
\newblock {The Roberge-Weiss endpoint in $N_f = 2$ QCD}.
\newblock {\em Phys. Rev. D}, 83:054505, 2011.

\bibitem{deForcrand:2017cgb}
Philippe de~Forcrand and Massimo D'Elia.
\newblock {Continuum limit and universality of the Columbia plot}.
\newblock {\em PoS}, LATTICE2016:081, 2017.

\bibitem{Karsch:2000kv}
F.~Karsch, E.~Laermann, and A.~Peikert.
\newblock {Quark mass and flavor dependence of the QCD phase transition}.
\newblock {\em Nucl. Phys. B}, 605:579--599, 2001.

\bibitem{Iwasaki:1996zt}
Y.~Iwasaki, K.~Kanaya, S.~Kaya, S.~Sakai, and T.~Yoshie.
\newblock {Finite temperature transitions in lattice QCD with Wilson quarks:
  Chiral transitions and the influence of the strange quark}.
\newblock {\em Phys. Rev. D}, 54:7010--7031, 1996.

\bibitem{Varnhorst:2015lea}
Lukas Varnhorst.
\newblock {The $N_f$=3 critical endpoint with smeared staggered quarks}.
\newblock {\em PoS}, LATTICE2014:193, 2015.

\bibitem{deForcrand:2007rq}
Philippe de~Forcrand, Seyong Kim, and Owe Philipsen.
\newblock {A QCD chiral critical point at small chemical potential: Is it there
  or not?}
\newblock {\em PoS}, LATTICE2007:178, 2007.

\bibitem{Karsch:2003va}
F.~Karsch, C.R. Allton, S.~Ejiri, S.J. Hands, O.~Kaczmarek, E.~Laermann, and
  C.~Schmidt.
\newblock {Where is the chiral critical point in three flavor QCD?}
\newblock {\em Nucl. Phys. B Proc. Suppl.}, 129:614--616, 2004.

\bibitem{Endrodi:2007gc}
G.~Endrodi, Z.~Fodor, S.D. Katz, and K.K. Szabo.
\newblock {The Nature of the finite temperature QCD transition as a function of
  the quark masses}.
\newblock {\em PoS}, LATTICE2007:182, 2007.

\bibitem{Ding:2011du}
H.-T. Ding, A.~Bazavov, P.~Hegde, F.~Karsch, S.~Mukherjee, and P.~Petreczky.
\newblock {Exploring phase diagram of $N_f=3$ QCD at $\mu=0$ with HISQ
  fermions}.
\newblock {\em PoS}, LATTICE2011:191, 2011.

\bibitem{Cuteri:2020yke}
Francesca Cuteri, Owe Philipsen, Alena Sch\"on, and Alessandro Sciarra.
\newblock {The deconfinement critical point of lattice QCD with $N_{\rm f}=2$
  Wilson fermions}.
\newblock 9 2020.

\bibitem{Ejiri:2019csa}
Shinji Ejiri, Shota Itagaki, Ryo Iwami, Kazuyuki Kanaya, Masakiyo Kitazawa,
  Atsushi Kiyohara, Mizuki Shirogane, and Takashi Umeda.
\newblock {End point of the first-order phase transition of QCD in the heavy
  quark region by reweighting from quenched QCD}.
\newblock {\em Phys. Rev. D}, 101(5):054505, 2020.

\bibitem{Kharzeev:2007jp}
Dmitri~E. Kharzeev, Larry~D. McLerran, and Harmen~J. Warringa.
\newblock {The Effects of topological charge change in heavy ion collisions:
  'Event by event P and CP violation'}.
\newblock {\em Nucl. Phys. A}, 803:227--253, 2008.

\bibitem{Skokov:2009qp}
V.~Skokov, A.Yu. Illarionov, and V.~Toneev.
\newblock {Estimate of the magnetic field strength in heavy-ion collisions}.
\newblock {\em Int. J. Mod. Phys. A}, 24:5925--5932, 2009.

\bibitem{Deng:2012pc}
Wei-Tian Deng and Xu-Guang Huang.
\newblock {Event-by-event generation of electromagnetic fields in heavy-ion
  collisions}.
\newblock {\em Phys. Rev. C}, 85:044907, 2012.

\bibitem{DElia:2010abb}
Massimo D'Elia, Swagato Mukherjee, and Francesco Sanfilippo.
\newblock {QCD Phase Transition in a Strong Magnetic Background}.
\newblock {\em Phys. Rev. D}, 82:051501, 2010.

\bibitem{Shovkovy:2012zn}
Igor~A. Shovkovy.
\newblock {\em {Magnetic Catalysis: A Review}}, volume 871, pages 13--49.
\newblock 2013.

\bibitem{Ding:2020hxw}
H.-T. Ding, S.-T. Li, A.~Tomiya, X.-D. Wang, and Y.~Zhang.
\newblock {Chiral properties of (2+1)-flavor QCD in strong magnetic fields at
  zero temperature}.
\newblock 8 2020.

\bibitem{Ding:2020inp}
Heng-Tong Ding, Christian Schmidt, Akio Tomiya, and Xiao-Dan Wang.
\newblock {Chiral phase structure of three flavor QCD in a background magnetic
  field}.
\newblock 6 2020.

\bibitem{Bali:2011qj}
G.S. Bali, F.~Bruckmann, G.~Endrodi, Z.~Fodor, S.D. Katz, S.~Krieg, A.~Schafer,
  and K.K. Szabo.
\newblock {The QCD phase diagram for external magnetic fields}.
\newblock {\em JHEP}, 02:044, 2012.

\bibitem{Bali:2012zg}
G.S. Bali, F.~Bruckmann, G.~Endrodi, Z.~Fodor, S.D. Katz, and A.~Schafer.
\newblock {QCD quark condensate in external magnetic fields}.
\newblock {\em Phys. Rev. D}, 86:071502, 2012.

\bibitem{Ilgenfritz:2013ara}
E.~M. Ilgenfritz, M.~Muller-Preussker, B.~Petersson, and A.~Schreiber.
\newblock {Magnetic catalysis (and inverse catalysis) at finite temperature in
  two-color lattice QCD}.
\newblock {\em Phys. Rev. D}, 89(5):054512, 2014.

\bibitem{Bornyakov:2013eya}
V.G. Bornyakov, P.V. Buividovich, N.~Cundy, O.A. Kochetkov, and A.~Sch\"afer.
\newblock {Deconfinement transition in two-flavor lattice QCD with dynamical
  overlap fermions in an external magnetic field}.
\newblock {\em Phys. Rev. D}, 90(3):034501, 2014.

\bibitem{Bali:2014kia}
G.S. Bali, F.~Bruckmann, G.~Endr\"odi, S.D. Katz, and A.~Sch\"afer.
\newblock {The QCD equation of state in background magnetic fields}.
\newblock {\em JHEP}, 08:177, 2014.

\bibitem{Tomiya:2019nym}
Akio Tomiya, Heng-Tong Ding, Xiao-Dan Wang, Yu~Zhang, Swagato Mukherjee, and
  Christian Schmidt.
\newblock {Phase structure of three flavor QCD in external magnetic fields
  using HISQ fermions}.
\newblock {\em PoS}, LATTICE2018:163, 2019.

\bibitem{DElia:2011koc}
Massimo D'Elia and Francesco Negro.
\newblock {Chiral Properties of Strong Interactions in a Magnetic Background}.
\newblock {\em Phys. Rev. D}, 83:114028, 2011.

\bibitem{Andersen:2014xxa}
Jens~O. Andersen, William~R. Naylor, and Anders Tranberg.
\newblock {Phase diagram of QCD in a magnetic field: A review}.
\newblock {\em Rev. Mod. Phys.}, 88:025001, 2016.

\bibitem{Kojo:2012js}
Toru Kojo and Nan Su.
\newblock {The quark mass gap in a magnetic field}.
\newblock {\em Phys. Lett. B}, 720:192--197, 2013.

\bibitem{Bruckmann:2013oba}
Falk Bruckmann, Gergely Endrodi, and Tamas~G. Kovacs.
\newblock {Inverse magnetic catalysis and the Polyakov loop}.
\newblock {\em JHEP}, 04:112, 2013.

\bibitem{Fukushima:2012kc}
Kenji Fukushima and Yoshimasa Hidaka.
\newblock {Magnetic Catalysis Versus Magnetic Inhibition}.
\newblock {\em Phys. Rev. Lett.}, 110(3):031601, 2013.

\bibitem{Ferreira:2014kpa}
M.~Ferreira, P.~Costa, O.~Louren\c{c}o, T.~Frederico, and C.~Provid\^encia.
\newblock {Inverse magnetic catalysis in the (2+1)-flavor Nambu-Jona-Lasinio
  and Polyakov-Nambu-Jona-Lasinio models}.
\newblock {\em Phys. Rev. D}, 89(11):116011, 2014.

\bibitem{Yu:2014sla}
Lang Yu, Hao Liu, and Mei Huang.
\newblock {Spontaneous generation of local CP violation and inverse magnetic
  catalysis}.
\newblock {\em Phys. Rev. D}, 90(7):074009, 2014.

\bibitem{Feng:2015qpi}
Bo~Feng, Defu Hou, Hai-cang Ren, and Ping-ping Wu.
\newblock {Bose-Einstein Condensation of Bound Pairs of Relativistic Fermions
  in a Magnetic Field}.
\newblock {\em Phys. Rev. D}, 93(8):085019, 2016.

\bibitem{Li:2019nzj}
Xiang Li, Wei-Jie Fu, and Yu-Xin Liu.
\newblock {Thermodynamics of 2+1 Flavor Polyakov-Loop Quark-Meson Model under
  External Magnetic Field}.
\newblock {\em Phys. Rev. D}, 99(7):074029, 2019.

\bibitem{Mao:2016lsr}
Shijun Mao.
\newblock {From inverse to delayed magnetic catalysis in a strong magnetic
  field}.
\newblock {\em Phys. Rev. D}, 94(3):036007, 2016.

\bibitem{Gursoy:2016ofp}
Umut G\"ursoy, Ioannis Iatrakis, Matti J\"arvinen, and Govert Nijs.
\newblock {Inverse Magnetic Catalysis from improved Holographic QCD in the
  Veneziano limit}.
\newblock {\em JHEP}, 03:053, 2017.

\bibitem{Xu:2020yag}
Kun Xu, Jingyi Chao, and Mei Huang.
\newblock {Spin polarization inducing diamagnetism, inverse magnetic catalysis
  and saturation behavior of charged pion spectra}.
\newblock 7 2020.

\bibitem{DElia:2018xwo}
Massimo D'Elia, Floriano Manigrasso, Francesco Negro, and Francesco Sanfilippo.
\newblock {QCD phase diagram in a magnetic background for different values of
  the pion mass}.
\newblock {\em Phys. Rev. D}, 98(5):054509, 2018.

\bibitem{Endrodi:2019zrl}
Gergely Endrodi, Matteo Giordano, Sandor~D. Katz, T.G. Kov\'acs, and Ferenc
  Pittler.
\newblock {Magnetic catalysis and inverse catalysis for heavy pions}.
\newblock {\em JHEP}, 07:007, 2019.

\bibitem{Bonati:2016kxj}
Claudio Bonati, Massimo D'Elia, Marco Mariti, Michele Mesiti, Francesco Negro,
  Andrea Rucci, and Francesco Sanfilippo.
\newblock {Magnetic field effects on the static quark potential at zero and
  finite temperature}.
\newblock {\em Phys. Rev. D}, 94(9):094007, 2016.

\bibitem{GellMann:1968rz}
Murray Gell-Mann, R.J. Oakes, and B.~Renner.
\newblock {Behavior of current divergences under SU(3) x SU(3)}.
\newblock {\em Phys. Rev.}, 175:2195--2199, 1968.

\bibitem{Gasser:1984gg}
J.~Gasser and H.~Leutwyler.
\newblock {Chiral Perturbation Theory: Expansions in the Mass of the Strange
  Quark}.
\newblock {\em Nucl. Phys. B}, 250:465--516, 1985.

\bibitem{Ahmad:2020jzn}
Aftab Ahmad, Adnan Bashir, Marco~A. Bedolla, and J.J. Cobos-Mart\'\i{}nez.
\newblock {Color, Flavor, Temperature and Magnetic Field Dependence of QCD
  Phase Diagram: Magnetic Catalysis and its Inverse}.
\newblock 8 2020.

\bibitem{Braguta:2019yci}
V.V. Braguta, M.N. Chernodub, A.~Yu Kotov, A.V. Molochkov, and A.A. Nikolaev.
\newblock {Finite-density QCD transition in a magnetic background field}.
\newblock {\em Phys. Rev. D}, 100(11):114503, 2019.

\end{thebibliography}
\bibliographystyle{unsrt}

\end{document}